\documentclass[twocolumn]{emulateapj}

\shorttitle{T Tauri stars in Taurus Observed with {\it Spitzer} IRS}
\shortauthors{Furlan et al.}
\submitted{To appear in ApJS, 2006 August}

\begin{document}

\title{A Survey and Analysis of {\it Spitzer} Infrared Spectrograph Spectra \\
of T Tauri stars in Taurus}


\author{E. Furlan\altaffilmark{1}, L. Hartmann\altaffilmark{2},
N. Calvet\altaffilmark{2}, P. D'Alessio\altaffilmark{3}, 
R. Franco-Hern{\'a}ndez\altaffilmark{3}, 
W. J. Forrest\altaffilmark{4}, \\ D. M. Watson\altaffilmark{4}, 
K. I. Uchida\altaffilmark{1}, B. Sargent\altaffilmark{4}, 
J. D. Green\altaffilmark{4}, L. D. Keller\altaffilmark{5},  
T. L. Herter\altaffilmark{1}
}

\altaffiltext{1}{Center for Radiophysics and Space Research, 208 Space Sciences Building,
Cornell University, Ithaca, NY 14853; furlan@astro.cornell.edu, kuchida@astro.cornell.edu,
tlh10@cornell.edu}
\altaffiltext{2}{Department of Astronomy, The University of Michigan, 500 Church St.,
830 Dennison Building, Ann Arbor, MI 48109; ncalvet@umich.edu, lhartm@umich.edu}
\altaffiltext{3} {Centro de Radioastronom{\'\i}a y Astrof{\'\i}sica, UNAM, Apartado Postal 
3-72 (Xangari), 58089 Morelia, Michoac\'an, M\'exico; p.dalessio@astrosmo.unam.mx}
\altaffiltext{4}{Department of Physics and Astronomy, University of Rochester, Rochester, 
NY 14627; forrest@pas.rochester.edu, dmw@pas.rochester.edu, bsargent@pas.rochester.edu,
joel@pas.rochester.edu}
\altaffiltext{5}{Department of Physics, Ithaca College, Ithaca, NY 14850; lkeller@ithaca.edu}


\begin{abstract}
We present mid-infrared spectra of T Tauri stars in the Taurus star-forming 
region obtained with the {\it Spitzer} Infrared Spectrograph (IRS). For the first time, 
the 5--36 $\mu$m spectra of a large sample of T Tauri stars belonging to the 
same star-forming region is studied, revealing details of the mid-infrared excess
due to dust in circumstellar disks. We analyze common features and differences
in the mid-IR spectra based on disk structure, dust grain properties, and the presence 
of companions. Our analysis encompasses spectral energy distributions from the
optical to the far-infrared, a morphological sequence based on the IRS spectra,
and spectral indices in IRS wave bands representative of continuum emission. 
By comparing the observed spectra to a grid of accretion disk models, we infer 
some basic disk properties for our sample of T Tauri stars, and find additional 
evidence for dust settling. 
\end{abstract}


\section{Introduction}

The Taurus-Auriga star-forming region is a relatively nearby 
\citep[140 pc;][]{kenyon94b,bertout99} molecular cloud complex, which is 
characterized by lower extinction and more
isolated, low-mass star formation than other star-forming regions at comparable
distances, like Ophiuchus \citep[][hereafter KH95]{kenyon95}. It is therefore fairly well-studied from
wavelengths ranging from the X-rays to the radio and used to test theoretical predictions
of the star formation process. Given the young age of the Taurus complex (1--2 Myr),
many of its pre-main-sequence stars are still surrounded by material left over from
their formation process and accrete gas and dust from circumstellar disks. 

Dust grains in these disks are heated by irradiation from the star and from the 
accretion shocks at the stellar surface, as well as by local viscous dissipation
\citep{dalessio98}, and emit excess emission, i.e.\ at flux levels above those 
expected from a stellar photosphere, at infrared wavelengths. Most disks have
flared disk surfaces, i.e. the disk thickness increases more steeply than just
linearly with radius; this flaring allows disks to absorb more radiation from
the central source than a flat, thin disk, and thus to become hotter
\citep[see, e.g.,][]{kenyon87}.

Over time, accretion rates decrease, and, while the disk material is depleted, some of the
dust and gas could be built into planets. Young, low-mass ($\lesssim 2 M_{\odot}$) 
stars surrounded by these accretion disks are referred to as classical T Tauri 
stars (CTTSs). Once accretion has virtually stopped and disks are either passive, 
i.e. only reprocessing stellar radiation, or already dissipated, a young star is called a 
weak-lined T Tauri star (WTTS). The distinction between CTTSs and WTTSs is based
on the equivalent width of their H$\alpha$ emission line (EW[H$\alpha$]):
CTTSs have broad and asymmetric H$\alpha$ line profiles, which are generated 
in the magnetospheric accretion flows \citep{muzerolle03a}, resulting in an 
EW[H$\alpha$] larger than 10 {\AA}, while WTTSs have an EW[H$\alpha$] 
of less than 10 {\AA}; in WTTSs, strong surface magnetic fields provide 
the energy for strong X-ray, UV, and H$\alpha$ emission.

Based on an evolutionary sequence derived from the shape of the spectral
energy distribution (SED) in the infrared \citep{lada84, lada87}, 
young, low-mass stars are divided into three classes. Class I objects have rising
SEDs over the infrared spectral range, while Class II objects have either flat or 
somewhat decreasing SEDs; Class III objects have little infrared excess, and their 
SEDs can be understood as reddened stellar photospheres for wavelengths 
$\lesssim$ 10 $\mu$m. Thus, Class II objects are characterized by the presence 
of mid-infrared excess emission due to dust in circumstellar disks, while Class III 
objects have, for the most part, already dissipated their disks, and therefore their 
infrared flux is close to photospheric levels.
Classical T Tauri stars fall into the Class II category; weak-lined T Tauri stars
can be considered Class II or Class III objects, depending on the presence of disks
around them. When a young star has almost completely dissipated its circumstellar 
material, it is usually classified as a Class III object. Therefore, all Class III objects
are WTTSs, but not all WTTSs are Class III objects, as defined by their infrared SEDs.

Mid-infrared spectra of T Tauri stars (mostly CTTSs) reveal the details of the 
thermal emission by dust grains in the inner regions of circumstellar disks, 
from a few tenths to several AU from the central object. By deriving the 
structure and composition of the dust in protoplanetary disks, we can study 
the conditions and processes leading to planetary formation. In particular, 
dust growth and settling are seen as key steps in the formation of larger bodies.  

The {\it} Spitzer Space Telescope \citep{werner04} enables us, for the first time, 
to obtain spectra from 5 to 36 $\mu$m of T Tauri stars covering a variety of flux 
densities, from tens of mJy to tens of Jy. As part of a larger Infrared 
Spectrograph\footnote{The IRS was a collaborative venture between Cornell 
University and Ball Aerospace Corporation funded by NASA through the Jet 
Propulsion Laboratory and the Ames Research Center.} \citep[IRS;][]{houck04} 
guaranteed-time observing program, we observed 150 young stellar objects in the 
Taurus-Auriga star-forming region, ranging in evolutionary state from the protostellar,
or Class I, stage, when a forming star is still enshrouded by an envelope, to the 
Class II and III stages, when disk material is accreted and dissipated. 

Our Taurus sample is largely based on the sample studied by 
KH95, who compiled and analyzed photometric 
data of known young stellar objects in the Taurus-Auriga star-forming region. 
Here we present 85 IRS spectra of Class II objects and 26 spectra of Class III 
objects; the remaining spectra of the Class I objects in our sample will be presented
in a subsequent paper. We analyze the spectral energy distributions and
IRS spectra of Class II and III objects in terms of common features and
peculiarities arising from their disk structure, the properties of dust grains
embedded within, and their surrounding environment. The large-scale
environment has little impact on objects in Taurus, unlike young stars in 
OB associations or young clusters, where the cluster members are subject 
to the influence of high-energy radiation and powerful winds from the OB stars.

In $\S$ \ref{obs_data_reduction} we introduce our {\it Spitzer} IRS
observations and data reduction; in $\S$ \ref{sample} we present spectral
energy distributions, from the optical to the far-infrared, of the Class II and III
objects in our sample, as well as a morphological sequence of the IRS spectra 
of our 85 Class II objects; in $\S$ \ref{analysis} we compare our data with
a set of accretion disk models to infer some disk properties, present a new median
of mid-infrared fluxes of Class II objects in Taurus, and estimate upper limits for
dust masses around our Class III objects. We discuss some of our results in 
$\S$ \ref{discuss}, and give our conclusions in $\S$ \ref{conclude}.

\section{Observations and Data Reduction}
\label{obs_data_reduction}

All but three of our Taurus targets were observed with the IRS on {\it Spitzer} 
during IRS observing campaigns 3 and 4, from 2004 February 6 to 8, and from 
2004 February 27 to March 5. The remaining three targets (FQ Tau, GK Tau, and
V830 Tau) were observed during IRS campaign 12, from 2004 August 30 to 31. 
For all targets the full mid-IR spectrum from 5 to 40 $\mu$m was obtained by 
either using the two low-resolution IRS modules (Short-Low [SL] and Long-Low [LL], 
5.2--14 $\mu$m and 14--38 $\mu$m, respectively, $\lambda$/$\Delta\lambda$ 
$\sim$ 90) or the SL module and the two high-resolution modules (Short-High [SH] 
and Long-High [LH], 10--19 $\mu$m and 19--37 $\mu$m, respectively, 
$\lambda$/$\Delta\lambda$ $\sim$ 600). The high-resolution modules 
were usually only used for observations of bright ($\gtrsim$ a few Jy) targets. 
While some targets were observed in IRS staring mode, most of them were 
observed in mapping mode; with the latter, a 2$\times$3 step map on the 
target was carried out, with three steps separated by three-quarters (for SL) or half 
(for the other modules) of the slit width in the dispersion direction and two steps 
of one third of a slit length in the spatial direction. For staring mode observations, 
the target was placed subsequently in two nod positions along the spatial 
direction of the slit, at 1/3 and 2/3 of the slit length, respectively.
 
As a first step in the data reduction process, we fixed all bad pixels in the arrays 
(identified by the pixel masks, and including the so-called ``rogue pixels'' 
in SH and LH) using a simple interpolation over neighboring, good pixels.
We then extracted and calibrated our spectra using the SMART software tool
\citep{higdon04}. When taking low-resolution spectra, the target is first placed
on the slit yielding the 2nd order spectrum and then on the slit yielding the 1st
order spectrum; each time both orders are observed. We subtracted the sky
background by using observations taken at the same nod position, but
in different orders. The spectra were extracted using a variable-width
column extraction, which varies with the width of the point-spread function,
and calibrated with $\alpha$ Lac (A1 V) and its template spectrum 
\citep{cohen03}. For the high-resolution spectra, the spectra of
target and calibrator, $\xi$ Dra (K2 III), were extracted using a full slit 
extraction. 

Even though we were not able to subtract any sky background
from our high-resolution spectra, the contribution from any background
emission was only noticeable in our LH spectra. To correct for this contribution,
we applied a scalar correction to each LH spectrum to achieve a smooth 
transition at 19 $\mu$m, where SH and LH meet. Our SH spectra usually
matched the flux level of our SL spectra in the overlap region between SL 
and SH (10-14 $\mu$m); given the smaller aperture and shorter wavelength
coverage of SH, any background contribution is expected to be negligible.
All of our high-resolution data presented in this paper were rebinned to 
a resolution of $\lambda$/$\Delta\lambda$ = 200, and the SH data
were truncated below 13 $\mu$m. The rebinning of our SH and LH spectra
resulted in a clearer representation of the shape of the continuum and in a 
more homogeneous set of spectra, independent of the modules used for 
$\lambda > $ 14 $\mu$m.

For mapping mode observations, we extracted and calibrated all map positions,
but the final spectrum was usually obtained by averaging the two observations
in the map's center. For staring mode observations, the final spectrum is an 
average of the two nod positions. 

We estimate our absolute spectrophotometric accuracy to be of the order 
of 10\%, in some cases as good as 5\%. The scatter of neighboring flux 
values determines our relative accuracy and thus our ability to reliably
identify spectral features. This relative accuracy depends not only on the 
brightness of the object, but also on the modules used: while in SL, and 
usually also in LL2 (14-21 $\mu$m) and in SH, spectral features above 
the noise level can be considered real, the LL1 (20-36 $\mu$m) and 
especially the LH spectra are often dominated by artifacts due to unresolved 
calibration issues, which makes the identification of real features difficult. 
Therefore, spectral features beyond 20 $\mu$m need careful checking 
before they are identified. 
In addition, the wavelength region between 13.5 and 14.5 $\mu$m, 
where SL and LL meet, is affected by larger uncertainty in the calibration 
at the order edges and by possible flux mismatches between SL and LL.
In some cases a spectral feature seems to appear in this wavelength region; 
at this point, we are not sure whether this is a real feature or just an artifact 
introduced by the order edges. 

It should be noted that many of our targets are members of close multiple 
systems, ranging in separation from tenths of an arcsecond to several
arcseconds (see Tables \ref{tab1} and \ref{tab3}). Given that the 
narrowest IRS slit width is 3{\farcs}6 (SL), several targets could not be 
resolved by our observations. For these targets, we show the combined 
spectrum of the unresolved components; however, in many cases one 
component clearly dominates the mid-IR spectrum. For a small group of 
targets, a companion entered only partially in the wider slits (4{\farcs}7 
for SH, 10{\farcs}6 for LL, 11{\farcs}1 for LH) and more fully at the
longer wavelength ends of these modules. For these objects we made 
sure that the contribution of the companion was small before interpreting 
the object's spectrum, or we excluded the ``contaminated'' part of the 
spectrum from the analysis.

A few objects presented problems in their reduction; they will be listed below. \\
{\it IT Tau.} This classical T Tauri star is a 2{\farcs}4 binary. Both components
entered the SL and LL slits, even though some flux of the B component was lost
at longer SL wavelengths due to the larger size of the point-spread function. 
The A component should dominate the short-wavelength part of the IRS spectrum;
however, it is likely that in LL ($>$ 14 $\mu$m) A and B contribute about equally, 
since IRAC measurements \citep{hartmann05a} indicate that the contribution of 
the B component to the total flux of the system increases from 15\% at 3.6 $\mu$m 
to $\sim$ 30\% at 8 $\mu$m. This is also suggested by the IRS spectrum, 
where we multiplied SL by 1.25 to match it with LL, thus creating a combined 
spectrum for A and B. Due to this scaling the SL spectrum of IT Tau carries a 
large uncertainty. \\
{\it CW Tau, V773 Tau.} These two objects are part of a mispointed mapping-mode 
observation. We created continuous IRS spectra by using the brightest map positions
for each star; however, due to likely slit losses the absolute flux levels and, to a lesser 
degree, the shape of the spectra could be off. \\
{\it CX Tau, FN Tau, FO Tau.} The spectra of these objects were also affected 
by mispointed observations, and therefore their IRS spectra are less accurate. \\
{\it MHO 3.} This young star is so bright that a small part of its LL spectrum 
(from about 22 to 26 $\mu$m) was saturated. \\
{\it AB Aur, RY Tau.} These objects are also very bright and partly saturated the 
central map position in SL. We recovered an unsaturated spectrum by using an 
off-center map position for SL and scaled it to match the SH and LH observations. \\
{\it V892 Tau.} Due to the brightness of this star, SL and LH were partly 
saturated. Similarly to RY Tau, we used off-center map positions for SL and LH
and matched them to the central map observation of SH.

\section{The Sample}
\label{sample}

\subsection{Characterization and Classification}

The targets of our Taurus sample were selected from the objects analyzed by 
KH95, and from objects in the {\it IRAS} Faint Source Catalogue, selected on the
basis of their infrared colors. For the latter, we used a filter on the brightness 
and on the 12/25, 25/60, and 60/100 $\mu$m flux ratios (the first one being 
the most important one): $12/25 < 1.0$,  $25/60 < 1.2$, and $0.1 < 60/100 
<1.5$, similar to \citet{beichman92}. Sources from the Faint Source Catalogue
that do not have common names are labeled with their {\it IRAS} identifier,
e.g. F04101+3103. 
 
Spectral indices have commonly been used to classify the SEDs of protostars
and pre-main-sequence stars \citep{lada87,adams87}. The slope of the SED
from the near- to the mid-infrared (2.2\,--\,25 $\mu$m) in a log-log plot is 
used to define Class I, II, and III objects: \begin{math} 
n \equiv d \log(\lambda  F_{\lambda})/d \log (\lambda),
\end{math}
where n $>$ 0 identifies a Class I object, $ -2 < n < 0 $
a Class II object, and $ n < -2 $ a Class III object. A value of $n$ of $-3$ 
is equivalent to a stellar photosphere. 
However, this is just a simple classification scheme, since it does not take into 
account the detailed shape of the SED in this wavelength range; it describes the 
shape of the SED from 2 to 25 $\mu$m as a simple power law $\lambda  
F_{\lambda} \propto \lambda^n$. For example, some objects identified as
Class I objects might actually be Class II objects with cleared-out inner disk regions,
which are characterized by little excess emission at near-IR wavelengths and a 
large mid-IR excess, resulting in a steep slope of the near- to mid-infrared SED. 
Only mid-IR spectra reveal spectral features that can aid in the detailed evolutionary
classification of a young stellar object, like the presence of the 10-$\mu$m 
silicate emission feature, which is indicative of a flared circumstellar disk.

We separated our sample into Class II and III sources based on the existence
of an infrared excess in the IRS spectral range. This subdivision resulted in
a reclassification of some objects previously identified based on their SED slope
between 2 and 25 $\mu$m. Since the 25-$\mu$m data points were
usually derived from {\it IRAS} measurements with limited spatial resolution and lower
sensitivity, many previous classifications were more uncertain. Reclassification
was done for FP Tau, V836 Tau, VY Tau, and ZZ Tau, which were originally 
classified as Class III objects (e.g., KH95), and for DI Tau and HBC 423, which were 
considered Class II objects. Also, we reclassified a few Class II objects as Class I 
objects due to the presence of ice and silicate absorption features in their IRS spectra, 
as well as their SED shape; among these objects are CoKu Tau/1, HL Tau, and LkHa 358,
which will be included in a subsequent paper.

Accretion signatures, like broad H$\alpha$ emission and an excess in the UV or 
blue spectral range, are an indication for the presence of inner disk 
material. Combined with magnetospheric accretion models, these signatures are
used to derive mass accretion rates \citep{gullbring98, hartmann98, muzerolle03a}. 
Classical T Tauri stars in Taurus have average mass accretion rates of $\sim$ 10$^{-8}$ 
M$_{\odot}$ yr$^{-1}$; accretion rates seem to increase with the mass of the 
star and also decrease over time \citep{calvet04, muzerolle05, hartmann98}.

However, H${\alpha}$ emission is not necessarily correlated with the presence or
absence of a disk; several stars identified as WTTSs do have substantial disk masses, 
as derived from their millimeter emission \citep[see, e.g.,][]{beckwith90}, or 
have considerable mid-IR excesses, as our sample shows. This observational
fact could be explained with the presence of inner disk holes: a star with a missing
inner disk might not be accreting, and thus be classified as a WTTS, but 
still possess a prominent outer disk \citep[e.g., CoKu Tau/4;][]{dalessio05}. 
On the other hand, all stars with inner disks, as inferred from their infrared 
excesses, are accreting and thus CTTSs. 
A few CTTSs \citep[e.g., TW Hya, GM Aur;][]{jayawardhana99, uchida04, calvet05} 
have low near-IR excesses, but prominent mid-IR excesses, which could be an indication 
of grain growth in the inner disk. An extreme example is DM Tau \citep{calvet05},
which is an accreting T Tauri star but has no near-IR excess.

Many, if not most, T Tauri stars in Taurus have companions. A list of our Class II objects 
belonging to multiple systems can be found in Table \ref{tab1}; many components 
are separated by fractions of an arcsecond, and usually they form a binary or a 
hierarchical quadruple system. The multiplicity of young stars adds complexity 
when their stellar and circumstellar properties are studied. Some systems are 
surrounded by circumbinary disks or rings \citep[e.g.,][]{krist02}, often in 
addition to a circumprimary and a circumsecondary disk. These disks can be 
tidally truncated \citep{artymowicz94}; in general, disks are found well within 
or outside binary orbits. Therefore, the components of a binary system can be 
either both CTTSs or both WTTSs, especially in close ($\lesssim$ a few arcseconds) 
binary systems, or form a mixed pair \citep{prato97, duchene99b}. A circumbinary 
reservoir can replenish both the circumprimary and circumsecondary disk; for wider 
companions, the circumprimary disk is preferentially replenished, resulting in a 
stronger infrared excess for the primary component \citep{white01}. Depending
on the presence of material in its inner disk and thus its accretion signatures, 
the primary may be classified as CTTS or WTTS; eventually, both components
will evolve into Class III objects (and thus WTTSs), once all the disk material has
dispersed.

\subsection{Spectral Energy Distributions}

Our full sample of Class II and III objects is listed in Tables \ref{tab2} and
\ref{tab3}, respectively, together with the adopted spectral type,
visual extinction A$_V$, and T Tauri star type (CTTS, WTTS), taken from 
the literature \citep[mainly KH95;][]{white01}. 
Due to the variability of these sources and their multiplicity (where applicable) 
some measurements are quite uncertain. Several spectral types are not known 
or are only determined with an uncertainty of a few subclasses, and ages and 
mass accretion rates depend on the models used to derive them, which can 
cause uncertainties of an order of magnitude. When derived using the 
same assumptions, these quantities usually agree within a factor of two. 
When separating the stars of our sample into CTTSs and WTTSs, we adopted 
the \citet{white03} criteria for CTTSs, where the traditional EW[H$\alpha$] 
boundary of 10 {\AA} between WTTSs and CTTSs is adjusted for spectral type, 
with a delimiting width smaller than 10 {\AA} for spectral types earlier than 
K7 and a larger width for spectral types later than M2.5.

For each object with known spectral type, we
checked whether the A$_V$ value from the literature or the A$_V$ value we
derived from observed V-I colors yielded dereddened optical photometry 
consistent with the intrinsic colors of a main-sequence star with the same 
spectral type as the object under consideration. If the dereddened flux values
were off, we adjusted the A$_V$ values, usually by less than a factor of two, 
until a value was found that brought the observed and expected colors into 
agreement. Several A$_V$ values remain uncertain, but since for most 
Taurus targets the extinction is low, changes in A$_V$ by less than a 
factor of two will only minimally affect the shape of the IRS spectrum
and will not affect the results of our analysis.
 
Figure \ref{fig_classII} shows the SEDs of the objects in our Class II sample, 
while figure \ref{fig_classIII} displays those of our Class III objects.
For all objects with measured extinction, their fluxes have been corrected 
for reddening using the object's $A_V$ from Tables \ref{tab2} and \ref{tab3}
and Mathis's reddening law \citep{mathis90} with R$_V=3.1$. Included are 
photometric data points from the optical to the 
mid-IR mostly from KH95, the 2MASS $J$, $H$, and $K_s$ fluxes, the {\it IRAS} 
12, 25, and 60 $\mu$m fluxes, the IRAC 3.6, 4.5, 5.8, and 
8.0 $\mu$m fluxes \citep{hartmann05a}, where available, and the IRS spectrum. 

Also shown is the stellar photosphere for those objects with known spectral types; 
photospheric fluxes were derived from colors of main-sequence stars given in KH95 
and normalized at the J-band flux of each object with a spectral type later than G,
and at the V-band flux of stars with an earlier spectral type (AB Aur, V892 Tau, 
RY Tau, and SU Aur). This normalization is based on the fact that for K and early 
M stars, most of the photospheric radiation is emitted around the J-band, but for
earlier spectral types the peak emission shifts to shorter wavelengths. For spectral
types earlier than G, the U-band would be better suited for normalization, but 
since U-band measurements are often poor and very sensitive to extinction,
the V-band is a better choice. For F04101+3103, which is of A1 spectral type, 
the J-band was used for normalization, since it might have an excess in the V-band.

The {\it IRAS} data shown in the SED plots are mainly from \citet{weaver92}, who 
used co-added {\it IRAS} data to determine fluxes even for faint objects 
(as many Class III objects are), which usually fell just below the sensitivity 
limit of {\it IRAS}. As can be seen in the comparison with our IRS data, the fluxes for
most Class III objects, as determined by \citet{weaver92}, were often 
overestimated, resulting in mid-IR excesses for objects whose emission 
is actually close to photospheric over the entire infrared range. For objects not
included in \citet{weaver92}, such as objects with identifiers from the {\it IRAS} Point
Source or Faint Source Catalogues, we used the {\it IRAS} fluxes from these catalogs,
mostly from the latter one.

\subsubsection{Class II SEDs}

All Class II objects, even the ones defined as WTTSs based on the equivalent width 
of their H$\alpha$ emission line, show a clear mid-IR excess, indicative
of the presence of circumstellar disks. The majority of our Class II objects are
classical T Tauri stars, either single or in multiple systems; since they show
a pronounced accretion signature, indicated by their H$\alpha$ emission or
UV excess, it is expected that the circumstellar material in their inner disks 
generates a substantial infrared excess. 

There are only two single WTTSs in our Class II sample: CoKu Tau/4 and IQ Tau.
Their SEDs are quite different from each other; IQ Tau has a substantial excess 
from the near- to the mid-IR, indicating that at least the disk's inner regions (out 
to a few AU) are still prominent, while CoKu Tau/4 has no excess up to about
8 $\mu$m. This would indicate that WTTS disks are not all characterized 
by inner disk clearings, which is anticipated given that they are no longer 
accreting. 
However, CoKu Tau/4 might be the only true single WTTS in our sample of Class II 
objects; it is a so-called transitional disk, where the inner disk has been cleared out, 
but a prominent outer disk is still present \citep{forrest04, dalessio05}. Even 
though IQ Tau is classified as a WTTS based on its H${\alpha}$ equivalent
width \citep[7.8 {\AA};][]{herbig88},
it has a U-band excess, which possibly indicates 
that is it still accreting (\citealt{hartmann98} derived a mass accretion rate
of $\sim 3 \times 10^{-8}$ M$_{\odot}$ yr$^{-1}$, while \citealt{white01}
only derived an upper limit of $\sim 5 \times 10^{-9}$ M$_{\odot}$ 
yr$^{-1}$ but noted its red K\,--\,L color). Therefore, WTTSs with prominent 
inner disks seem to be rare, confirming previous results of a short transition 
time from the CTTS state to a diskless WTTS state \citep[e.g.,][]{simon95b}. 

When comparing the SEDs of single and of multiple T Tauri stars, there are
no obvious systematic differences. However, since the IRS spectral range is sensitive
to disk emission from a few tenths to a few AU from the star, we only expect to 
see any effects of multiplicity in close multiple systems ($\lesssim$ 0{\farcs}5, 
or 70 AU at the distance of Taurus). Disks can surround each component or 
the entire multiple system; companions will truncate both circumstellar disks
(at outer radii 0.2--0.5 times the semimajor axis of the binary) and circumbinary disks
(at inner radii 2--3 times the semimajor axis of the binary), depending on the
properties of the binary system \citep{artymowicz94}. In wider systems, inner 
disks are not influenced by the presence of the other components and are thus 
expected to have the same properties and lifetimes as disks around single stars. 

For most close multiple systems, the SED shows the combined photometry and 
spectrum of all components, and therefore the mid-IR flux could be generated 
by circumstellar dust from disks around each component or from a disk 
surrounding the multiple system. Given that most binaries consist of CTTS pairs,
both components will likely contribute to the IRS spectrum. 

By using ground-based near-IR photometry (K-band or, where 
available, L-band) that resolved both components \citep[e.g.,][]{white01}, 
we estimated each object's contribution to the IRS spectrum (see Table \ref{tab1}).
There is no clear correlation with binary separation: while in some close 
($\lesssim$ 1\arcsec) systems each component contributes roughly to equal 
amounts to the IRS spectrum (e.g., GN Tau), in some other close systems one 
component is the dominating one (e.g., V955 Tau). In the few mixed-pair binaries
of our sample, where one component is a CTTS and the other a WTTS, the primary
component is usually the CTTS and dominates the mid-IR spectrum. Therefore, 
the IRS spectrum shows emission mostly from the circumprimary disk. 
This implies that the disk around the CTTS is either preferentially replenished 
or was formed as a more massive disk; the companion ends up with a smaller disk
and a lower-mass star. This would also indicate that each system is strongly shaped by 
its initial conditions, i.e.\ the mass of each cloud fragment that formed a binary 
component, its kinematics, and whether other components formed nearby.

Almost all of the Class II objects display pronounced 10- and 20-$\mu$m 
emission features due to silicates. Usually the 10-$\mu$m feature is stronger
than the 20-$\mu$m feature, and it also varies from object to object due
to different grain composition and sizes.  A smooth 10-$\mu$m feature
centered around 9.8 $\mu$m is indicative of amorphous silicates (similar
to the interstellar medium), while a structured 10-$\mu$m feature (see,
e.g., the IRS spectrum of DH Tau) reveals the presence of crystalline silicates,
likely processed from amorphous silicates in the circumstellar environment
\citep{sargent06,forrest04}. In a separate paper,
we explore the correlations between disk properties and the structure and
strength of the 10-$\mu$m silicate feature (D. M. Watson et al. 2006, 
in preparation).

In addition to the almost ubiquitous presence of silicate emission features, 
we note the close to complete absence of PAH emission features in
our objects of spectral type later than about G1, with the possible exception
of UX Tau A (see section \ref{notes_object}). 
PAH emission is typical for the more massive counterparts of
T Tauri stars, the Herbig Ae/Be stars \citep[see, e.g.,][]{meeus01}, and
their absence in the circumstellar environment of lower mass stars is probably
a result of the weaker UV radiation field of these stars.

\subsubsection{Class III SEDs}

All Class III objects in our sample have narrow H${\alpha}$ equivalent
widths -- all of them are WTTSs -- and are not accreting within the
detection limits, indicative for dispersal of their inner disks. 
Our IRS spectra support this finding: 
almost all Class III mid-IR spectra are photospheric. Even though the
dispersal of disk material would suggest that these objects are more
evolved, and thus older, the ages of Class III objects do not seem to be 
systematically larger than those of Class II objects, which are supposedly 
in an earlier phase of evolution \citep[see, e.g.,][]{white01}.
It appears that the timescales for circumstellar disk evolution are rather 
determined by the initial conditions and the specific environment around a 
young star than set by some universal processes. Young Class III objects 
($\lesssim$ 1-2 Myr) apparently dispersed their circumstellar material 
faster than some long-lived Class II objects; multiplicity could play a role 
\citep[see, e.g.,][]{meyer97} but is likely not the only factor. 

Three of our Class III objects have nearby companions with infrared excesses
that partly entered the 10${\farcs}$6 wide LL slit but not the narrower SL
slit: DI Tau, HBC 423, and HP Tau/G2. DI Tau is separated by 15\arcsec\ from
DH Tau, which is a classical T Tauri star (see figure \ref{fig_classII} for the 
SED of DH Tau). In the case of HBC 423, V955 Tau, which is 10{\arcsec} away, 
and, to a small degree, HBC 422, which is 26{\arcsec} away, contribute to the 
flux past 14 $\mu$m (see figure \ref{fig_classII} for the SED of 
V955 Tau). HP Tau/G2 is 10{\arcsec} from HP Tau/G3; both stars contribute 
to the LL flux, and in addition light of HP Tau (see figure \ref{fig_classII}), which 
is 21{\arcsec} from HP Tau/G2, probably entered in LL, especially toward the 
longer wavelengths. 
Given that the fluxes of DI Tau, HBC 423, and HP Tau/G2 are at photospheric 
levels over the SL wavelength range (5--14 $\mu$m), but show a 
sharp increase where LL starts (at 14 $\mu$m), it is very likely that all of their 
apparent excess is caused by their companion(s). For these sources, we only show 
their SL spectra.
 
Two more Class III objects show a similar behavior to the three previously 
mentioned objects: HBC 427 and V819 Tau. These two stars do not have any 
known companions with infrared excesses; however, for each of them the 
2MASS K-band image shows a nearby source that likely contaminated the LL 
observation. V819 Tau is separated by about 10\arcsec\ from a faint source 
to the south, whose flux in the K-band is about 30 times smaller than that of
V819 Tau. HBC 427 has a ``companion'' about 15\arcsec\ to the southeast, 
which is about half as bright as HBC 427 in the K-band.  Given that either 
``companion'' did not fully enter the LL slit, but still likely contributed to the 
flux in this module due to the sharp rise in flux at the SL-LL boundary, they 
might have substantial mid-IR excesses. Since the mid-IR flux of 
those nearby sources is not known, we cannot determine which fraction of 
the infrared excess we observe is due to the target itself; 
the almost photospheric flux levels of HBC 427 and V819 Tau at wavelengths 
shorter than 14 $\mu$m indicates that the intrinsic infrared excess is likely small.

V410 X-ray 3 seems to have an excess at wavelengths longer than 20 $\mu$m, 
but since its spectrum is very faint in that wavelength region ($\sim$ 2 mJy), it 
could be an artifact of the extraction. This object is one of the lowest-mass T Tauri
stars in our sample; its spectrum was first presented in \citet{furlan05a}.

HBC 356, Hubble 4, and L1551-51 were too faint beyond 14 $\mu$m, 
so only their SL spectra could be extracted. For HBC 392, LkCa 1, and LkCa 21, 
the longer-wavelength part of LL (beyond 21 $\mu$m) was too faint, so their 
spectra include only a part of the LL spectrum.

\subsection{Notes on Individual Objects}
\label{notes_object}
 
\subsubsection{04303+2240, 04370+2559, 04385+2550, CoKu~Tau/3, FV~Tau, 
Haro~6-13, MHO~3, V410~Anon~13} 
These objects are all highly reddened T Tauri stars with an A$_V$ larger than about
5 mag (see Table \ref{tab2}); in addition, 04303+2240 and Haro 6-13 have relatively 
high mass accretion rates \citep{white04}. 
04370+2559 has a companion that is likely a substellar object \citep{itoh99,itoh02}. 
04385+2550, also known as Haro 6-33, has been considered as a Class I object by 
some authors based on its SED shape and bolometric temperature 
\citep[e.g.,][]{motte01, young03}. It is detected at sub-mm and mm 
wavelengths and is somewhat extended in the former, but not in the latter
wavelength range \citep{motte01, young03}. On the other hand, the high extinction 
could indicate that 04385+2550 is a Class II object seen edge-on.
V410 Anon 13 is a very low-mass classical T Tauri star, whose IRS spectrum was 
already studied in \citet{furlan05a}. 

The objects 04370+2559, CoKu Tau/3, and FV Tau have a CO$_2$ ice 
absorption feature at 15.2 $\mu$m in their spectra. Since the lack of other ice 
features, the presence of a prominent silicate emission feature, and the SED shape 
suggest that these objects are T Tauri stars and not protostars surrounded by an 
envelope, the CO$_2$ ice absorption feature is likely due to dense cold material 
along the line of sight to these objects. The relatively high extinction towards 
all of these objects would support the idea of molecular cloud material causing 
the CO$_2$ ice feature. 

\subsubsection{CZ~Tau~B, DG~Tau, DR~Tau, FS~Tau (Aa+Ab), HN~Tau~A, 
XZ~Tau~B} 
According to \citet{white01}, these T Tauri stars have veiled 
optical spectra, unusually red K\,--\,L colors ($>$\,1.4 mag), and, where measured,
large H$\alpha$ equivalent widths, indicative of high accretion rates. 
Therefore, their luminosities and temperatures are uncertain, and thus extinctions, 
ages, and mass accretion rates are not well determined. These stars might be 
experiencing an episodic phase of high accretion, which could be accompanied 
by a larger amount of extinction \citep{white01}. 

{\it CZ Tau.} CZ Tau has a peculiar IRS spectrum, which has a positive slope 
from 5 to 20 $\mu$m, but falls sharply beyond 20 $\mu$m. Based on their 
K- and L-band flux ratios \citep{white01}, the two components of this 
0${\farcs}$32 binary are probably contributing in equal measure to the 
IRS spectrum, but the B component is in a high-accretion state and therefore 
likely suffering from larger extinction. This could explain the shorter-wavelength 
slope of the IRS spectrum, but not the longer-wavelength drop; for the latter, 
disk truncation might be responsible. 

{\it DG Tau, DR Tau.} These two objects are the only single stars in this
sample of high-accretion-rate stars; 04303+2240 could be added to this
subgroup, too. It seems that for these objects the 
excess emission from the accretion shock strongly veils the emission of the 
photosphere, possibly dominating the emission from the ultraviolet to the optical 
\citep[see, e.g.,][]{gullbring98, gullbring00}. Since it is difficult to 
determine the extinction and the flux level of the photosphere, we did not add 
stellar photospheres in their SED plots; we also did not attempt to adjust the 
extinction, but instead adopted the values published in the literature. 

DG Tau is a ``flat-spectrum'' T Tauri star, i.e.\ it has a flat SED over the infrared
spectral range. Like other ``flat-spectrum'' sources, it is likely surrounded by an infalling
envelope \citep[e.g.,][]{calvet94}. \citet{wooden00} reported variations from
emission to absorption behavior in the 10-$\mu$m silicate feature, in addition
to changes in dust mineralogy, on the timescales of months to years. Our IRS spectrum,
taken in 2004 March, shows a very weak 10-$\mu$m silicate emission feature whose
shape indicates that crystalline silicates might be present. We note that the flux level 
of the IRS spectrum is much lower than the {\it IRAS} fluxes at 12 and 25 $\mu$m. 
This is partly an aperture effect, since the {\it IRAS} beam included flux contributions 
from DG Tau, DG Tau B (a Class I object about half as bright as DG Tau), and 
FV Tau (a Class II source, 0.25\,--\,0.5 times fainter than DG Tau), while the IRS
observed only DG Tau (with slit sizes of 3{\farcs}6, 4{\farcs}7, and 11{\farcs}1). 
Part of the discrepancy could also be caused by intrinsic variability of the target(s). 

{\it FS Tau.} FS Tau A and B, which are separated by 20\arcsec, are both surrounded by 
reflection nebulosities \citep{krist98, white01}; in our IRS observations, only 
the sub-arcsecond binary FS Tau Aa+Ab was included. Given the smaller aperture 
of the IRS, we note that the IRS fluxes are considerably lower than the {\it IRAS} 
fluxes of FS Tau. 

{\it HN Tau and XZ Tau.} In the HN Tau and XZ Tau binary systems, both 
components are classical T Tauri stars, but the one component with the high 
mass accretion rate is clearly dominating the IRS spectrum (HN Tau A and XZ 
Tau B, respectively). 

\subsubsection{AB Aur, F04101+3103, V892 Tau} 
These three objects are Herbig Ae/Be stars and therefore more massive and 
brighter than classical T Tauri stars. For AB Aur, we adopted the spectral type 
(A0) and extinction (A$_V$=0.25 mag) from \citet{dewarf03}. F04101+3103
was assigned an A1 spectral type by \citet{kenyon90}; in its SED plot, we
used the $BVRI$ photometry of \citet{iyengar97}.

The spectral type of V892 Tau published in the literature varies from B9, with 
an extinction of A$_V$=8.85 mag, to A6 and an extinction of 4 mag 
\citep[e.g.,][]{strom94,kenyon95}. Here, we adopted the earlier 
spectral type and higher extinction (adjusted somewhat to A$_V$=8.0), 
as it yields a more realistic SED. The high extinction is also supported by 
evidence of a nearly edge-on disk around V892 Tau \citep{haas97}.
In addition, for V892 Tau we used the B-, V-, and R-band measurements 
from the Herbig-Bell Catalog \citep[][hereafter HBC]{herbig88}, and the R-I color from 
\citet{strom94}, since KH95 only list a V-band magnitude. The R and I 
photometry of \citet{strom94} is about 1 mag brighter than the HBC 
measurements, probably a sign of variability or an aperture effect.

In the SED plots, the photospheres of AB Aur and V892 Tau were 
normalized at V; both objects already have an excess in the J-band.
Even though the near-IR excess suggests the presence of an inner
disk around V892 Tau, it is classified as a WTTS. This could be explained 
by its early spectral type, for which the identification as WTTS based on 
its narrow H$\alpha$ emission line profile might not be valid 
\citep{calvet04}. 

We note that V892 Tau is a 60 mas binary, and it also has a faint T 
Tauri star companion 4{\farcs}1 away \citep{smith05}. Our IRS 
observations included all three stars, but the subarcsecond binary likely 
dominates the flux of the system. 

\subsubsection{GG Tau, UX Tau, UZ Tau, V773 Tau} 
All of these four objects form quadruple systems, the highest-order multiple 
systems in our sample. 

{\it GG Tau.} GG Tau is a hierarchical quadruple system, consisting of two 
pairs of binaries: components Aa and Ab are separated by 0{\farcs}25, Ba and Bb
by 1{\farcs}48, and components A and B by $\sim$ 10\arcsec. Our observations
were pointed at the A binary; while in LL some flux of the B components also entered
the slit, they are so much fainter than A that their contribution is negligible. In fact,
the GG Tau B components are of late M spectral type, and Bb is most likely a brown 
dwarf \citep{white99}. All four stars exhibit accretion signatures in the form of
strong H$\alpha$ emission and are therefore surrounded by accretion disks; 
additionally, GG Tau A is surrounded by a circumbinary ring 
\citep[e.g.,][]{mccabe02}. 

{\it UX Tau.} UX Tau consists of component A, separated by 2${\farcs}$6 from 
C, and by 5${\farcs}$9 from B, which itself is a subarcsecond binary. Both A and C 
entered the SL slit, and in LL all four components contributed to the IRS spectrum. 
However, the A component is likely to dominate the mid-IR spectrum; it
is the only classical T Tauri star of the system (the other components are 
classified as WTTSs), and it is brighter than components B and C by a factor 
of 7 and 24, respectively, in the L-band \citep[3.6 $\mu$m;][]{white01}. 
To only contribute substantially beyond 13 $\mu$m, but not at 10 $\mu$m
and shorter wavelengths, the other components would have to have very red spectra.

The peculiar IRS spectrum of UX Tau suggests the presence of some material
in the inner disk and a substantial outer disk. Since the other stars are a few 
arcseconds away, they are unlikely to influence the disk around the primary. 
UX Tau A has a very weak 10-$\mu$m silicate emission feature, but a 
prominent 20-$\mu$m feature, as well as some small peaks beyond about
20 $\mu$m possibly indicating the presence of crystalline silicates. 
Interestingly, UX Tau A shows a PAH feature at 11.3 $\mu$m (forsterite has
a feature centered at slightly shorter wavelengths that is wider than the feature 
we observe); the 6.2 and 7.7 $\mu$m PAH features could be hidden by the 
strong continuum. Since UX Tau A's spectral type is K5, and we usually do not 
observe any PAH features in stars later than G, it would be the latest spectral 
type in our sample to show any PAH features. This latter fact might partly be 
due to the weak silicate emission of UX Tau, which allows us to easily identify 
the 11.3 $\mu$m PAH feature. 

{\it UZ Tau.} In the UZ Tau system, component A, a spectroscopic binary, is 
separated by 3{\farcs}5 from the $\sim$ 0{\farcs}4 binary Ba-Bb. All four components 
entered the IRS slits. UZ Tau A is surrounded by an accreting circumbinary disk; 
even though the B components are also surrounded by accretion disks, their 
outer disks are likely truncated \citep{jensen96,simon00,hartigan03}. 
Therefore, the A component likely dominates the IRS spectrum. We note that 
in the SED plot the IRS fluxes are below the IRAC data points, 
which are the sum of the measured fluxes of UZ Tau e (i.e., A) and UZ Tau w 
(i.e., Ba+Bb); this could indicate that the IRS observations were somewhat 
mispointed and therefore missing part of the flux of the brighter
component (which is $\sim$ 5--7 times brighter than the
B components in the IRAC bands). 

{\it V773 Tau.} V773 Tau is a very tight quadruple system: all four stars are found 
within 0{\farcs}25 from each other. V773 Tau AB is a spectroscopic binary
and a WTTS, V773 Tau C is a CTTS, and the fourth component, discovered by
\citet{duchene03}, is an optically faint, and therefore deeply embedded, 
``infrared companion''. V773 Tau D gradually increases in brightness from 
2 $\mu$m, where it is the faintest object in the system, to 4.7 $\mu$m, 
where it is the brightest source \citep{duchene03}. Thus, component D is 
likely the dominating source in the IRS spectrum. 

The SED of V773 Tau is somewhat similar to that of V807 Tau, which
consists of a primary CTTS and a close binary WTTS, with separations of 
less than 0{\farcs}5 between the components: both systems have
little excess emission in the near-IR, a very weak 10-$\mu$m silicate feature,
but a prominent 20-$\mu$m feature, suggesting that their inner disks
have been partially cleared, possibly due to the action of the close components
in these systems. 

\subsubsection{HK Tau} 
HK Tau consists of two components, separated by 2{\farcs}3;
while the primary is surrounded by a face-on disk, the secondary's disk is
oriented edge-on \citep{stapelfeldt98}. We were not able to resolve the
binary in our IRS observations, but the primary is likely much brighter than the
secondary \citep[flux ratios of 21 and 30 in the K- and L-band, 
respectively;][]{white01}. 

\subsubsection{RY Tau and SU Aur} 
These two stars are the only early G-type T Tauri stars in our sample. As with 
the SED plots of the Herbig Ae/Be stars, we normalized their photospheres at V. 
The IRS spectra of both objects display pronounced 10- and 20-$\mu$m silicate 
features; the smoothness of the 10-$\mu$m feature indicates that the dust is 
mostly amorphous. 

SU Aur has an infrared excess already in the J-band, which would point to the 
presence of inner disk regions. Previous determinations of its EW[H$\alpha$] 
of 2 {\AA} indicated that it likely is a WTTS \citep{kenyon98}, in apparent 
conflict with the existence of an inner disk; however, given its G1 spectral type,
the classification as a CTTS or WTTS based on EW[H$\alpha$] might not
be appropriate \citep[see][]{white03}. Recent measurements by 
\citet{calvet04} suggest that it has a wide H$\alpha$ equivalent line width, 
typical of CTTSs. 
Also, its UV excess reveals that it is accreting; it has an inferred mass accretion 
rate of $\sim 5 \times 10^{-9}$ M$_{\odot}$ yr$^{-1}$ \citep{calvet04}. 
Therefore, SU Aur is probably a classical T Tauri star. However, it has a peculiarity
in its spectrum: it displays PAH emission features like those seen in Herbig Ae/Be stars 
\citep{meeus01, acke04, sloan05} but not in the late-type young stars of our sample.

\subsubsection{T Tau} 
T Tau is a triple system; the northern and southern component are 
separated by 0{\farcs}7, while the southern binary has an orbital separation of 
0{\farcs}1. The northern component, which is more massive and the source 
detected in the optical, is not very veiled, while the southern components are 
heavily extinguished and might therefore be surrounded by an envelope 
\citep{calvet94,koresko97}. 
In the infrared, the southern components vary in brightness due to variable
extinction; in addition, the brightness ratio of the two components is changing
on timescales of several months to a year \citep{beck04}. 

T Tau N and S probably contribute about equally to the mid-infrared flux, even 
though the southern components could be brighter especially at the longer 
wavelengths \citep[see][]{ghez91}. 
The IRS spectrum of the T Tau system is the only spectrum in our sample of 
T Tauri stars that shows a 10-$\mu$m absorption feature. Since it was 
observed before that the northern component likely has a 10-$\mu$m emission 
feature, while the southern component displays a silicate absorption feature at 
10 $\mu$m \citep{ghez91,vancleve94}, the fact that we observe a 
10-$\mu$m absorption feature in the spectrum of the T Tau system 
suggests that T Tau S is the brighter source already at 10 $\mu$m 
and is indeed surrounded by an envelope.

\subsubsection{FP Tau, V836 Tau, VY Tau, ZZ Tau} 
These four objects were
classified as Class III objects in KH95. However, our IRS spectra clearly
show that they are Class II objects; their SEDs show excess above photospheric
flux levels roughly starting at the L-band (3.5 $\mu$m) -- V836 Tau, whose
L-band flux is likely erroneous, already has some excess at K (2.2 $\mu$m) -- 
and their SED slopes are shallower than expected for Class III objects. 
The lack of near-IR excess emission in FP Tau, VY Tau, and ZZ Tau could be 
explained by cleared-out inner disk regions and/or very low mass accretion rates.
In fact, the spectrum of ZZ Tau is reminiscent of the spectrum of St 34 
\citep{hartmann05b}, where a spectroscopic binary caused partial clearing
of the inner disk regions, and dust growth and settling are likely responsible
for weak silicate features and low dust continuum levels. ZZ Tau is also a close
binary, with a separation of only 0{\farcs}04 between the two components.
Like St 34, it has little excess emission below about 8 $\mu$m, but it does
have a prominent silicate feature with a narrow peak at about 9.2 $\mu$m,
most likely due to silica \citep[see, e.g.,][]{bowey02}. 

The dust excess of VY Tau is particularly interesting.  \citet{herbig77} emphasized
that VY Tau has had a remarkable history of outbursts, varying in photographic
magnitude from a lower base of $m_{pg} \sim 14$ up to $\sim$ 10 or brighter,
on timescales of a year or less.  These intermittent, highly-irregular outbursts may then 
plausibly be attributed to events of accretion from a disk, possibly mostly cleared in 
its inner regions.  This object should be monitored more extensively for further 
insights into its behavior.

\subsection{Morphological Sequence}
\label{morph_seq}

\begin{figure}
\includegraphics[angle=90,scale=0.75]{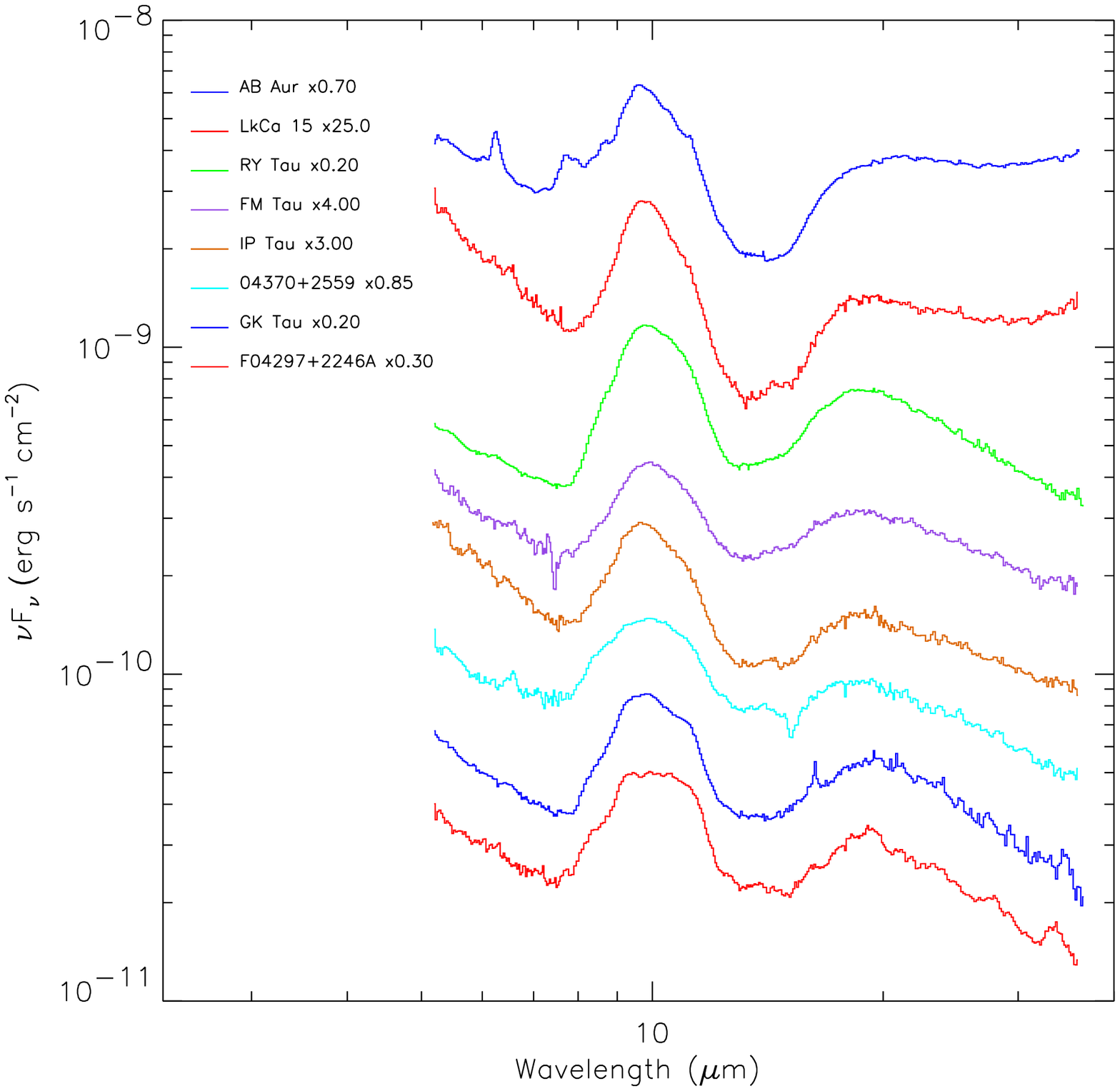}
\caption{Morphological sequence of Class II objects: group A. \label{fig_classII_multi1}}
\end{figure}

\begin{figure}
\includegraphics[angle=90,scale=0.75]{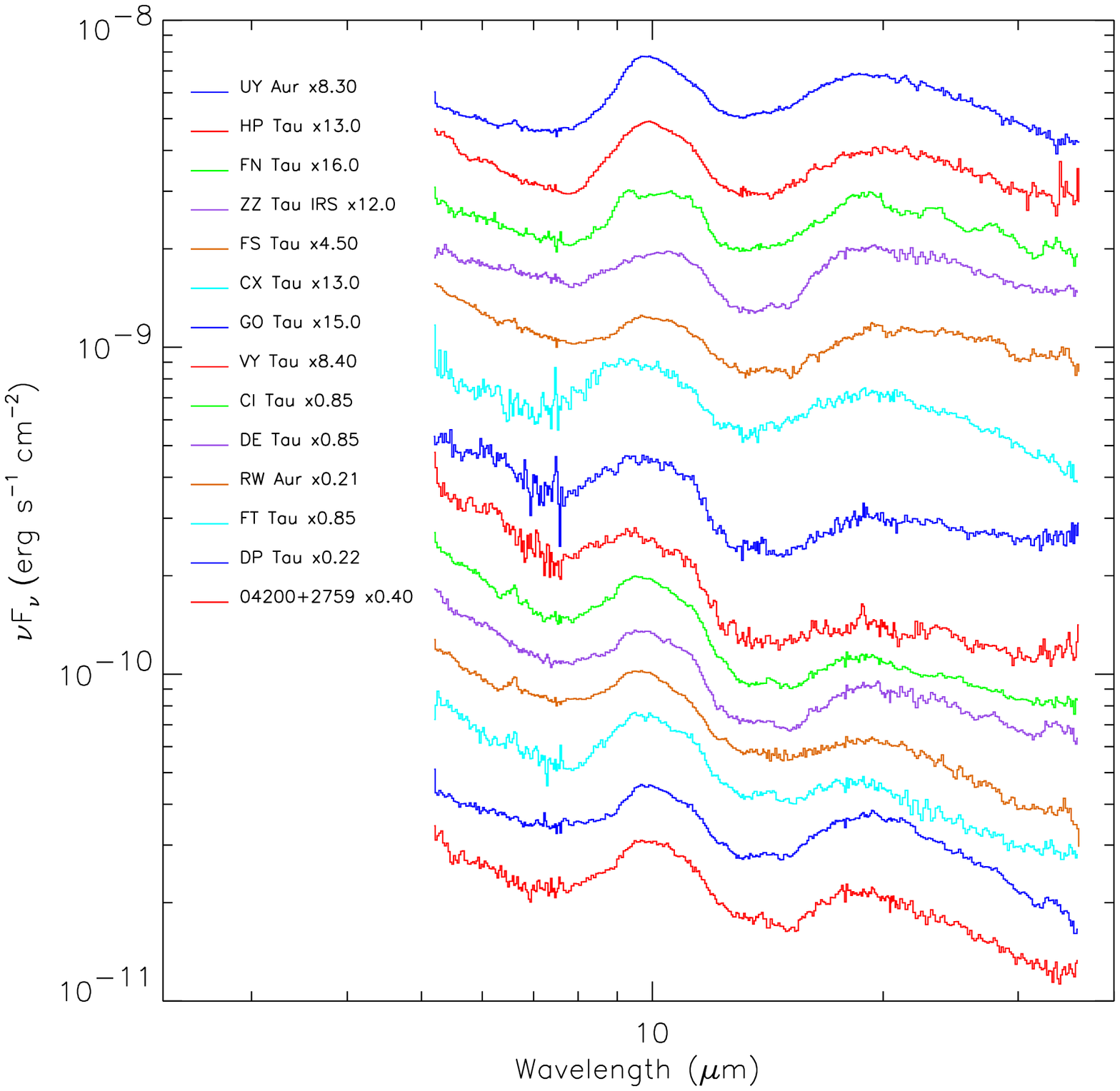}
\caption{Morphological sequence of Class II objects: group B. \label{fig_classII_multi2}}
\end{figure}

\begin{figure}
\includegraphics[angle=90,scale=0.75]{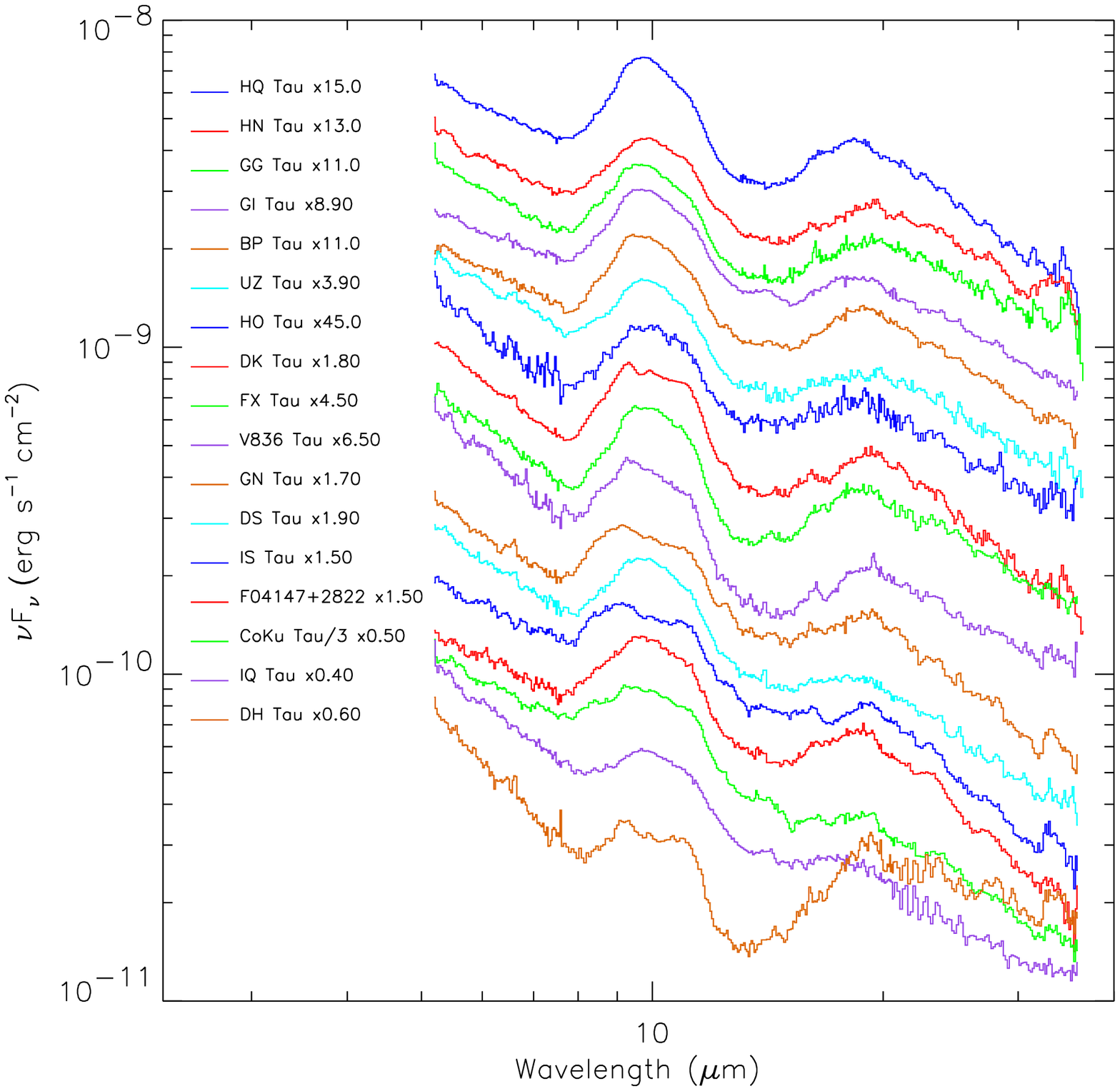}
\caption{Morphological sequence of Class II objects: group C. \label{fig_classII_multi3}}
\end{figure}

\begin{figure}
\includegraphics[angle=90,scale=0.75]{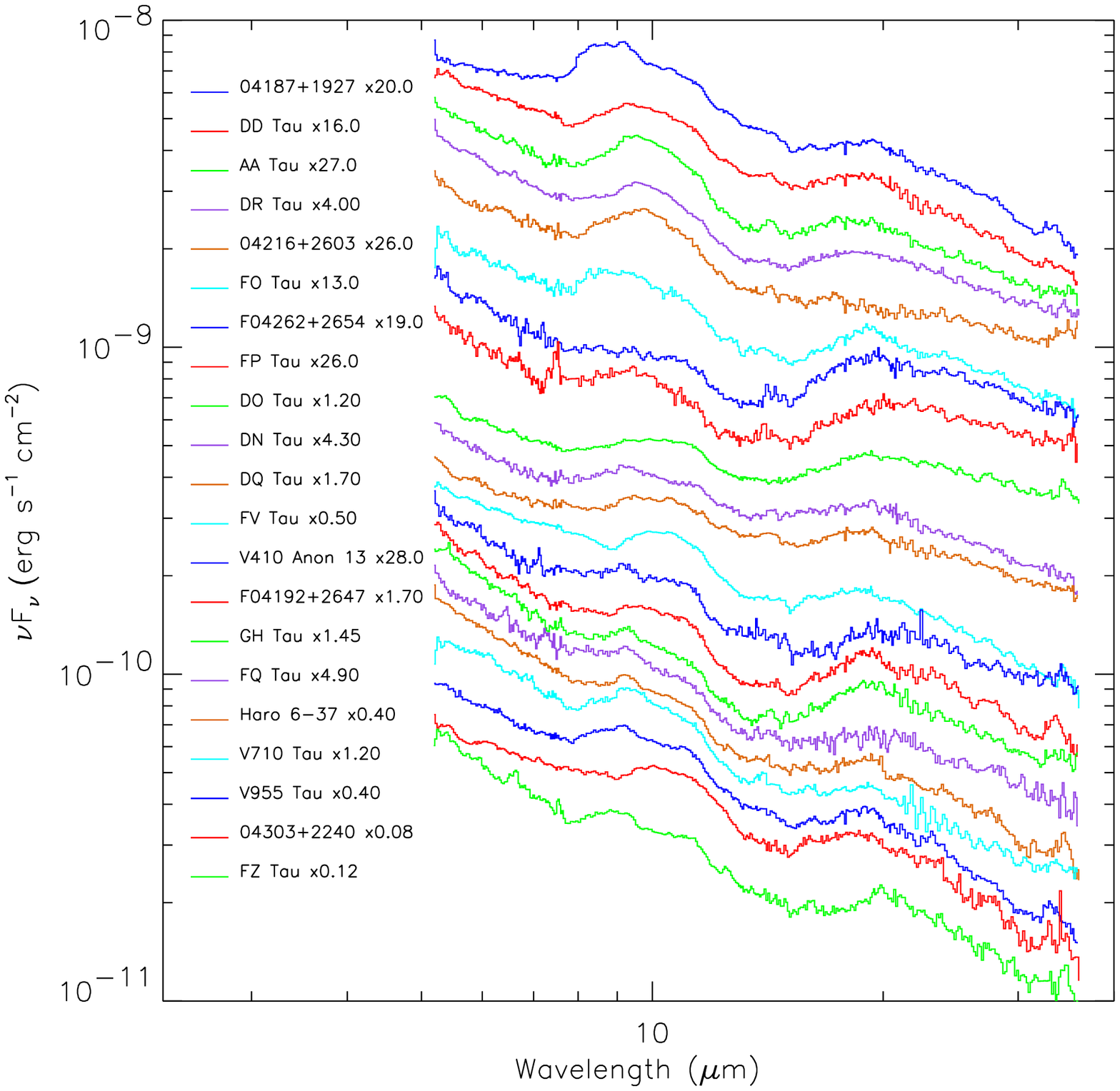}
\caption{Morphological sequence of Class II objects: group D. \label{fig_classII_multi4}}
\end{figure}

\begin{figure}
\includegraphics[angle=90,scale=0.75]{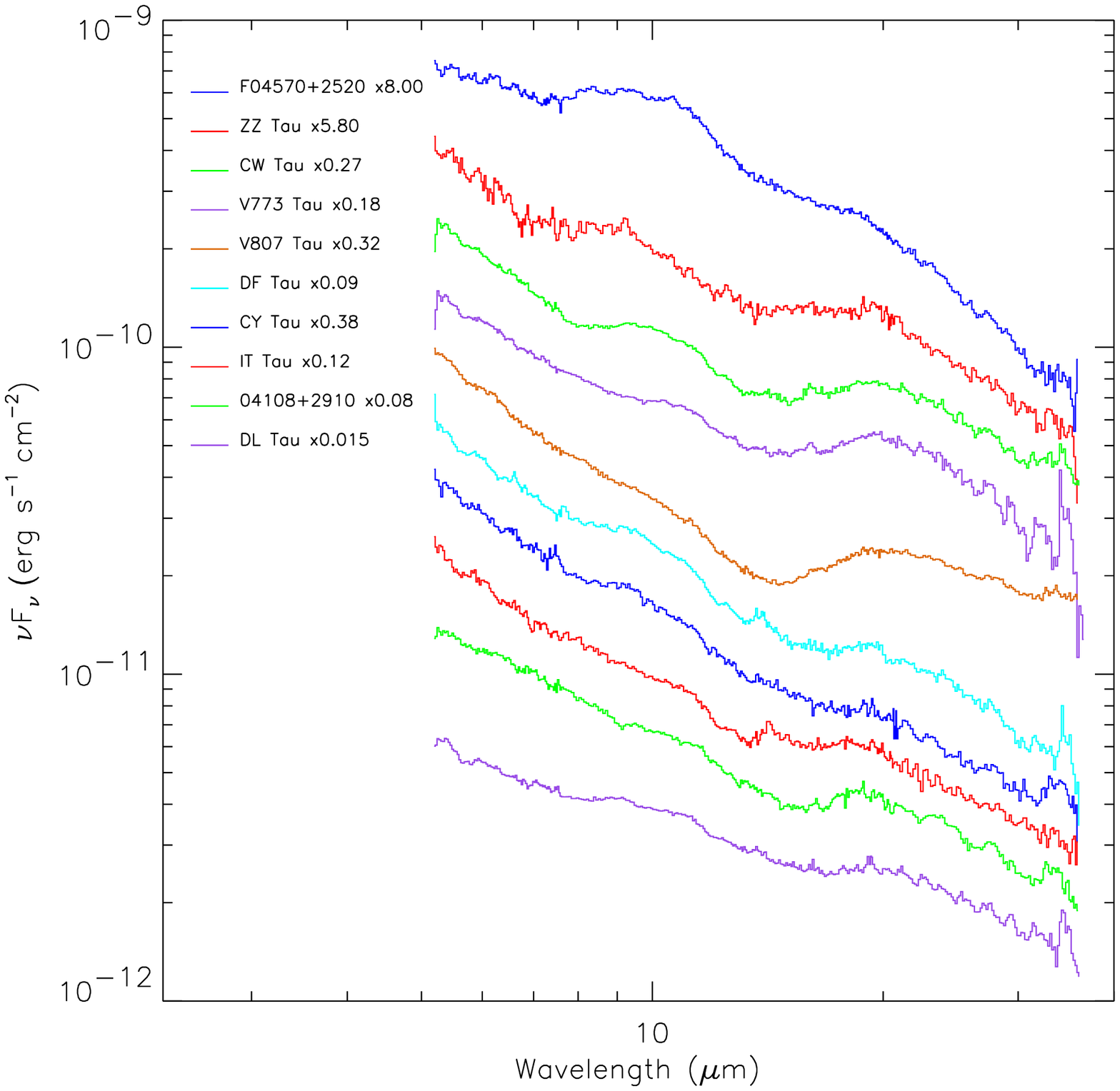}
\caption{Morphological sequence of Class II objects: group E. \label{fig_classII_multi5}}
\end{figure}

\begin{figure}
\includegraphics[angle=90,scale=0.75]{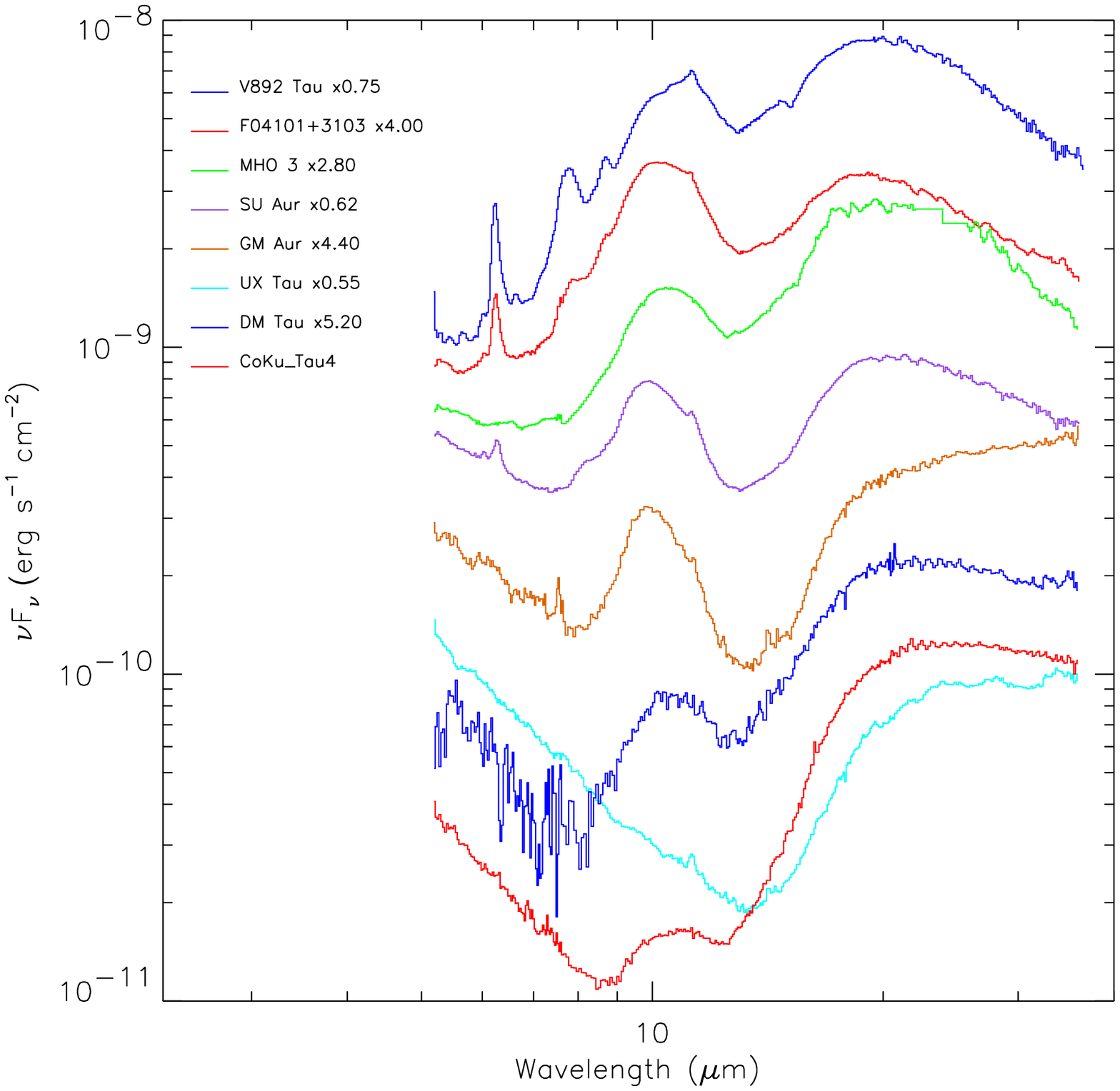}
\caption{Morphological sequence of Class II objects; some of the ``outliers'' of 
the morphological sequence: Herbig Ae/Be stars, Class II objects with 
rising SEDs over the IRS spectral range, and the so-called transitional disks. 
\label{fig_classII_multi6}}
\end{figure}

\begin{figure}
\includegraphics[angle=90,scale=0.75]{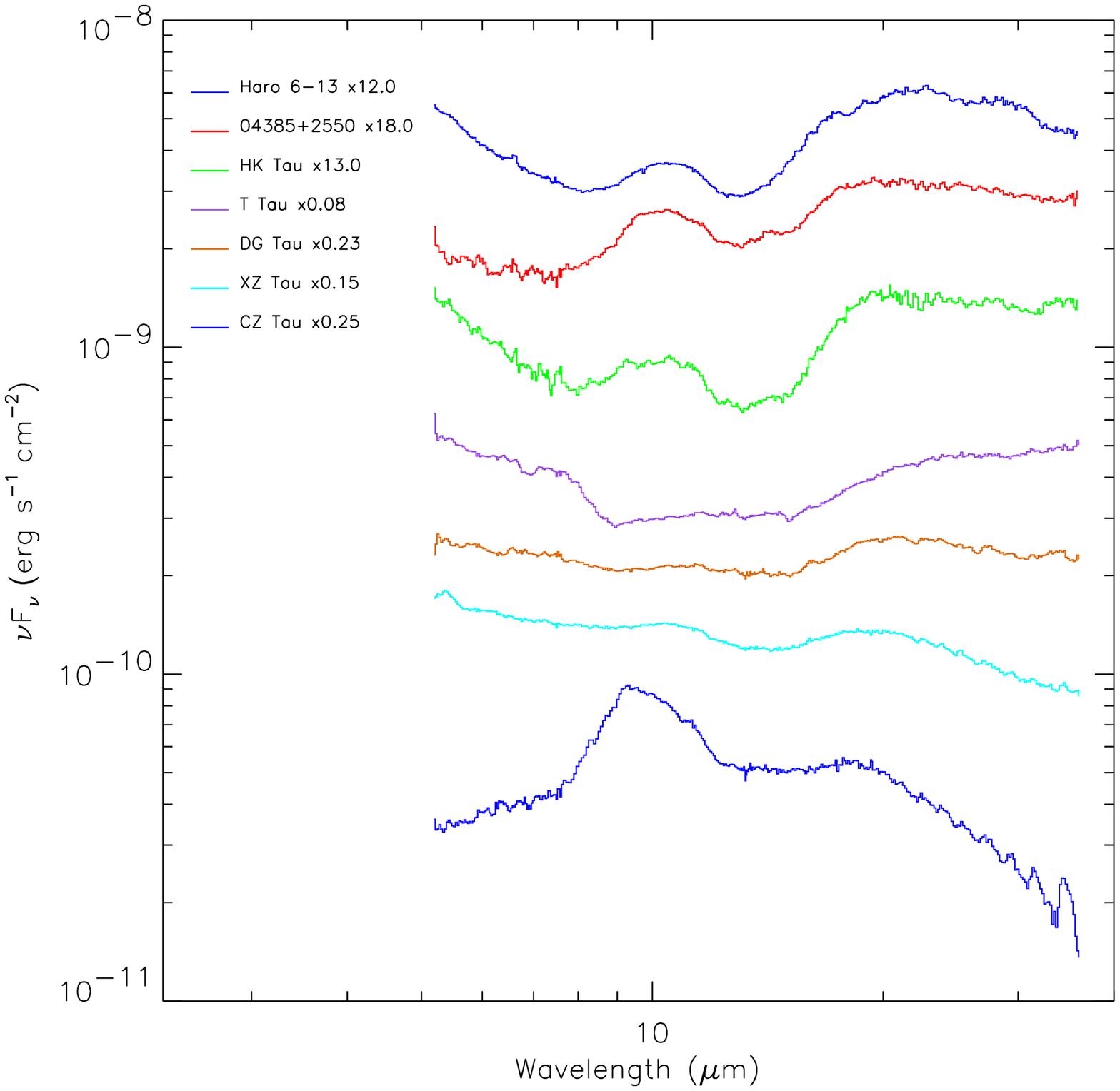}
\caption{Morphological sequence of Class II objects; remaining ``outliers'' of 
the morphological sequence: embedded/heavily reddened T Tauri stars, and CZ Tau.
\label{fig_classII_multi7}}
\end{figure}

We arranged the IRS spectra of our 85 Class II objects
into a morphological sequence according to the shape of their SED and the strength 
of their 10- and 20-$\mu$m silicate features (see Figures \ref{fig_classII_multi1} 
to \ref{fig_classII_multi7}). In each figure the spectra are ordered so that objects 
with stronger silicate features are at the top and objects with a steeper SED and weaker
silicate feature are at the bottom; however, this sequence is not strictly followed
to prevent overlaps of spectra in the figures. In addition, the spectra in figures
\ref{fig_classII_multi6} and \ref{fig_classII_multi7} should be considered
as ``outliers'', whose peculiar IRS spectra cannot easily be assigned to any group
of our sequence.
To avoid erroneous reddening corrections due to uncertain extinctions A$_V$, 
which could result in more pronounced 10-$\mu$m features for large values 
of A$_V$, figures \ref{fig_classII_multi1} to \ref{fig_classII_multi7} show 
the IRS spectra without any reddening correction. Since few T Tauri stars in 
our sample have large extinctions, most spectra show the actual shape of 
the SED in this wavelength range even without reddening correction.

Our first group of objects (Fig.\ \ref{fig_classII_multi1}), group A, shows 
pronounced silicate features and a flat or somewhat decreasing SED beyond 
about 20 $\mu$m. The next group (Fig.\ \ref{fig_classII_multi2}), group B, 
is very similar to group A but has somewhat weaker silicate features.
Group C (Fig.\ \ref{fig_classII_multi3}) is also characterized by more or less
prominent silicate features, but it has clearly decreasing SEDs beyond 
18 $\mu$m. Among these objects, DH Tau has a somewhat peculiar SED: the
slope of its spectrum between 5 and 8 $\mu$m and between 20 and 35 $\mu$m
matches that of the other spectra in this group, while there is a steep increase in
flux between 14 and 20 $\mu$m. Also, its silicate feature is relatively wide and
shows substructure characteristic of crystalline silicates \citep[see][]{sargent06}.

The spectra in groups A, B, and C display a large variety in their silicate feature,
which does not seem to be correlated with the slope of the spectrum. Some 
10-$\mu$m features are smooth and narrow, indicating amorphous silicates
(e.g., UY Aur), while other silicate features are wider and show distinct peaks,
which is a sign of grain growth and the presence of crystalline silicates (e.g.,
DK Tau).

The fourth group (Fig.\ \ref{fig_classII_multi4}), group D, is characterized by 
a weak 10 $\mu$m feature and an overall negative SED slope over the IRS 
range. Many weak silicate features are relatively square and wide and show 
characteristic peaks of crystalline silicates; trends between the relative strengths
of crystalline and amorphous silicates and the FWHM of the silicate feature will 
be explored by D. M. Watson et al. (2006, in preparation).
In some spectra of group D, the 20-$\mu$m silicate feature is somewhat more prominent 
than the feature at shorter wavelengths. This trend becomes more pronounced 
in our group E (Fig.\ \ref{fig_classII_multi5}), where almost no discernible 
10-$\mu$m feature is left in the spectrum (F04570+2520 being the exception), 
and the SED is decreasing steeply. 

Figures \ref{fig_classII_multi6} and \ref{fig_classII_multi7}
show spectra that are not part of our morphological sequence.
The spectra in figure \ref{fig_classII_multi6} display an overall 
rising SED over the IRS spectral range and prominent 10-$\mu$m emission features.
V892 Tau, F04101+3103, and SU Aur are of earlier spectral types than most
of the Class II objects in our sample and, as opposed to T Tauri stars of spectral type
later than G1, have prominent PAH features in their spectra. 
CoKu Tau/4, DM Tau, and GM Aur are so-called transitional disks 
\citep{rice03, forrest04, quillen04, dalessio05, calvet05}, characterized by a 
lack or decrease of excess emission shortward of about 8 $\mu$m due to inner 
disk clearings and a steep SED beyond 13 $\mu$m. UX Tau A would seem like a 
transitional disk due to its steep increase in flux at about 13 $\mu$m, but it has 
a substantial excess above photospheric flux levels beyond 2 $\mu$m and only
a very weak 10-$\mu$m silicate emission feature. As mentioned in section 
\ref{notes_object}, UX Tau is a quadruple system, but the A component is 
likely the dominating source; some disk evolution has likely occurred in this
system.

Figure \ref{fig_classII_multi7} shows a few sources with roughly flat SEDs 
from 5 to 35 $\mu$m, as well as the spectrum of CZ Tau, whose SED is 
peculiar, unlike those of any other object in our sample (see section 
\ref{notes_object}). The flat-spectrum 
sources in this figure are objects associated with Herbig-Haro jets and nebulosities 
in their surroundings \citep[][ and references therein]{white04, krist98}.  
These objects are probably in transition between Class I and II stage, when 
their envelopes have not fully dissipated yet. These ``young'' T Tauri stars 
include, for example, DG Tau, T Tau, and XZ Tau.

\section{Analysis}
\label{analysis}
\subsection{Dust Growth and Settling}

The mid-IR excess emission from Class II objects can be divided into optically thin 
emission from sub-micron sized dust grains in the flared disk surface layer (i.e., the upper
layers of the disk atmosphere), whose most prominent signature is the silicate emission 
feature at 10 $\mu$m, and a continuum component, which originates from deeper, 
optically thick layers of the disk atmosphere.
As disks evolve in time, dust grains grow and settle towards
the midplane \citep{weidenschilling97, dullemond04}; the upper disk layers 
close to the star (within a few AU) are depleted on a very short timescale, much 
less than the $\sim$ 10 Myr lifetime of a protoplanetary disk \citep{dullemond04},
with larger grains settling faster than the small grains. The larger grains  
around the midplane of the disk generate more emission at mm wavelengths, 
while the depletion of small grains in the upper disk layers causes a decrease in the 
continuum emission from the mid- to the far-IR, but these small grains still 
generate an emission band at 10 $\mu$m. Therefore, the effects of dust settling,
and thus the extent of disk evolution, can be seen clearly at mid-IR wavelengths
\citep{miyake95, furlan05b, dalessio06}. 

However, the idea of dust evolving from small grains to larger grains and 
eventually to planetesimals over the course of a few Myr is probably too 
simplistic: recently, \citet{dullemond05} found that dust growth by 
coagulation occurs on very short timescales (less than $\sim$ 1 Myr, 
the typical age of a T Tauri star) in their simulations, and argued that small 
grains must therefore be replenished to produce the observed infrared excess. 
If grain growth and fragmentation both occur at the same time, grain sizes 
cannot be used to infer the evolutionary state of a disk, i.e.\ the presence
of large grains in disks does not necessarily mean that a disk is older and thus
more evolved.

Over the last several years, it has become increasingly clear 
that disk models with dust settling are needed to explain the observed SEDs
 of T Tauri stars: maximum grain sizes and dust-to-gas 
mass ratios are different for the disk interior and the disk surface, with larger 
grains ($\lesssim$ 1 mm) close to the midplane and small, ISM-like dust in 
the disk surface layer. To quantify the amount of dust settling, we use a
variable, $\epsilon$, which is the ratio of the dust-to-gas mass ratio in the 
disk atmosphere and the standard dust-to-gas mass ratio of the interstellar
medium (1:100) \citep{dalessio06}.
A value of $\epsilon$=1 means that no settling has taken place, while a small value
of $\epsilon$ indicates the presence of larger grains towards the midplane
of the disk and fewer small grains in the upper layers.

The morphological sequence from group A to E could be partly understood in 
terms of disk evolution: as dust grows, the 10-$\mu$m silicate feature 
becomes wider and flatter \citep{przygodda03}, and as it settles, the mid-IR 
excess decreases, causing a steeper negative slope of the mid-IR part of the SED
\citep{dalessio06}. 
Due to small dust grains still present in the disk atmosphere, the 10-$\mu$m feature 
can still be strong relative to the continuum; it is more pronounced than the 
20-$\mu$m feature due to the wavelength-dependent emissivity of the dust.
A decrease in the 10- to 20-$\mu$m flux ratio could indicate a decrease of 
emission from inner disk regions relative to the outer parts, since the 10-$\mu$m 
emission arises mostly from inner regions of the disk, while the main 20-$\mu$m 
emission region lies further out. Given that dust growth and settling is fastest in 
the inner parts of the disk \citep{hayashi85, weidenschilling97}, it is expected
that the silicate emission from the inner disk would decrease first. 
A more detailed, quantitative study, which is beyond the scope of this paper, 
is required to tie our morphological sequence to disk evolution; here, we
tentatively suggest that dust has grown and settled to a larger extent in the 
disks of groups D and E than in groups A, B, and C.

\subsection{Settled Disk Models}
\label{settling_models}

To analyze our Class II objects in a more quantitative way, we compare them
to a grid of accretion disk models that include dust settling \citep{dalessio06}. 
The vertical disk structure is derived self-consistently, as in \citet{dalessio98, 
dalessio99, dalessio01}; the resulting flared disk extends from the dust sublimation 
radius to an outer radius of 300 AU, which was set arbitrarily, but does not affect 
the shape of the mid-infrared spectrum. At the inner disk edge, the ``wall'', light 
from the star is incident parallel to the surface normal. The disk is heated by 
viscous dissipation and stellar irradiation. The central source is a typical T Tauri star 
in Taurus: T=4070 K, R=1.864 R$_{\odot}$, M=0.8 M$_{\odot}$, 
L=0.79 L$_{\odot}$. 

Silicates and graphite make up the dust grains in
the disk \citep[optical constants from][]{draine84}; their sizes range
from 0.005 to 0.25 $\mu$m in the upper disk layers (i.e., similar
to the interstellar medium), while in the lower disk layers, the maximum grain
size is set at 1 mm. The dust  particles follow a size distribution of the form 
$n(a) da = a^{-3.5} da$ between the minimum and maximum grain sizes.
Models with either isotropic scattering of the incident radiation by dust grains or 
perfect forward scattering were considered, with the true case lying in between 
the two limiting cases. The value of $\epsilon$ was also varied; to include
the effect of both dust settling and grain growth, a higher depletion of small
grains in the upper layers is accompanied by an increase of large grains in
the lower disk layers. Thus, while the dust is depleted in the disk atmosphere,
the amount of dust is increased in the layers around the midplane, conserving
the total amount of dust.
The models of the grid vary in accretion rate
(10$^{-8}$ and 10$^{-9}$ M$_{\odot}$ yr$^{-1}$), inclination angle
along the line of sight (75{\fdg}5, 60\degr, 50\degr, 40\degr, 30\degr, 
20\degr, and 11{\fdg}5), amount of 
dust settling ($\epsilon$= 1, 0.1, 0.01, and 0.001), and dust composition. 
As we did in \citet{furlan05b}, we use the spectral coverage of the IRS 
combined with models to estimate the degree of dust settling.

\subsection{Spectral Indices}
\label{spec_indices_section}
\subsubsection{Definition}

To characterize the SEDs of the T Tauri stars in our sample, we computed spectral 
indices at different wavelengths in the IRS spectral range that represent continuum
emission, similar to our analysis in \citet{furlan05b}. We integrated the flux of 
our dereddened spectra in the following wave bands: 
5.4--6.0 $\mu$m (center wavelength, ${\lambda}_c$, at 5.7 $\mu$m), 
12.5--14.0 $\mu$m (${\lambda}_c$ = 13.25 $\mu$m), and 23.5--26.5 
$\mu$m (${\lambda}_c$ = 25 $\mu$m). Then we divided the result by the 
width of the respective wave band, which yielded a flux density in wavelength 
units ($F_{\lambda}$) for each band. To obtain spectral indices, $n$, we computed 
\begin{displaymath}
n = \log \left(\frac{{\lambda}_2  F_{{\lambda}_2}}{{\lambda}_1  
F_{{\lambda}_1}} \right)/\log \left( \frac{\lambda_2}{\lambda_1} \right),
\end{displaymath}
first between 5.7 and 13.25 $\mu$m ($n_{6-13}$), then between 13.25 and 
25 $\mu$m ($n_{13-25}$). 

Furthermore, to quantify the properties of our morphological sequence, we
used our IRS spectra not corrected  for reddening and calculated a spectral 
index covering most of the IRS spectral range, from 6 to 25 $\mu$m 
($n_{6-25}$), and estimated the strength of the 10-$\mu$m silicate feature. 
For the latter, we interpolated the continuum from data between 5.0
and 7.5 $\mu$m and between 13.0 and 16.0 $\mu$m with a third-order 
polynomial, subtracted it from the silicate feature defined between 8.0 
and 12.4 $\mu$m, integrated the flux of the continuum-subtracted silicate feature, 
and normalized it to the continuum integrated over the same wavelength range.
Due to uncertainties in the continuum fit, this procedure might not always yield
accurate measurements of the 10-$\mu$m silicate feature strength, but for the 
majority of objects it will result in reasonable estimates.

\subsubsection{Results: IRS Data}

\begin{figure}
\includegraphics[angle=90,scale=0.75]{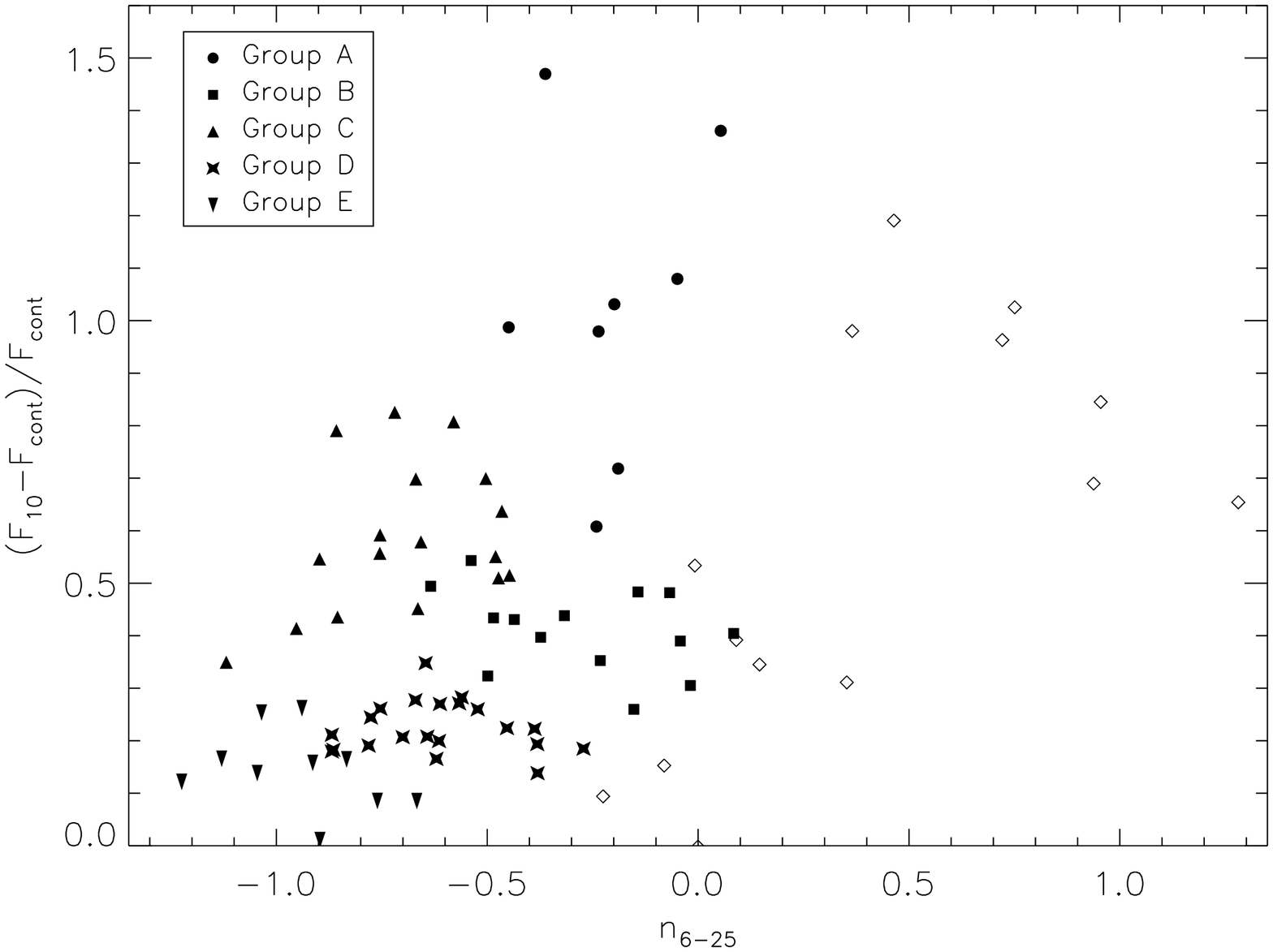}
\caption{Strength of the continuum-subtracted 10-$\mu$m feature, normalized to
the continuum, vs.\ the spectral index between 6 and 25 $\mu$m, $n_{6-25}$.
The data points belonging to the groups defined in our morphological sequence are 
identified by different plotting symbols: group A -- {\it circles}, group B -- {\it squares}, 
group C -- {\it upward facing triangles}, group D -- {\it stars}, group E -- {\it downward
facing triangles}. The open diamonds identify the ``outliers'' of the morphological sequence. 
\label{fig_groups_sp_index}}
\end{figure}

Figure \ref{fig_groups_sp_index} displays the continuum-subtracted, integrated
flux of the 10-$\mu$m feature, normalized to the continuum, versus the $n_{6-25}$
spectral index. The data points belonging to spectra from groups A to E of our 
morphological sequence are identified by different plotting symbols; the data for 
the plot can be found in Table \ref{tab_sp_ind_10mic}.
As indicated by our morphological sequence, objects with the smallest values of 
$n_{6-25}$ in our sample ($n_{6-25}<-1.0$) have weak 10-$\mu$m silicate 
features. The spectra of groups D and E have the smallest 10-$\mu$m feature 
strengths, but SED slopes in the range of those of group C, which has stronger
silicate features. Only objects with $n_{6-25}>-0.5$ have very strong features, 
but for these values of the spectral index there is a large spread in silicate feature
strength. 

\begin{figure}
\includegraphics[angle=90,scale=0.75]{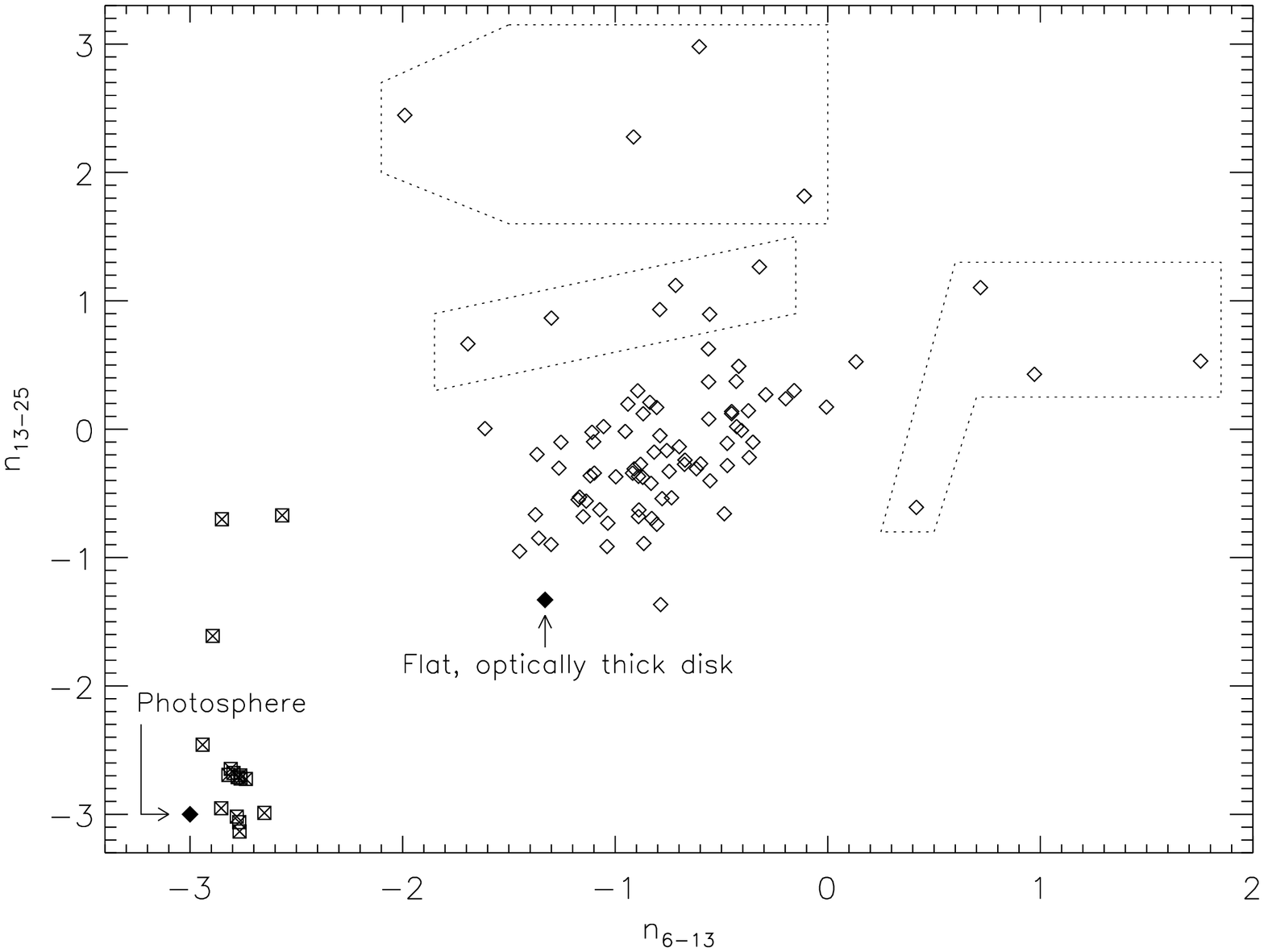}
\caption{Spectral index $ n \equiv d \log(\lambda  
F_{\lambda})/d \log (\lambda) $ evaluated between 13 and 25 $\mu$m
vs.\ the spectral index between 6 and 13 $\mu$m for our sample of Class II 
({\it open diamonds}) and Class III ({\it crossed squares}) objects, together with 
values for a stellar photosphere in the Rayleigh-Jeans limit 
($\lambda$F$_{\lambda}$ $\propto$  ${\lambda}^{-3}$, i.e.\
$n=-3$) and for a geometrically thin, optically thick disk ($\lambda$F$_{\lambda}$ 
$\propto$  ${\lambda}^{-4/3}$, i.e.\ $n=-4/3$).
The dotted regions delinate outliers in the plot discussed in the text. \label{fig_sp_index_all}}
\end{figure}

Figure \ref{fig_sp_index_all} shows a plot of $n_{13-25}$ versus $n_{6-13}$, 
computed using the IRS spectra of the Class II and III objects in our sample 
({\it open diamonds and crossed squares, respectively}), and for two extremes, a flat, 
passive disk and a naked photosphere; the data for the plot are also given in
Table \ref{tab_sp_ind}. A photosphere in the 
mid-IR has a spectral index of $-3 $ (Rayleigh-Jeans limit), while a geometrically 
thin, optically thick disk has an index of $-4/3$. Due to some spectrophotometric
uncertainty in our IRS data, the typical error bars for the spectral indices are
about $\pm$0.05. The dashed boxes delineate outliers and will be discussed
below.

\begin{figure}
\epsscale{0.75}
\plotone{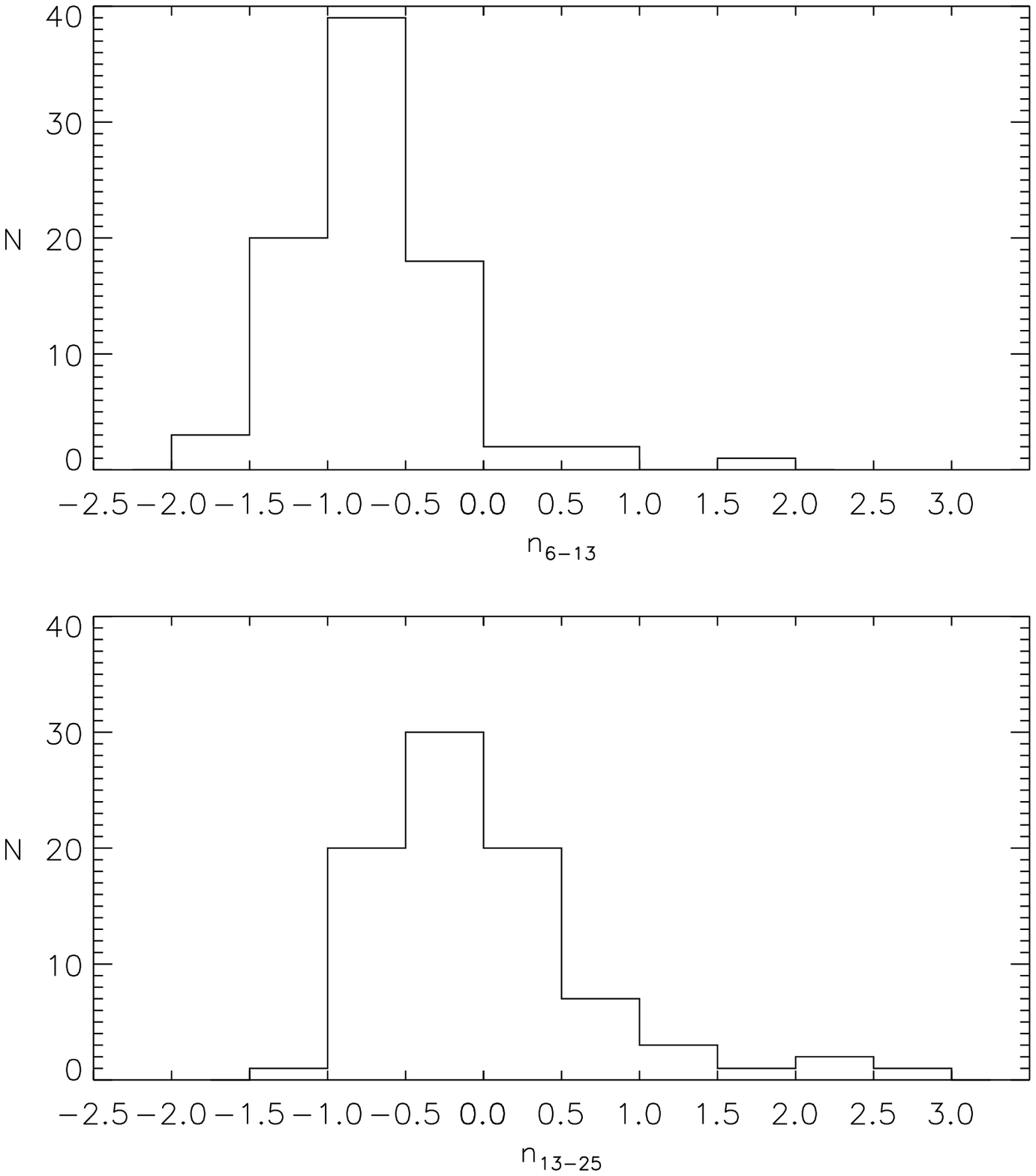}
\caption{Frequency distribution of the spectral indices $n_{6-13}$ ({\it top})
and $n_{13-25}$ ({\it bottom}) for our sample of Class II objects. \label{fig_freq_spindex}}
\end{figure}

As noted in \citet{furlan05b}, the fact the Class II objects in Fig. \ref{fig_sp_index_all} 
have spectral indices that are larger than the value expected for a flat, optically thick disk
means that their disks must be flared. Only one object in our sample, F04570+2520,
has an $n_{13-25}$ spectral index that is slightly smaller than that of a flat, 
optically thick disk; its SED shows only a weak 10-$\mu$m silicate feature and
a steep decrease in flux beyond it, probably indicating an evolved disk in which
a large fraction of the dust has grown and settled. 
For most objects, the $n_{6-13}$ and $n_{13-25}$ spectral indices seem
to be correlated; a steeper slope of the spectrum between 6 and 13 $\mu$m is
usually accompanied by a steeper slope from 13 to 25 $\mu$m. As can be
seen from the frequency distribution of both indices in Figure \ref{fig_freq_spindex},
the distribution peaks at values between $-1.5$ and $0$ for $n_{6-13}$, and between
$-1$ and $0.5$ for $n_{13-25}$. While negative spectral slopes larger than $\sim -1$
prevail for both wavelength ranges, which is typical for Class II objects, the 
slope measured between 13 and 25 $\mu$m tends to be flatter than the 
one measured between 6 and 13 $\mu$m. 

In addition, there is a small number of objects with spectral indices $n_{13-25}$
larger than about 0.5, which is represented by the seemingly randomly scattered 
points at positive $n_{13-25}$ values in Figure \ref{fig_sp_index_all}, and framed 
by dotted boxes in the figure. These ``outliers" can be divided up into three groupings:
the 6 objects in the lower left box, the 4 objects in the right box, and the 4 objects in 
the uppermost box. 

The first-mentioned grouping contains DH Tau, LkCa 15, AB Aur, HK Tau, 
Haro 6-13, and SU Aur. These Class II objects are characterized by a steep rise in 
slope starting at 13 $\mu$m, indicating a strong 20-$\mu$m silicate feature, and a 
relatively flat SED past 20 $\mu$m. The right box frames F04101+3103, 
MHO 3, V892 Tau, and, at a negative $n_{13-25}$ value, CZ Tau. These objects 
also have a strong 20-$\mu$m silicate feature, in addition to a pronounced 
10-$\mu$m emission feature, but their SEDs decrease past 20 $\mu$m.
Also, their emission from 5 to 8 $\mu$m is lower than the emission at longer 
wavelengths, resulting in a positive index $n_{6-13}$. The decrease of the SED 
beyond about 20 $\mu$m is especially dramatic for CZ Tau, which explains the 
low value of $n_{13-25}$ of this object. As mentioned earlier, CZ Tau's outer 
disk might be truncated, resulting in a substantial decrease of its far-IR flux.

The final grouping includes the known transitional disks in Taurus, CoKu Tau/4, 
GM Aur, and DM Tau, as well as UX Tau A at $n_{6-13} \sim -2$. These
``outliers'' have negative spectral indices from 6 to 13 $\mu$m and
positive spectral indices from 13 to 25 $\mu$m, the latter due to 
a steep increase in flux at about 13 $\mu$m. 

Also included in Figure \ref{fig_sp_index_all} are the spectral indices 
$n_{13-25}$ versus $n_{6-13}$ for those Class III objects in our sample
whose 25 $\mu$m flux was detected. 
However, in many cases the 25-$\mu$m flux is very weak and 
the LL spectrum noisy, resulting in a larger uncertainty for the index
calculation. Nevertheless, most of the indices of our Class III objects 
are close to values of $-3$, as expected for close-to-photospheric 
fluxes. One object, V410 Tau, which was not included in the plot, lies 
at spectral indices smaller than $-3$, which is mainly a result of its 
very noisy LL spectrum; its IRS spectrum seems to follow the slope of 
a photosphere, with a possible slight excess. 
V410 X-ray 3 lies at $n_{13-25}=-1.6$ and $n_{6-13}=-2.9$; 
HBC 427 and V819 Tau, which have an apparent infrared excess in LL, 
can be found at $n_{13-25}=-0.7$ and $n_{6-13}=-2.8$ and $-2.6$, 
respectively. If at least part of their infrared excess is real, the latter three
objects would constitute the transition phase between Class II and Class III, 
which seems to be very rapid due to the lack of objects with 
spectral indices between $-1$ and $-2$. This result is similar to what is 
found when the K\,--\,L colors are plotted versus the K\,--\,N colors 
\citep[see, e.g.,][]{kenyon95}: there is a continuous transition between 
Class I and II objects, with the latter displaying bluer IR colors, but a gap 
between the Class II sources with the smallest infrared excess and the 
Class III sources, which have almost photospheric colors.

\subsubsection{Results: IRS Data and Models}
\label{IRS_sp_indices_models}

\begin{figure}
\epsscale{1.1}
\plottwo{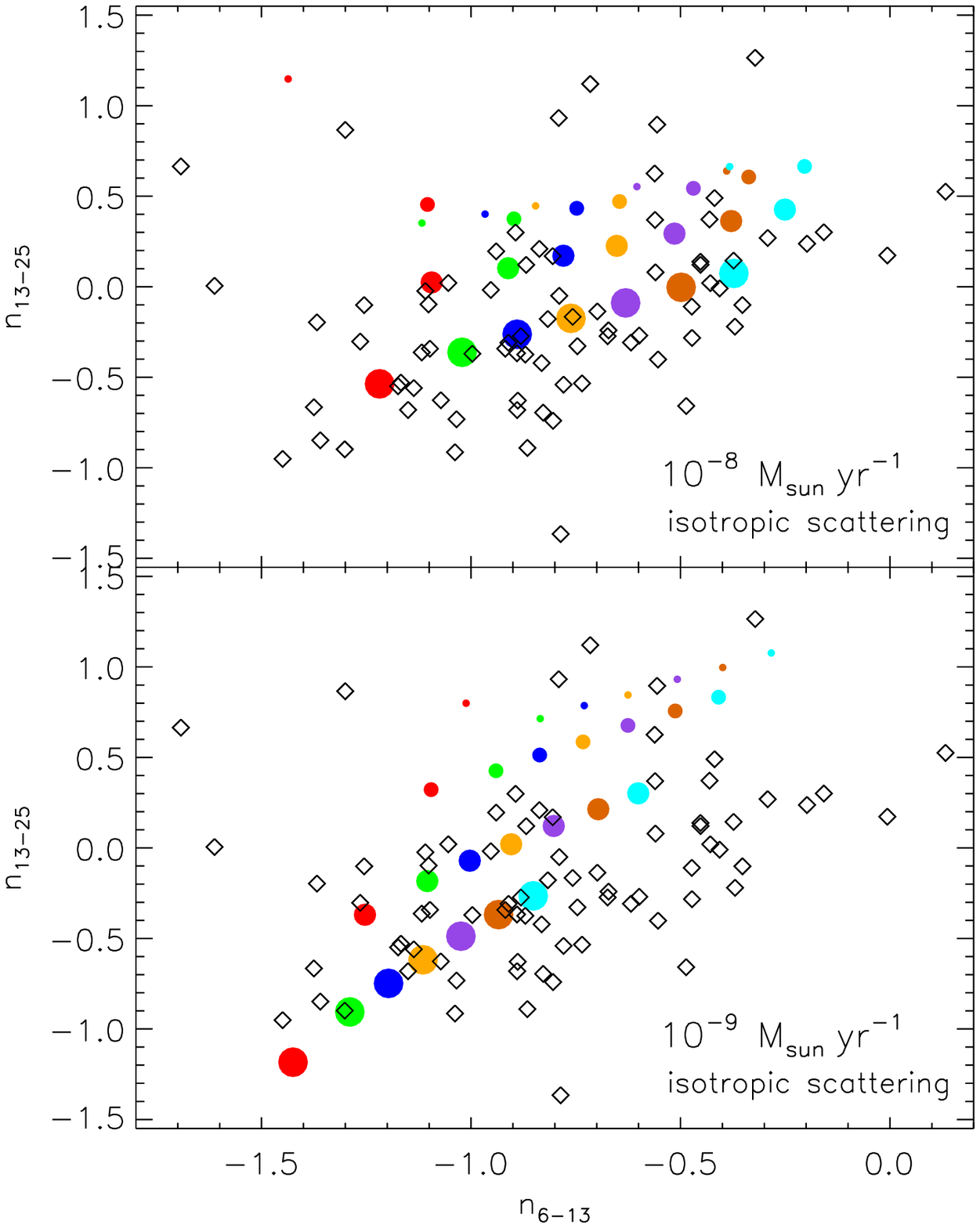}{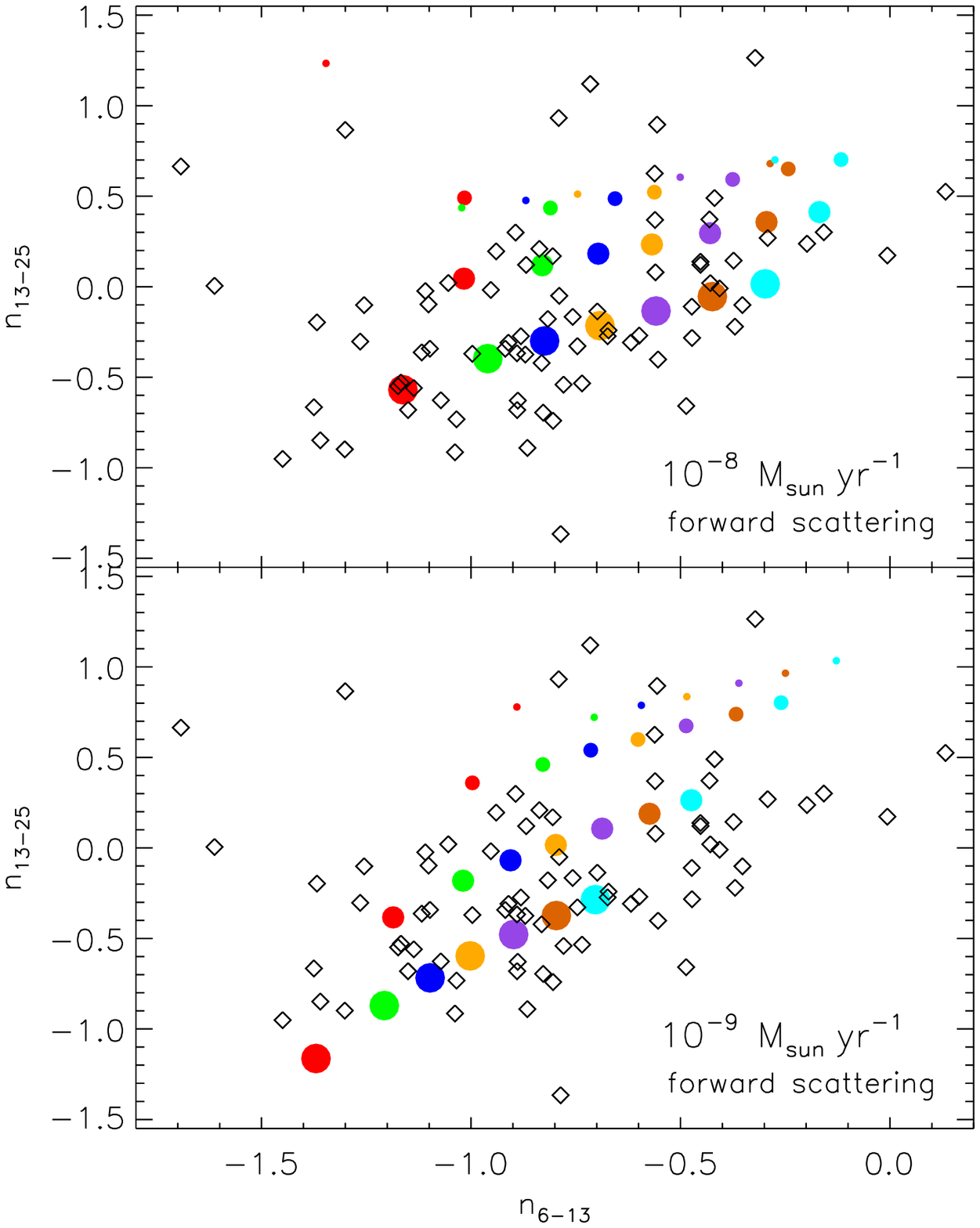}
\caption{Spectral index between 13 and 25 $\mu$m vs.\ the spectral index between
6 and 13 $\mu$m for our sample of Class II objects ({\it open diamonds}) and for 
accretion disk models ({\it filled circles}) with 10$^{-8}$ M$_{\odot}$ yr$^{-1}$ 
({\it upper panels}) and 10$^{-9}$ M$_{\odot}$ yr$^{-1}$ ({\it lower panels}).
The models on the left-hand side were computed assuming isotropic scattering, while
the models on the right-hand side included only perfectly forward scattering dust grains.
The colors of the filled circles represent the following inclination angles: red -- 75{\fdg}5,
green -- 60$\degr$, dark blue -- 50$\degr$, yellow -- 40$\degr$, purple -- 30$\degr$, 
orange -- 20$\degr$, and light blue -- 11{\fdg}5. The sizes of the filled circles 
represent a depletion factor $\epsilon$ of 0.001, 0.01, 0.1, and 1 from largest to smallest, 
respectively. \label{fig_spindex_models}}
\end{figure}

We computed analogous spectral indices ($n_{13-25}$ vs. $n_{6-13}$) 
for the grid of accretion disk models and overplotted the results from IRS 
data and disk models in Figure \ref{fig_spindex_models}. We show indices
computed from two sets of disk models, one with isotropic scattering of the
incident stellar radiation by dust grains and the other with perfect forward 
scattering (which has the same effect as no scattering, i.e.\ albedo=0); they 
represent the two limiting cases of scattering. 
There is only a slight difference between 
these two sets; spectral indices are not very sensitive to the degree of 
scattering in the models. Each model set contains models computed with
two mass accretion rates, 10$^{-8}$ and 10$^{-9}$ M$_{\odot}$ 
yr$^{-1}$, which are typical for T Tauri stars in Taurus 
\citep[see, e.g.][]{gullbring98, hartmann98}.

The spread in the $n_{6-13}$ index is partly due to different accretion rates, 
but mostly an effect of varying inclination angles. This can be understood 
in terms of the wall emission, which dominates the flux at wavelengths shortward
of about 6 $\mu$m; the wall emission increases as the inclination angle increases,
resulting in a larger increase of the 6-$\mu$m flux than of the 13-$\mu$m
flux, while the disk emission, which dominates at 25 $\mu$m, decreases 
\citep{dalessio06}.
The different extents of dust settling in the models, which are described by 
$\epsilon$, cause a spread in both $n_{6-13}$ and $n_{13-25}$. 
Since for mass accretion rates of 10$^{-8}$ and 10$^{-9}$ M$_{\odot}$ 
yr$^{-1}$ irradiation heating represents the main heating source 
\citep{dalessio99,dalessio06}, 
more settling causes a decrease in the mid-IR continuum emission;
the disk intercepts less radiation from the central star if its outer parts are less
flared. This affects the 25-$\mu$m flux more than the flux at shorter wavelengths,
resulting in smaller spectral indices (i.e., steeper SED slopes) for smaller 
$\epsilon$. 

For example, AA Tau, which has a mass accretion rate of 6.5$\times$ 
10$^{-9}$ M$_{\odot}$ yr$^{-1}$ \citep{white01}, lies at 
$n_{6-13}=-0.87$ and $n_{13-25}=-0.37$, which is close to the 
model with an accretion rate of 10$^{-8}$ M$_{\odot}$ yr$^{-1}$, 
${\epsilon}=0.001$, and inclination angle of 50$\degr$. 
\citet{bouvier99} derived a lower limit for the inclination angle 
of AA Tau of 53$\degr$, but favored $i=75\degr$ to explain the peculiar
variability of this object. Our comparison would suggest a somewhat lower 
inclination angle, and in addition a large amount of settling in the disk of 
AA Tau.
As another example, BP Tau, which has an inclination angle of 39$\degr$ 
\citep{muzerolle03b} and a mass accretion rate of 1.3$\times$ 10$^{-8}$ 
M$_{\odot}$ yr$^{-1}$ \citep{white01} lies at $n_{6-13}=-0.67$ 
and $n_{13-25}=-0.24$, which is in between the model points for 
$\dot{M}=10^{-8}$ M$_{\odot}$ yr$^{-1}$, ${\epsilon}=0.001$, 
and inclination angles of 30$\degr$ and 40$\degr$, respectively.

Most Class II objects can be described with models that include substantial settling, 
with an $\epsilon$ between 0.01 and 0.001, which means dust depletions of factors
of 100 to 1000 in the upper layers. The spectral indices for accretion 
disks with $\epsilon < 1$ are lower than for those in which no settling has taken 
place, since the reduced mid-IR continuum causes a steeper slope, towards the value of 
$-4/3$ of a flat, optically thick disk. The settling of dust does not affect the silicate 
emission features, which we see in virtually all of our Class II spectra; these features 
are generated by the small grains still present in the upper hot layers of the disk atmosphere. 
As indicated earlier with a smaller sample of CTTSs \citep{furlan05b}, we 
conclude with this larger sample of CTTSs and WTTSs that the majority of T Tauri 
stars have disks in which dust has started to grow and settle towards the midplane,
thus affecting the SEDs of these young (1--2 Myr) pre-main-sequence stars. This
is consistent with the simulations of \citet{dullemond05}, which predict dust
growth on timescales of $\sim$ 1 million years or less.

Given that all of our T Tauri stars are at about the same age, the difference 
in the spectra of various objects must be due to different initial conditions and 
different circumstellar environments influenced, for example, by the presence 
of close companions. Thus, the extent to dust settling in a disk, at least in its
initial stages, might be less a function of age and more related to the object's
formation history and current environment. 

We caution that the comparison of our data with a grid of accretion disk
models is valid to gauge trends in our data, but that additional parameters,
in particular the composition of the dust, need adjustments before any
detailed interpretations can be carried out reliably. For example, choosing
glassy pyroxene or olivine instead of ``astronomical'' silicates and graphite
will generate different dust
emission, both in the continuum and in the emission bands, and 
different spectral indices will result. Therefore, the detailed composition
of the dust adds additional scatter to the spectral indices plot, which has
to be considered in addition to the effects of inclination and dust settling.
The values we derived for $\epsilon$ should thus be
treated as rough estimates only.
The effect of dust composition on the models, as well as models of 
individual objects, will be presented in a future paper (P. D'Alessio et al.
2006, in preparation).

\subsection{Taurus Median}

In order to determine a typical SED range for T Tauri stars in Taurus, we 
computed a median SED from 1.25 to 34 $\mu$m using the 2MASS J, H, 
and K$_s$ photometry and the IRS spectra of all our 85 Class II objects,
dereddened using the A$_V$ values listed in Table \ref{tab2}.
To obtain a clearer representation of the flux values in the IRS spectral range,
we integrated the IRS spectra over narrow wave bands and divided by the width 
of the band, creating the equivalent of narrowband photometric measurements at
5.7, 7.1, 8.0, 9.2, 9.8, 11.3, 12.3, 13.25, 16.25, 18.0, 21.0, 25.0, 30.0, and 34.0
$\mu$m. The width of the bands ranged from 0.6 $\mu$m at the shorter
wavelengths to 3 $\mu$m at the longer wavelengths; the width of the silicate 
bands, centered at 9.8 and  18 $\mu$m, was chosen to be 2 $\mu$m. 
As in \citet{dalessio99}, we normalized all flux values at H (1.65 $\mu$m)
before computing median values, since, for most objects, the H-band flux is
photospheric, and thus normalization at H is equivalent to normalizing to the
stellar luminosity. Since for most of the T Tauri stars in our sample mass 
accretion rates are low enough ($\lesssim$ 10$^{-7}$ M$_{\odot}$ 
yr$^{-1}$) that the star is the main heating source of the disk, the
infrared excess of the disk should scale with stellar luminosity. 
By computing the median values, we also obtained the quartiles, which define
the range around the median where 50\% of all flux values lie. 

\begin{figure}
\includegraphics[angle=90,scale=0.7]{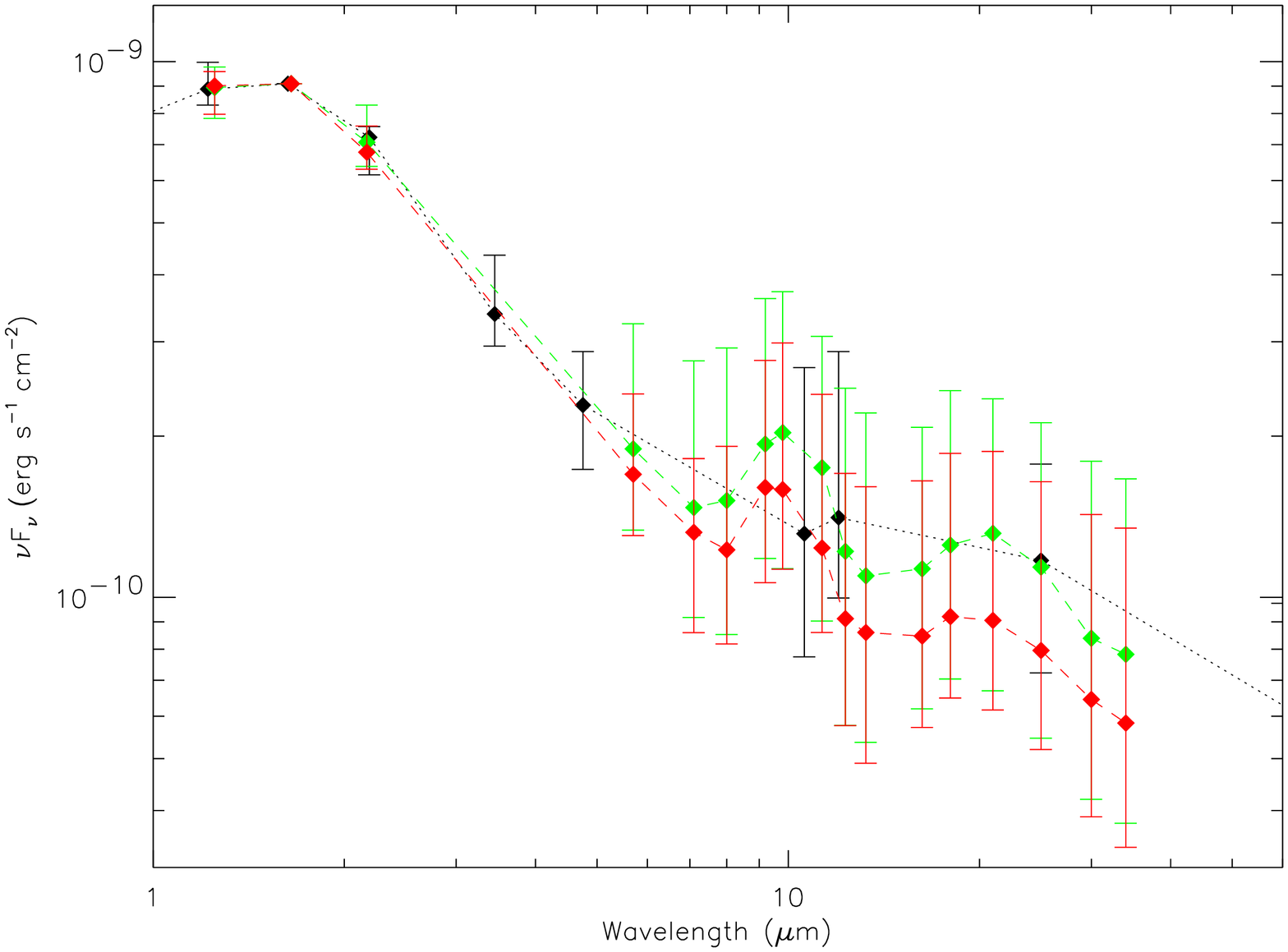}
\caption{Median SED from 1.25 to 34 $\mu$m computed using 2MASS
photometry and the IRS spectrum of all 85 Class II objects ({\it green})
and of those 55 Class II objects with spectral types between K5 and M2 
({\it red}). Also shown is the median SED from \citet{dalessio99} 
({\it black}). The error bars define the quartiles, i.e.\ the range around 
the median where 50\% of all flux values lie. \label{median_Taurus}}
\end{figure}

Figure \ref{median_Taurus} shows our median with the quartiles (indicated by 
the error bars) and that obtained by \citet{dalessio99}. We computed the median
using all 85 Class II objects in our sample ({\it upper, green data points}), and by including
only stars with spectral types between K5 and M2 ({\it lower, red data points}). The latter
data set constitutes a more homogeneous sample of stars with similar effective 
temperatures and allows a more direct comparison to the median from 
\citet{dalessio99}; it is also given in table \ref{tab_median}.

With the IRS data, we are able to resolve the silicate emission features at 10 
and 20 $\mu$m; the median SED clearly shows that most objects have 
pronounced silicate emission and thus small grains in the hot upper layers of the 
disk atmospheres. It is remarkable 
that our median SED at 25 $\mu$m almost coincides with the one from 
\citet{dalessio99}, who used {\it IRAS} 25 $\mu$m flux measurements.
The median SED for K5\,--\,M2 stars is lower than the one for all Class II 
objects in our sample, as expected for a sample of stars that  excludes earlier,
and thus hotter, spectral types. In particular, the decrease in the 13--25 
$\mu$m range is more pronounced than that at shorter wavelengths, 
indicating that the silicate feature at 20 $\mu$m decreases more 
than the 10-$\mu$m silicate feature for late spectral types.  

\begin{figure}
\includegraphics[angle=90,scale=0.7]{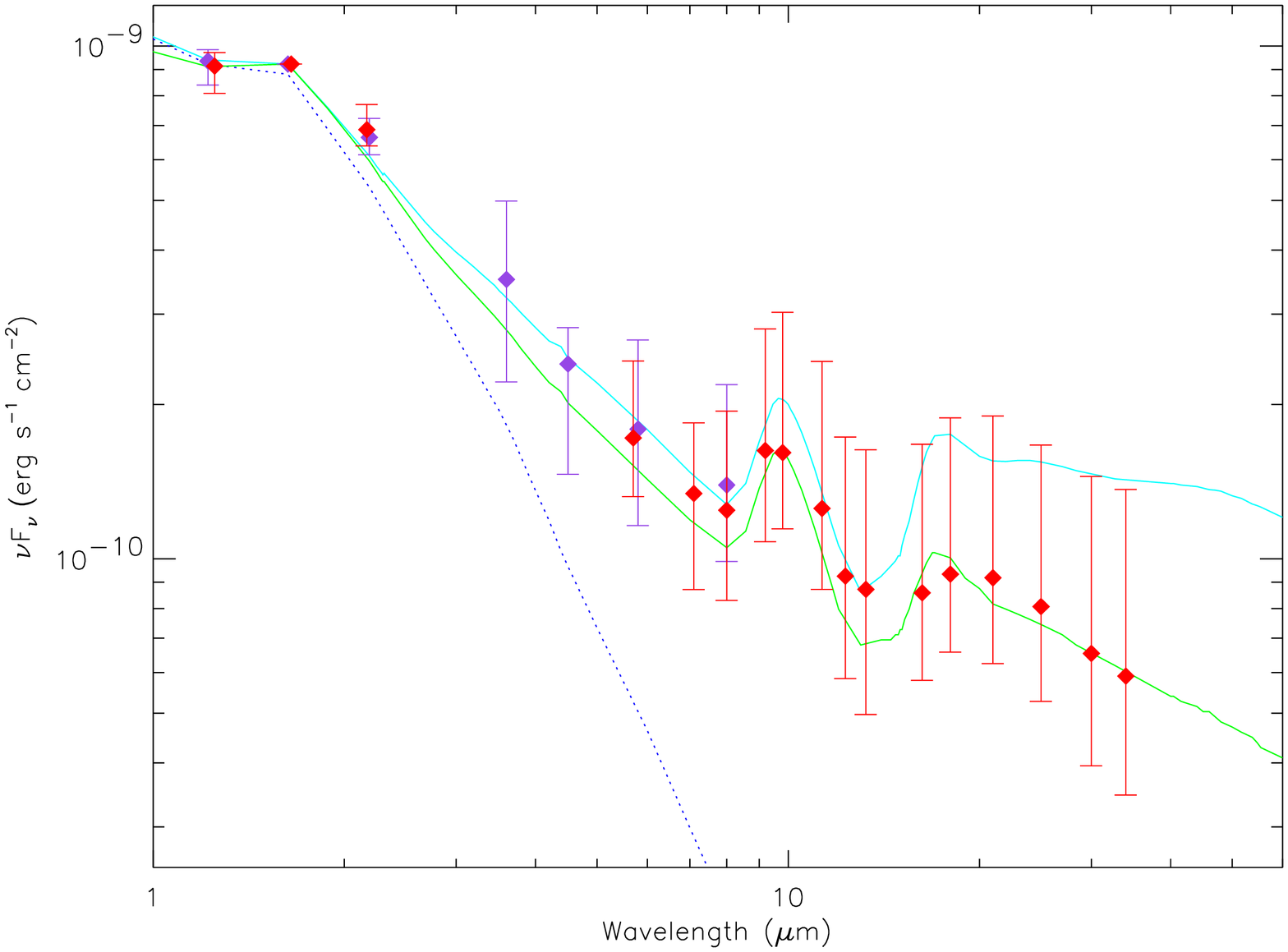}
\caption{Median SED from 1.25 to 34 $\mu$m computed using 2MASS
photometry and the IRS spectrum of only those Class II objects with spectral
types between K5 and M2 ({\it red}), as well as the median SED constructed 
using IRAC data from \citet{hartmann05a} ({\it purple}), compared to two
models with $\dot{M}=3 \times 10^{-8}$ M$_{\odot}$ yr$^{-1}$, 
i=60$\degr$, dust consisting of amorphous pyroxene and graphite, and 
two values of settling parameter $\epsilon$ ({\it upper, light blue line},
$\epsilon=0.1$; {\it lower, green line}, $\epsilon=0.001$). The dotted line 
represents the photosphere (SED of the WTTS HBC 427). 
\label{median_Taurus_models}}
\end{figure}

Figure \ref{median_Taurus_models} displays the IRS median SED for 
spectral types K5\,--\,M2 ({\it red}), the median computed from IRAC data 
\citep[][{\it purple}]{hartmann05a}, and the photospheric component ({\it dotted line}), 
represented by the SED of the WTTS HBC 427 \citep{hartmann05a}. 
We compare these to SEDs of two irradiated accretion disk models with 
different degrees of dust settling. The photospheric component is normalized 
at J, while all the other SEDs are normalized at H. 

The models shown in figure \ref{median_Taurus_models} were constructed 
according to the methods of \citet{dalessio06} and are described in section 
\ref{settling_models}, with a few differences: accretion shock irradiation
is included as heating mechanism for the disk (but it does not make a 
significant difference for $\dot{M} \lesssim 10^{-8}$ M$_{\odot}$ 
yr$^{-1}$), and the dust mixture consists of spherical particles of 
amorphous pyroxene \citep[glassy ${\rm Mg_{0.7} \, Fe_{0.3} \, SiO_3}$, 
optical constants from][]{dorschner95} and graphite \citep[optical constants 
from][]{draine84}. This dust composition results in higher silicate emission at
10 and 20 $\mu$m and therefore is a better fit for the Taurus median
SED (see also discussion in $\S$ \ref{IRS_sp_indices_models}). 
The models were computed for an inclination angle of 60$\degr$, a mass 
accretion rate of $3 \times 10^{-8}$ M$_{\odot}$ yr$^{-1}$, and two 
values of the settling parameter $\epsilon$, 0.1 and 0.001. 
Comparison of the models with the median and quartiles suggests large degrees 
of depletion in the upper layers of disks in Taurus: about 50\% of the disks have 
a dust depletion larger than a factor of 1000 ($\epsilon \le 0.001$), while 
about 25\% of the disks can be characterized with degrees of depletion between 
a factor of 10 and 1000. Only about 25\% of the disks have between a factor
of 10 and no depletion of dust in the upper disk layers.

\subsection{Class III Objects: Upper Limits for Disk Masses}

We estimated upper limits for the mass of small, warm dust grains (size
$\lesssim$ 10 $\mu$m, temperature $\gtrsim$ 100 K) in the disks around the 
26 Class III objects we observed, assuming optically thin dust emission
from ``astronomical'' silicates \citep[optical constants from][]{draine84}.
Since none of the Class III objects considered here show silicate emission 
bands at 10 or 20 $\mu$m, larger grains ($\gtrsim$ 3--4 $\mu$m),
if any, must be present in their disks; for this reason we included only grains between 
sizes of 3 and 10 $\mu$m (the mid-infrared is not sensitive to emission from
larger particles). 
First, we calculated the equilibrium temperature
for each grain size at a distance $r$ from the star by solving the following equation
numerically:
\begin{equation}
\left(\frac{R_{\ast}}{r}\right)^2 \int_0^{\infty} Q_{abs}(\nu)
B_{\nu}(T_{\ast}) d{\nu} = 4 \int_0^{\infty} Q_{abs}(\nu)
B_{\nu}(T_d) d{\nu}, 
\label{equ_temp}
\end{equation}
where $R_{\ast}$ is the stellar radius, $Q_{abs}$ is the absorption
efficiency, and $B_{\nu}(T)$ is the Planck function, evaluated at the
stellar effective temperature $T_{\ast}$ and at the dust grain temperature
$T_d$, respectively. Assuming that the disk is axially symmetric and 
geometrically thin, we parametrized the optical depth as
\begin{equation}
\tau(r,a) = \Sigma_0(r_0) \left( \frac{r}{r_0} \right)^{-\gamma}
\left( \frac{a}{a_0} \right)^{-3.5} \pi a^2 Q_{abs}(a, \nu)
\label{equ_tau},
\end{equation}
where $\Sigma_0(r_0)$ is the reference surface number density of dust grains
of size $a_0$ at a distance $r_0$ from the star; this column density varies
with radius as $r^{-\gamma}$, and the grain size distribution varies as
$a^{-3.5}$. We chose $r_0=1$~AU, $a_0=1$~${\mu}$m, and, besides
for a few exceptions, ${\gamma}=1$.
The flux emitted from the disk at each wavelength is obtained by integrating 
over annuli and summing over each dust grain size:
\begin{equation}
F_{\lambda} = \sum_{a_i} \left( \frac{1}{D^2} \int_{r_{in}}^{r_{out}}
2 \pi r \tau(r,a_i) B_{\nu}(T_d(r,a_i)) dr \right),
\label{equ_flux}
\end{equation}
where $\tau(r,a_i)$ is defined in equation (\ref{equ_tau}), $T_d(r,a_i)$
is obtained from equation (\ref{equ_temp}), and $D$ is the distance to the
star, assumed to be 140 pc, the distance to the Taurus star-forming region.
The total dust mass in the disk is given by
\begin{equation}
M_d = \sum_{a_i} m_d \int_{r_{in}}^{r_{out}} 2 \pi r \Sigma(r) dr,
\end{equation}
where $m_d$ is the mass of a single dust grain.
To obtain upper limits for the dust masses in our Class III objects, we
adjusted the parameter $\Sigma_0(r_0)$ (and in some cases also 
$\gamma$), until the disk emission from equation (\ref{equ_flux}) 
matched the emission of each Class III object after photosphere
subtraction. The photospheres are represented by AMES-dusty models 
of \citet{allard01} for effective temperatures up to 5000 K; for hotter
stars, we approximated the photospheres by blackbodies in the Rayleigh-Jeans
limit. We integrated the flux from 0.1 to 50 AU and summed over 11 grain 
sizes from 3 to 10 $\mu$m radius. 

With all the assumptions mentioned above, our model fits result in 
typical upper limits for warm, small dust grains in any circumstellar disks 
around our Class III objects of a few times 10$^{-4}$ lunar masses. 
Assuming that the infrared excess of HBC 427 and V819 Tau can be fully 
attributed to disks around these objects, the maximum amount of dust mass 
in these disks would 
amount to a few times 10$^{-3}$ lunar masses. In addition, since their 
excess starts rather abruptly at 14 $\mu$m, their disks would require inner 
holes of a few AU in size to explain the lack of shorter-wavelength excess emission. 
HP Tau/G2 seems to already have a slight excess shortward of 14 $\mu$m; 
again assuming all the excess emission originates in a disk around this object, 
the upper limit to the dust mass would amount to somewhat less than 10$^{-4}$ lunar 
masses. A dust mass of a few times 10$^{-4}$ lunar masses is comparable 
to that found around debris disks \citep{habing01}, although closer to the 
lower mass values.

\section{Discussion}
\label{discuss}

The multi-wavelength SEDs, which weight the spectrum in terms of the relative
power emitted at each wavelength, provide an immediate initial estimate of
the relative luminosities of the star and the disk. In most cases, the luminosity of
the disk is less than that of the star because most of the disk heating and thus 
disk emission are due to absorption and reprocessing of stellar radiation. 
However, of the 85 Class II systems shown in Figure \ref{fig_classII},
XZ Tau, UY Aur, HN Tau, FS Tau, 04187+1927, 04200+2759, DG Tau, and DP Tau 
appear to have anomalously large IRS excesses relative to the central star.  
As noted in section \ref{notes_object}, a few of these objects have envelopes;
UY Aur has an IR companion separated by 0{\farcs}9 from the primary and likely
seen through the disk of the primary, given its larger reddening \citep{hartigan03}.
Together with the other $\sim 9$ objects with very high and therefore uncertain $A_V$ 
mentioned in section \ref{notes_object}, there might be about 17 candidates for 
edge-on disks. In addition, HK Tau B \citep{stapelfeldt98} and likely also AA Tau
\citep{menard03,bouvier03} have edge-on disks, bringing the total of T Tauri stars
likely seen at an inclination angle close to 90{\degr} to 19. Thus, 22\% of the Class II
objects in our sample (or 17\% of the T Tauri stars of our Taurus sample) are likely 
seen edge-on, which is roughly what is expected from statistical considerations 
for objects with inclinations between 77{\degr} and 90{\degr}.
In addition, these objects do not show any deep silicate absorption features,
as expected from models \citep{dalessio99}; therefore, either these highly inclined
disks are not massive enough, or we are observing the 10-$\mu$m thermal emission
from the inner disk, scattered by the dust in the outer disk atmosphere 
\citep[see][]{mccabe03}.  

A few of our Class II objects show a decrease or lack of IR excess at shorter IRS 
wavelengths, indicative of inner disk clearings or gaps.  Some of these sources 
(GM Aur, DM Tau, CoKu Tau/4) have been described as transition disks and 
are the subject of other papers \citep{dalessio06, calvet05}. Only 4\% of 
the Class II objects in our sample, or 3\% of the 111 T Tauri stars we observed 
in Taurus, are in this transitional disk phase, which would indicate that this 
stage is short-lived. 

If their longer-wavelength excess proves to be real, HBC 427, V819 Tau, and 
possibly also V410 X-ray 3, could also be considered as objects in transition,
even though their excess emission is so weak that it is probably not optically
thick disk emission. On the other hand, they might
be regarded as debris disks, where dust is generated by collisions of larger bodies
and is optically thin at all wavelengths. If that were the case, they would be among
the youngest debris disks discovered so far.
 
UX Tau A has a very peculiar spectrum, with significant excess at shorter wavelengths, 
but a sharp transition at 20 $\mu$m. Even though it is a multiple system, the other
components are at distances that are unlikely to affect the inner disk we are observing 
with the IRS. Also, their flux contribution in the mid-IR is expected to be small. Thus,
the IRS spectrum of UX Tau A suggests that it might be in a pre-transition state, 
when most dust in the inner disk has already settled and grown, and thus the lifetime 
of the inner disk is coming to an end. 

Other objects that could be in a pre-transition state are 04385+2550, Haro 6-13, 
HK Tau, and LkCa 15; their 20-$\mu$m silicate feature 
is stronger or comparable to the 10-$\mu$m feature, reminiscent of transitional 
disks, but they have substantial excess emission in the near-IR. In these objects, 
a gap might be opening, but a significant inner disk is still present.

\section{Conclusions}
\label{conclude}

We presented the mid-infrared spectra of 85 Class II and 26 Class III 
objects in Taurus, obtained with the Infrared Spectrograph on board 
the {\it Spitzer Space Telescope}. This large sample of pre-main-sequence
stars of similar age, belonging to the same star-forming region, allows
us to draw the following conclusions about circumstellar disk evolution:

${\bullet}$ All CTTSs and some WTTSs display mid-IR excesses, which are
generated in the inner (a few tenths to a few AU) parts of circumstellar
disks. Accretion signatures like the H$\alpha$ equivalent line width do 
not necessarily indicate the presence or absence of disks, but they usually
reveal whether the innermost disk regions are still present. While accreting
T Tauri stars have dusty disks extending from the dust sublimation radius
outward and thus generate excess emission starting at near-IR wavelengths, 
WTTSs often have inner disk holes and thus lack near-IR excess emission.
Since there are few WTTSs with disks, the transition period between an 
active (i.e., accreting) and a passive (i.e., purely reprocessing) disk stage 
seems short-lived.

${\bullet}$ The details of the mid-IR excess, and thus the distribution and
properties of the dust in circumstellar disks, do not seem to be correlated
with the multiplicity of the young star system. A sub-arcsecond ($\lesssim$ 140 AU 
at the distance of Taurus) companion does not necessarily influence the inner
disk; even spectroscopic binaries, like DQ Tau, seem to have substantial inner
disks. We note that our IRS spectra cannot separate the contribution of
close multiple systems, and therefore, in some cases, they show the combined
mid-IR spectrum of circumprimary, circumsecondary, and circumbinary disk. The mid-IR
spectra of multiple systems do not have any peculiarities that would set them
apart from single stars; this could be explained if most disks in these systems are 
not influenced by the presence of companions and either the circumprimary disk 
clearly dominates the system or all components have similar disks, or if a 
circumbinary disk mimicked the behavior of a circumstellar disk around
a single star. 

${\bullet}$ Almost all Class II objects show 10- and 20-$\mu$m silicate
emission features in their mid-IR spectra, generated in the optically thin
disk surface layer. There is a wide variety of feature strengths and crystallinity,
independent of the spectral type and multiplicity of the objects. However,
objects with weak 10-$\mu$m silicate features usually have a decreasing
SED in the mid-IR range, which could be an indication of dust growth and
settling. In objects where the 10- and 20-$\mu$m features are suppressed, 
dust grains must exceed radii of about 4 $\mu$m.

${\bullet}$ After comparing the spectral indices 
$d \log(\lambda  F_{\lambda})/d \log (\lambda)$ between 6 
and 13 $\mu$m and between 13 and 25 $\mu$m of our Class II 
objects with accretion disk models that include dust settling, we infer that our
data are consistent with these models. Within the assumptions of the adopted 
models, most T Tauri stars seem to have experienced substantial dust settling 
(and likely also dust growth), with depletions of factors of about 100--1000 of the 
standard dust-to-gas mass ratio in the disk atmosphere. 
Thus, we are able to estimate the amount of dust settling for a large sample of 
T Tauri stars, using the high sensitivity and spectral coverage of the IRS. 
This result will aid in the refinement of models predicting the evolution
of dust in protoplanetary disks.

${\bullet}$ Most WTTSs in our sample are Class III objects, which have 
nearly photospheric fluxes in the infrared, indicating a dispersal of dust in their 
circumstellar disks. The upper limit for the mass of small, warm dust grains 
in these disks lies around 10$^{-4}$
lunar masses, which is similar to the dust mass of low-mass debris disks.
In some cases nearby companions ($\sim$ 10\arcsec), which seem 
to have an infrared excess, contaminated the LL part of our Class III sources.
It is interesting that objects of roughly the same age, formed in close proximity
to each other, do evolve on different time scales. This result, as well as the
diversity in mid-IR spectra observed for our Class II objects, indicates that
age does not seem to be the dominating factor determining disk evolution, 
but rather the initial conditions of the star-forming core, like its mass and
angular momentum. 

While there likely is an evolutionary sequence from a classical T Tauri star 
with an accretion disk to a weak-lined T Tauri star with a passive disk to a 
T Tauri star with little or no disk left, there does not seem to be an absolute, 
evolutionary timescale regulating when the transitions from one stage to the 
next occur. Therefore, we can witness circumstellar disk evolution in young 
star-forming regions such as Taurus and infer the major steps in the process 
by analyzing a large sample of pre-main-sequence stars; this paper is a 
contribution to this effort by presenting and analyzing unprecedented data from the
{\it Spitzer Space Telescope}.


\acknowledgments
We thank our anonymous referee for helpful comments and suggestions.
This work is based on observations made with the {\it Spitzer Space Telescope}, 
which is operated by the Jet Propulsion Laboratory, California Institute of Technology, 
under NASA contract 1407. Support for this work was provided by NASA through 
contract number 1257184 issued by JPL/Caltech. N.C. and L.H. acknowledge support
from NASA grants NAG5-13210 and NAG5-9670, and STScI grant AR-09524.01-A.
P.D. acknowledges grants from PAPIIT, UNAM and CONACyT, M\'exico.
This publication makes use of data products from the Two Micron All Sky Survey, 
which is a joint project of the University of Massachusetts and the Infrared Processing 
and Analysis Center/California Institute of Technology, funded by the National 
Aeronautics and Space Administration and the National Science Foundation. It has also 
made use of the SIMBAD and VizieR databases, operated at CDS (Strasbourg, France),
NASA's Astrophysics Data System Abstract Service, and of the NASA/ IPAC Infrared 
Science Archive operated by JPL, California Institute of Technology (Caltech), under 
contract with NASA.




\clearpage


\begin{figure}
\epsscale{0.75}
\plotone{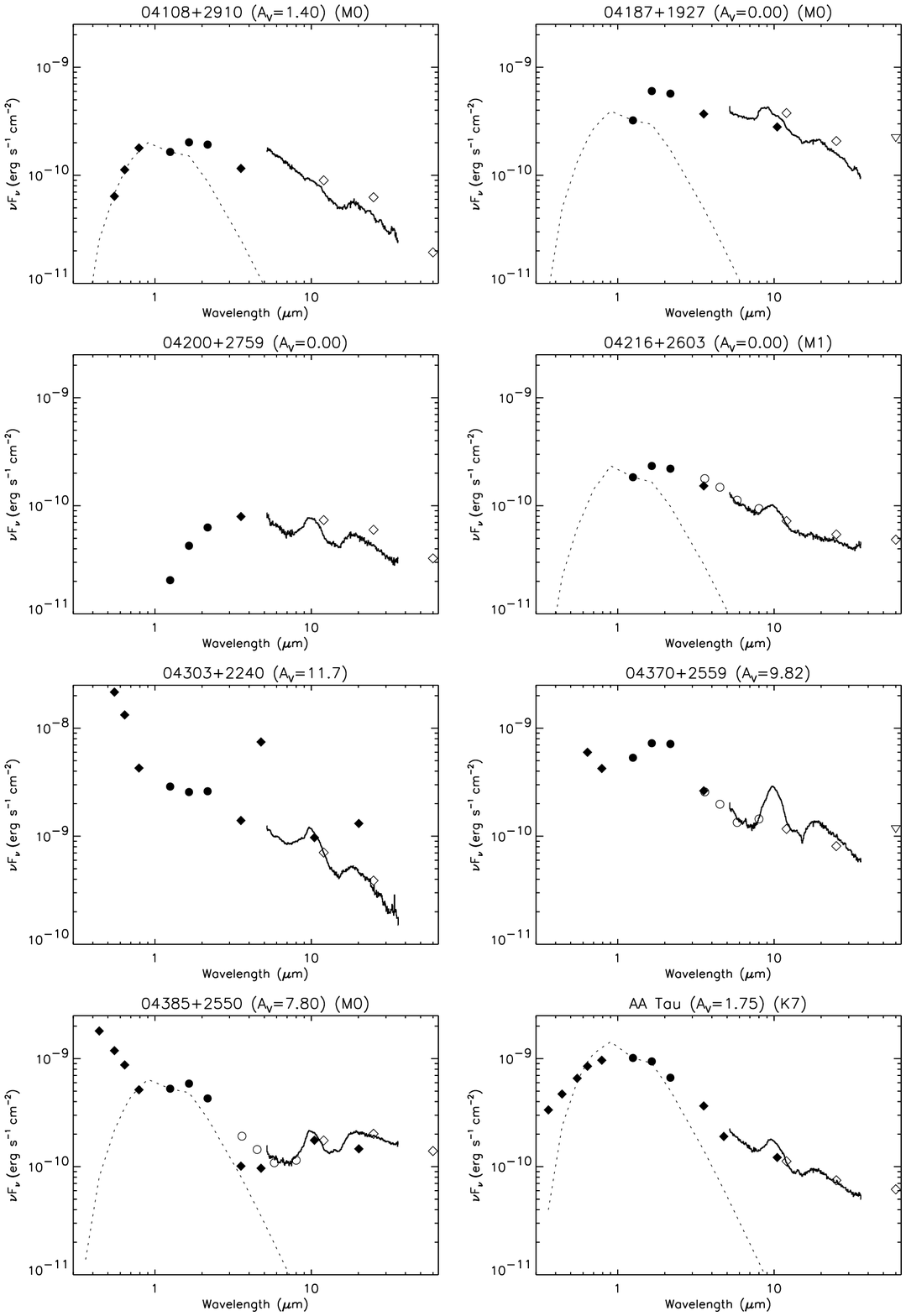}
\caption{SED plots of the Class II objects in our sample, ordered alphabetically by their 
name. In addition to the IRS spectrum, we include optical to mid-IR, ground-based 
photometry ({\it filled diamonds}) mostly from KH95, the 2MASS
$J$, $H$, and $K_s$ fluxes ({\it filled circles}), the IRAC 3.6, 4.5, 5.8, and 8.0 $\mu$m
fluxes ({\it open circles}) from \citet{hartmann05a}, where available, and the
{\it IRAS} 12, 25, and 60 $\mu$m fluxes ({\it open diamonds}, or {\it open triangles}
if upper limit) mostly from \citet{weaver92}. The photosphere is also sketched in for 
objects with known spectral type (see text for details). All fluxes, including the IRS 
spectrum, were dereddened using Mathis's reddening law \citep{mathis90} and 
the extinctions listed in table \ref{tab2}. \label{fig_classII}}
\end{figure}

\clearpage 

\begin{figure}
\epsscale{0.85}
\plotone{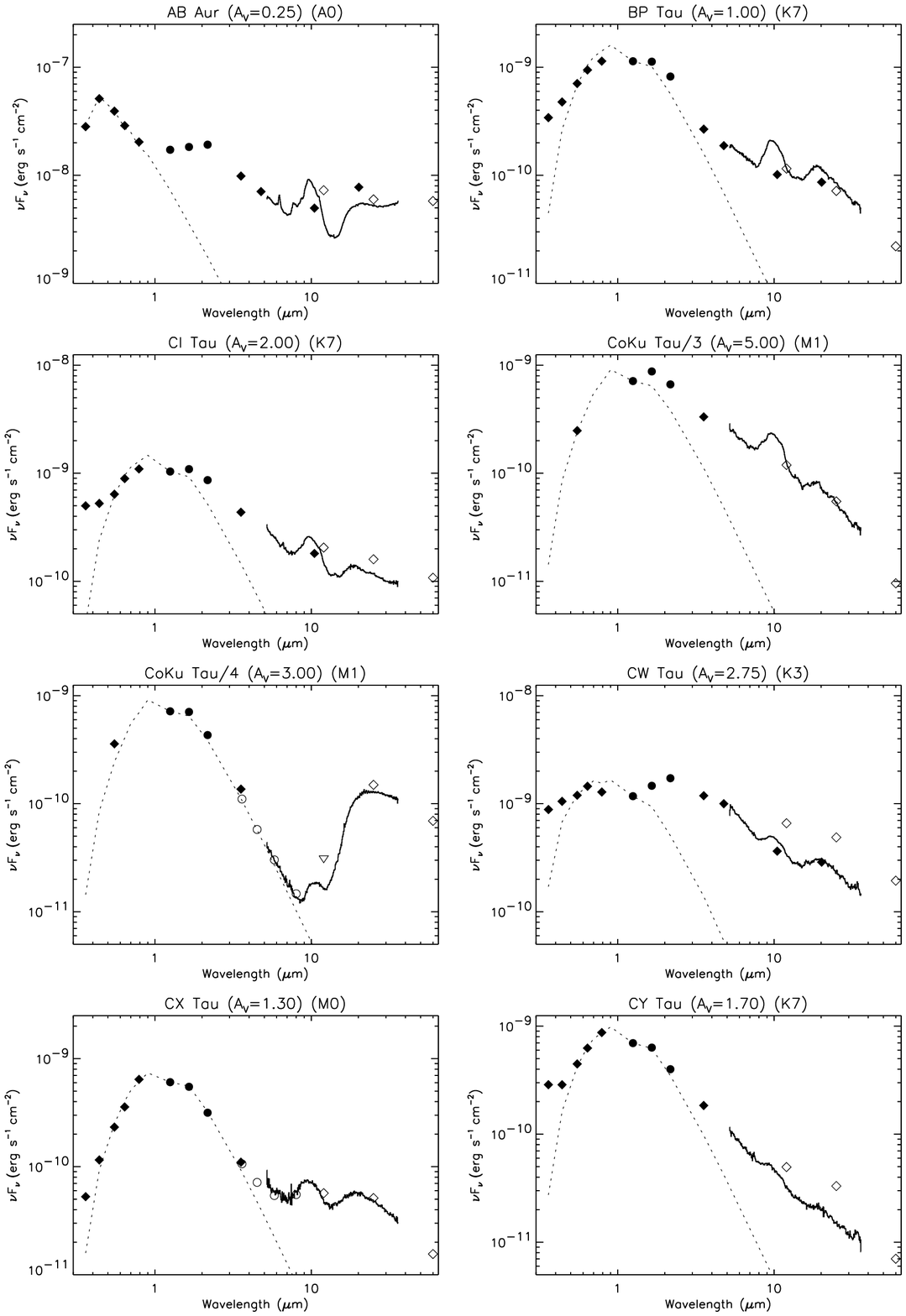}
\figurenum{\ref{fig_classII}}\caption{continued.}
\end{figure}

\clearpage 

\begin{figure}
\epsscale{0.85}
\plotone{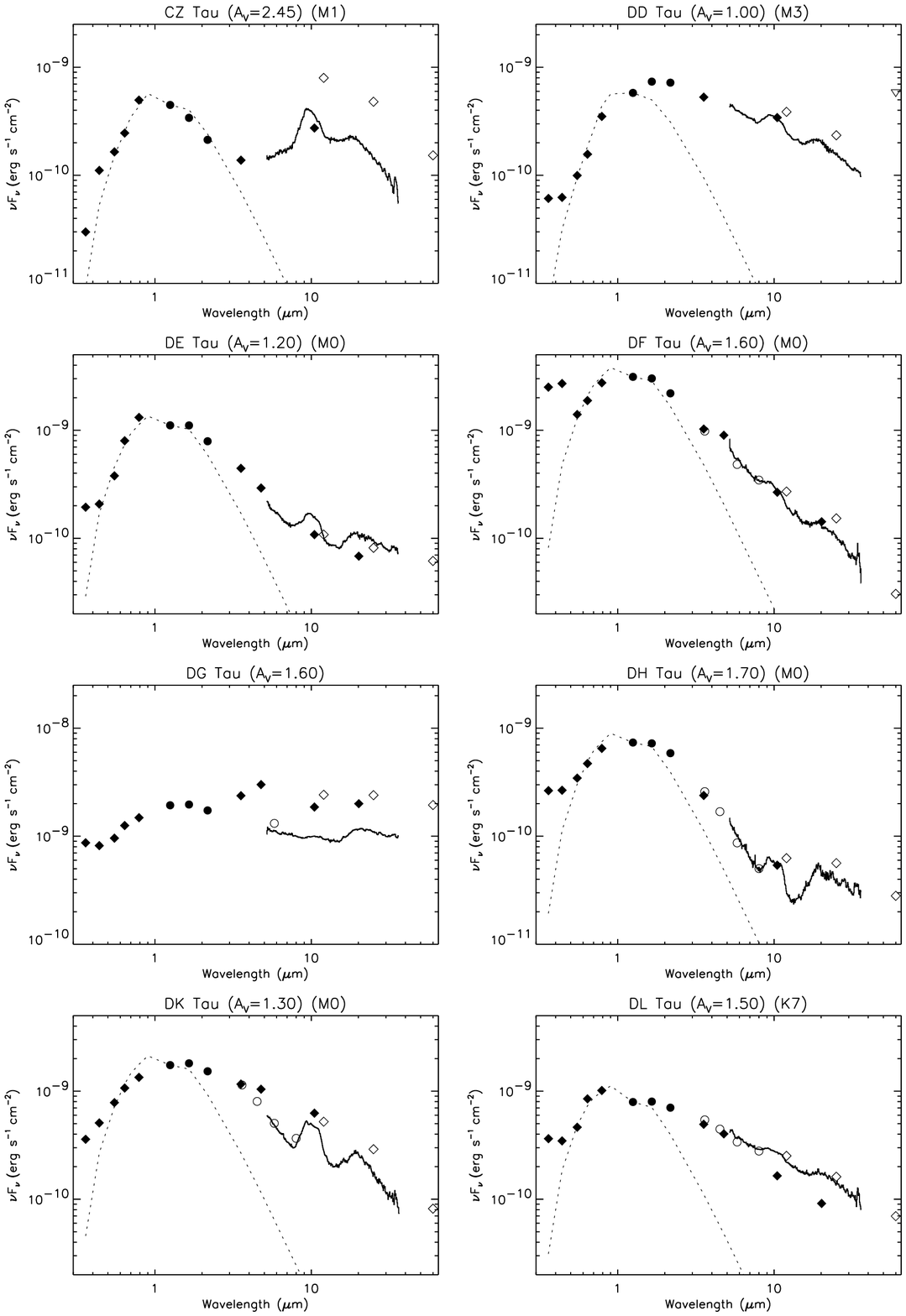}
\figurenum{\ref{fig_classII}}\caption{continued.}
\end{figure}

\clearpage 

\begin{figure}
\epsscale{0.85}
\plotone{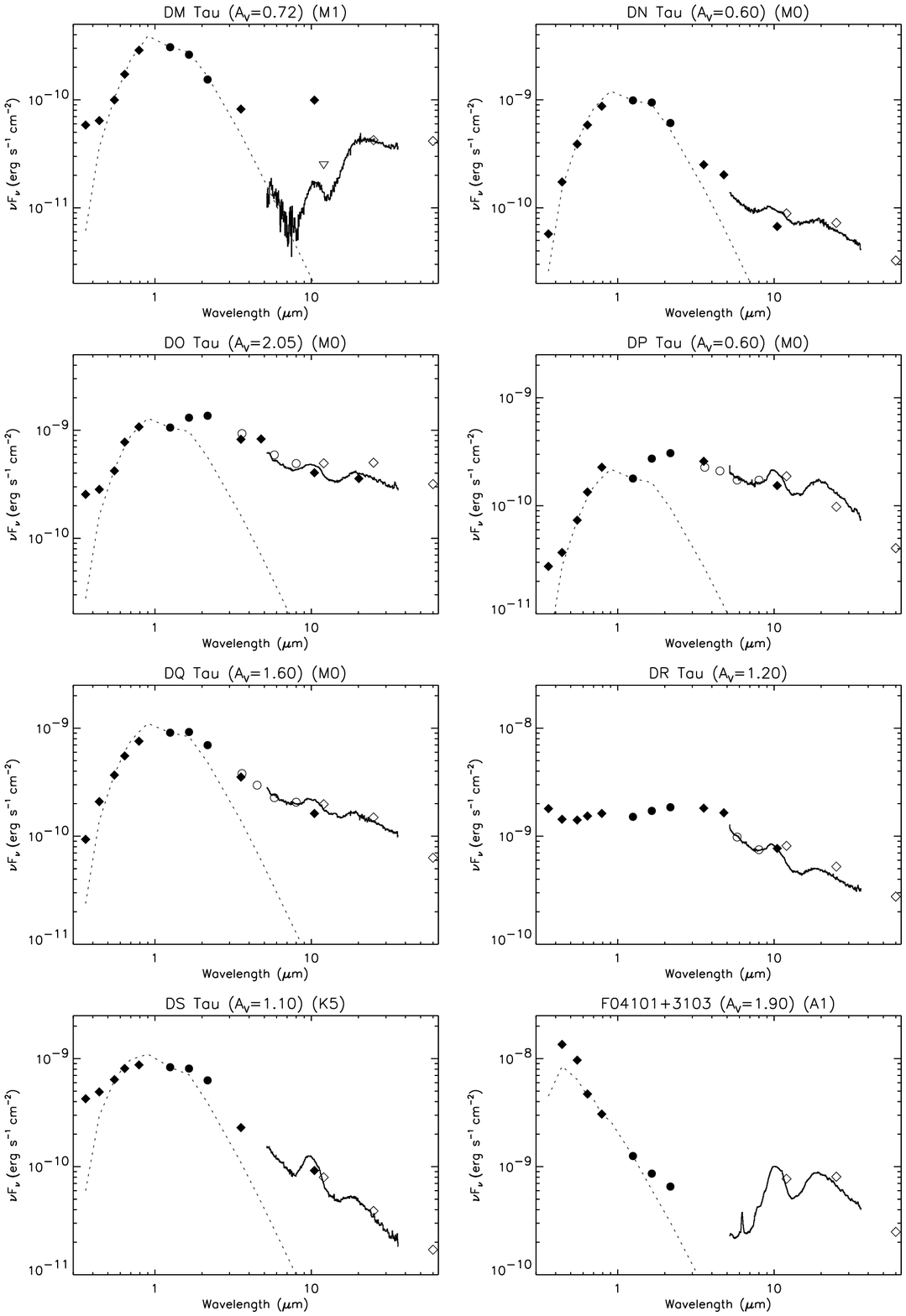}
\figurenum{\ref{fig_classII}}\caption{continued.}
\end{figure}

\clearpage 

\begin{figure}
\epsscale{0.85}
\plotone{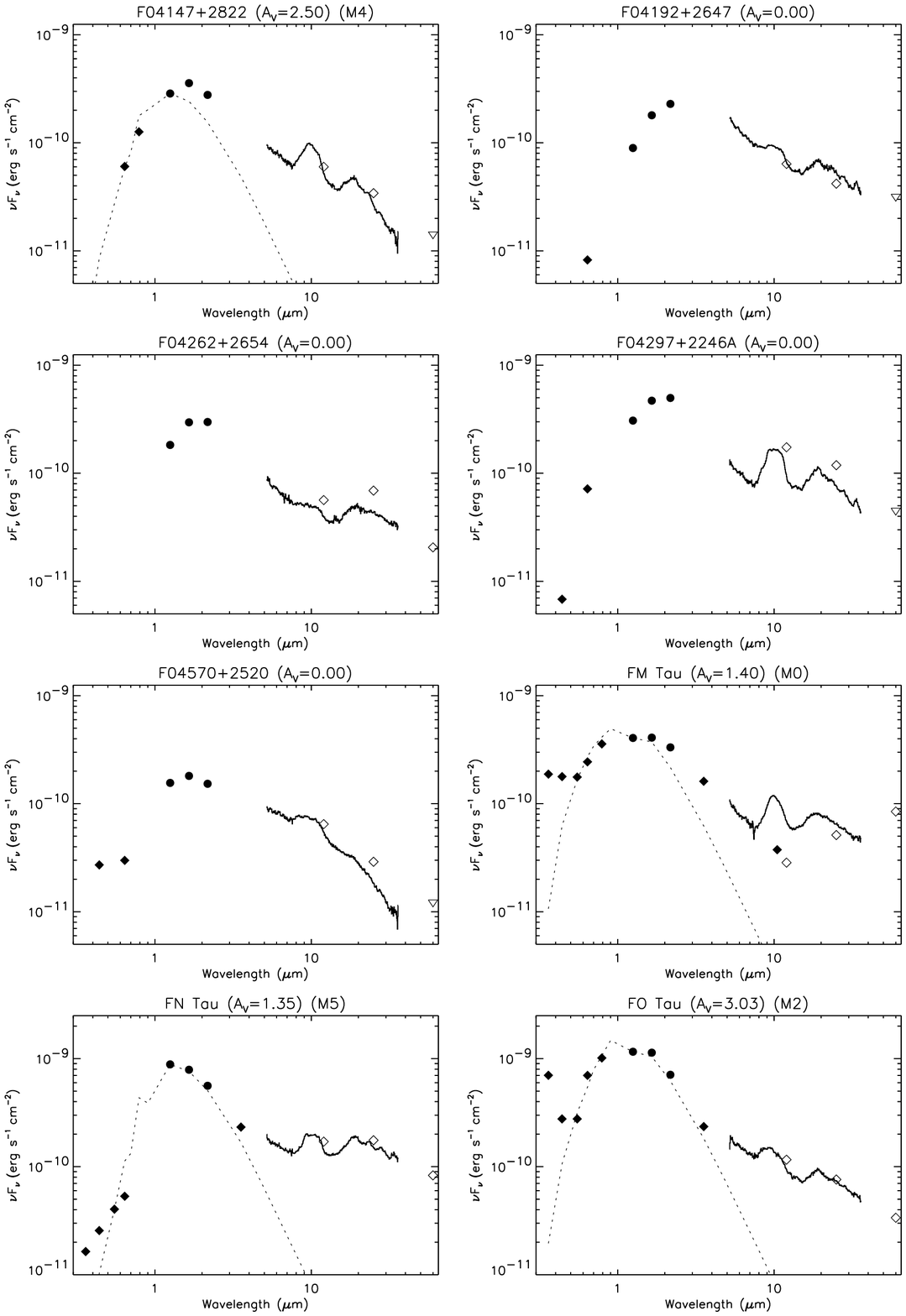}
\figurenum{\ref{fig_classII}}\caption{continued.}
\end{figure}

\clearpage 

\begin{figure}
\epsscale{0.85}
\plotone{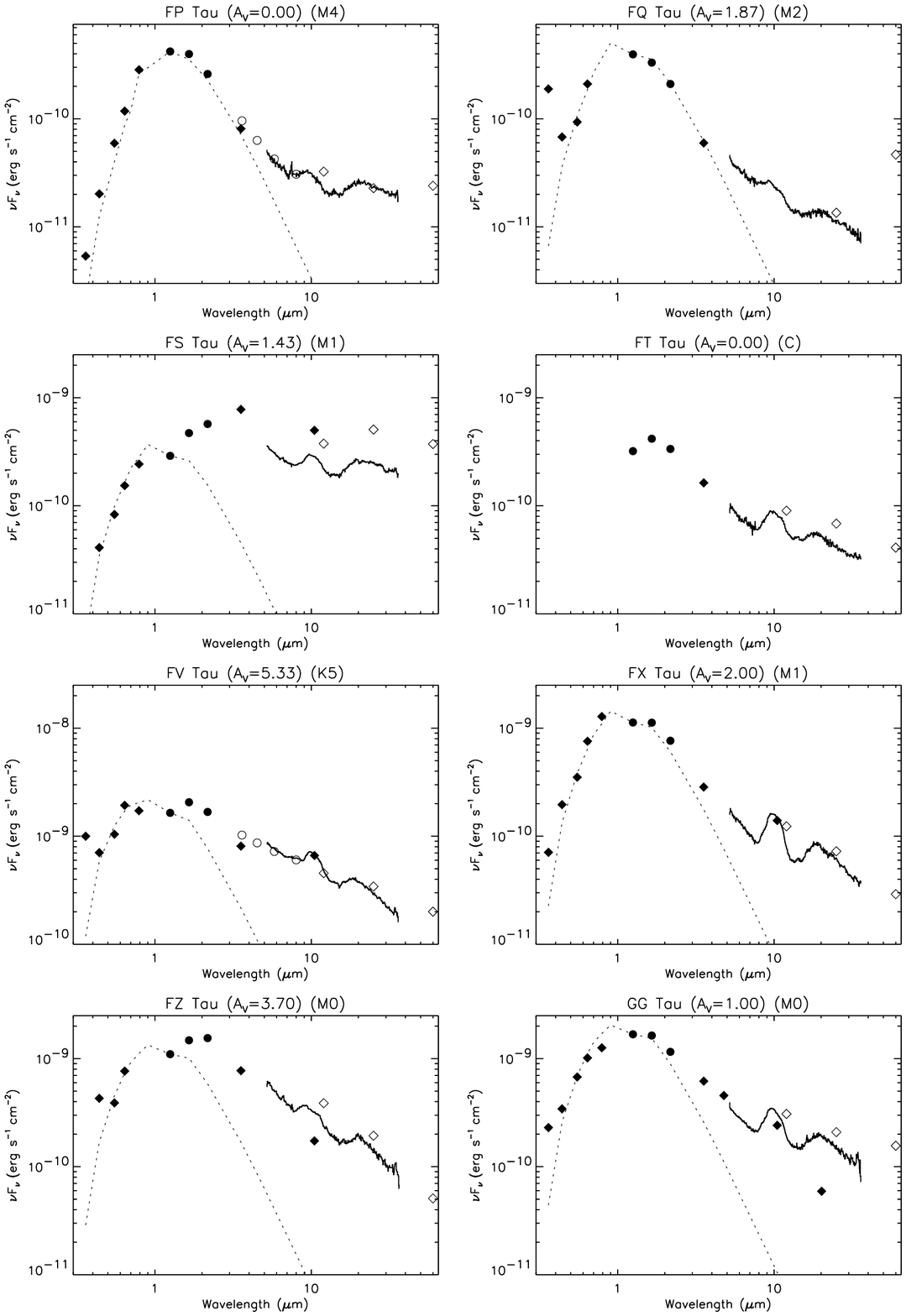}
\figurenum{\ref{fig_classII}}\caption{continued.}
\end{figure}

\clearpage 

\begin{figure}
\epsscale{0.85}
\plotone{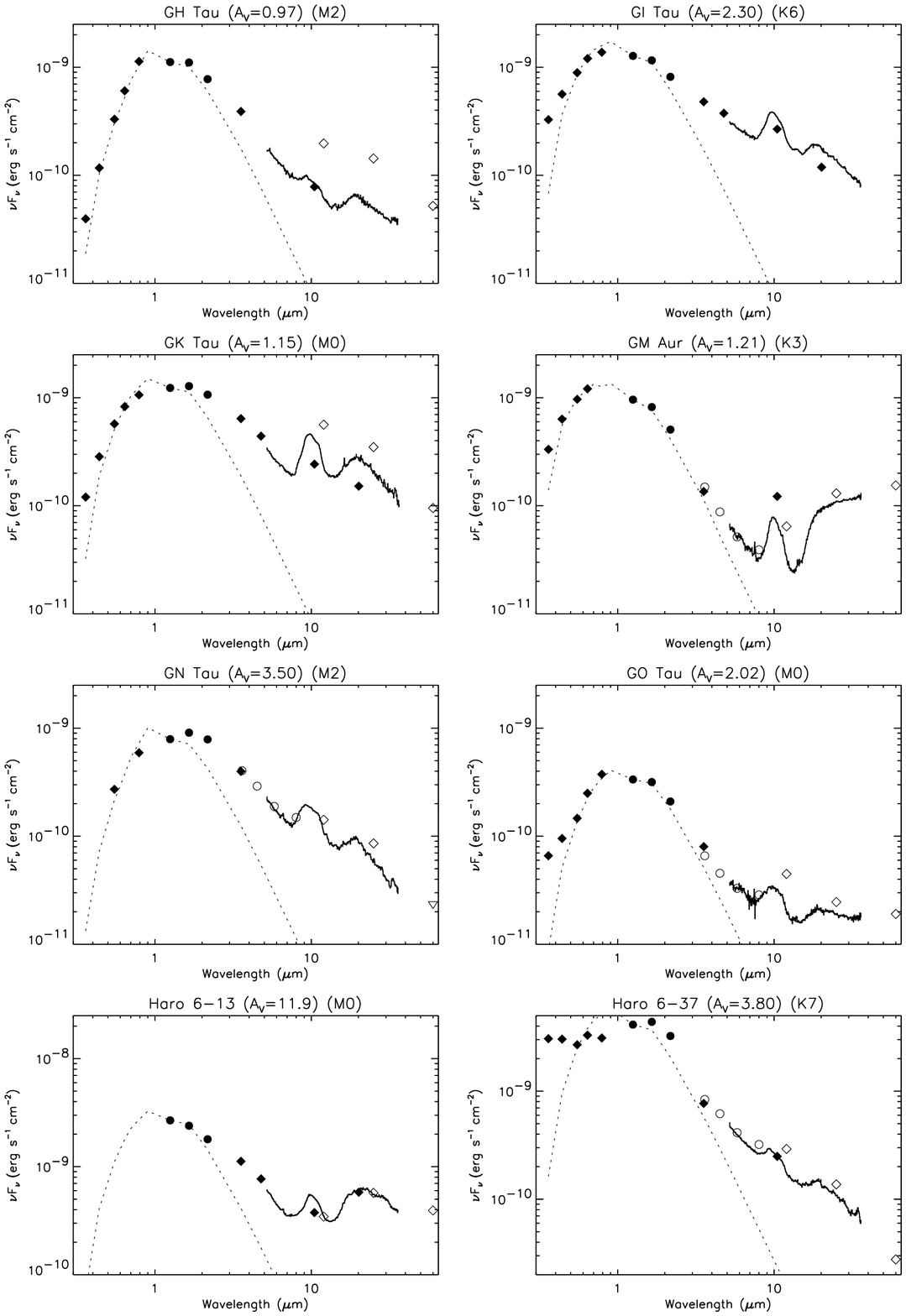}
\figurenum{\ref{fig_classII}}\caption{continued.}
\end{figure}

\clearpage 

\begin{figure}
\epsscale{0.85}
\plotone{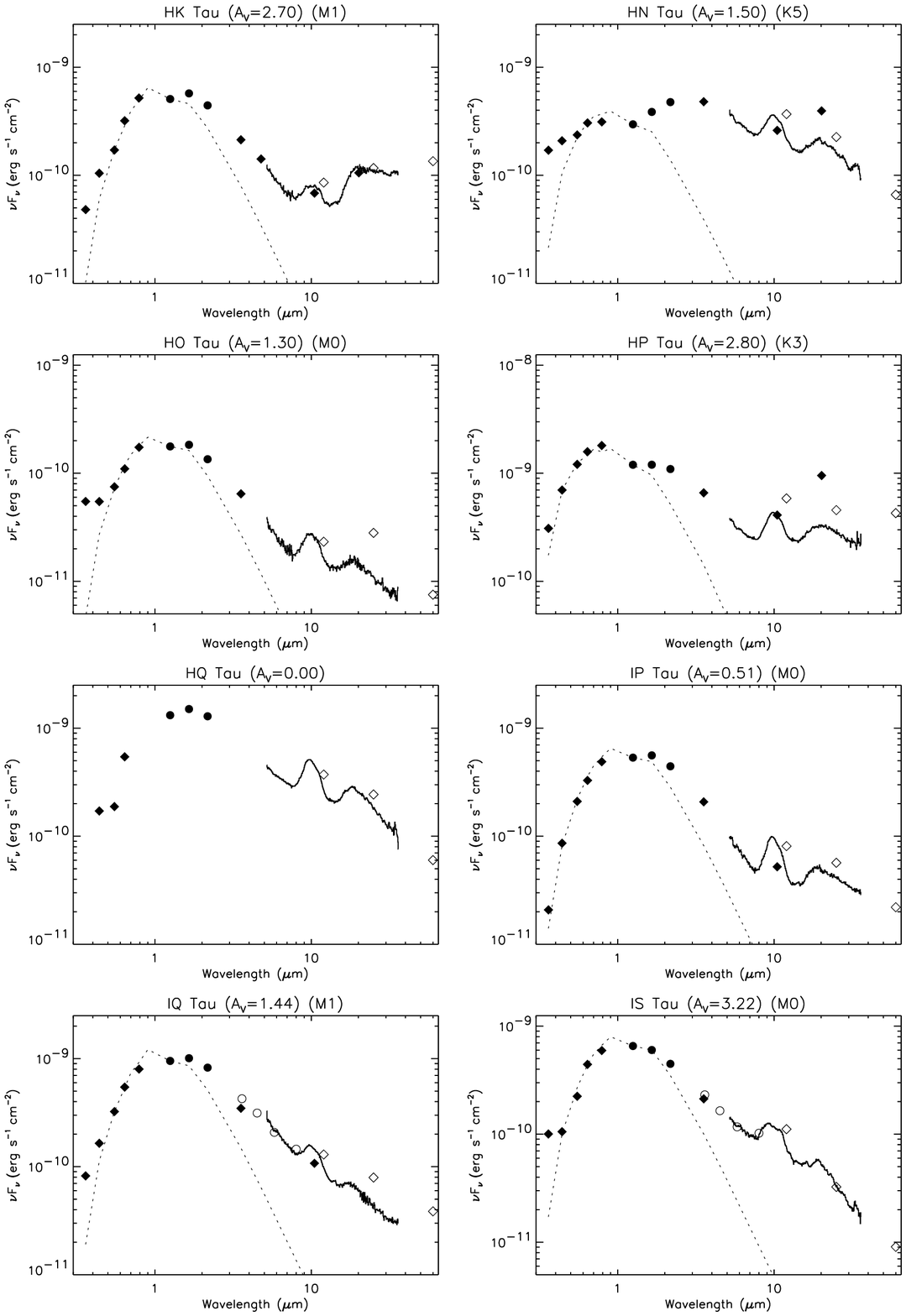}
\figurenum{\ref{fig_classII}}\caption{continued.}
\end{figure}

\clearpage 

\begin{figure}
\epsscale{0.85}
\plotone{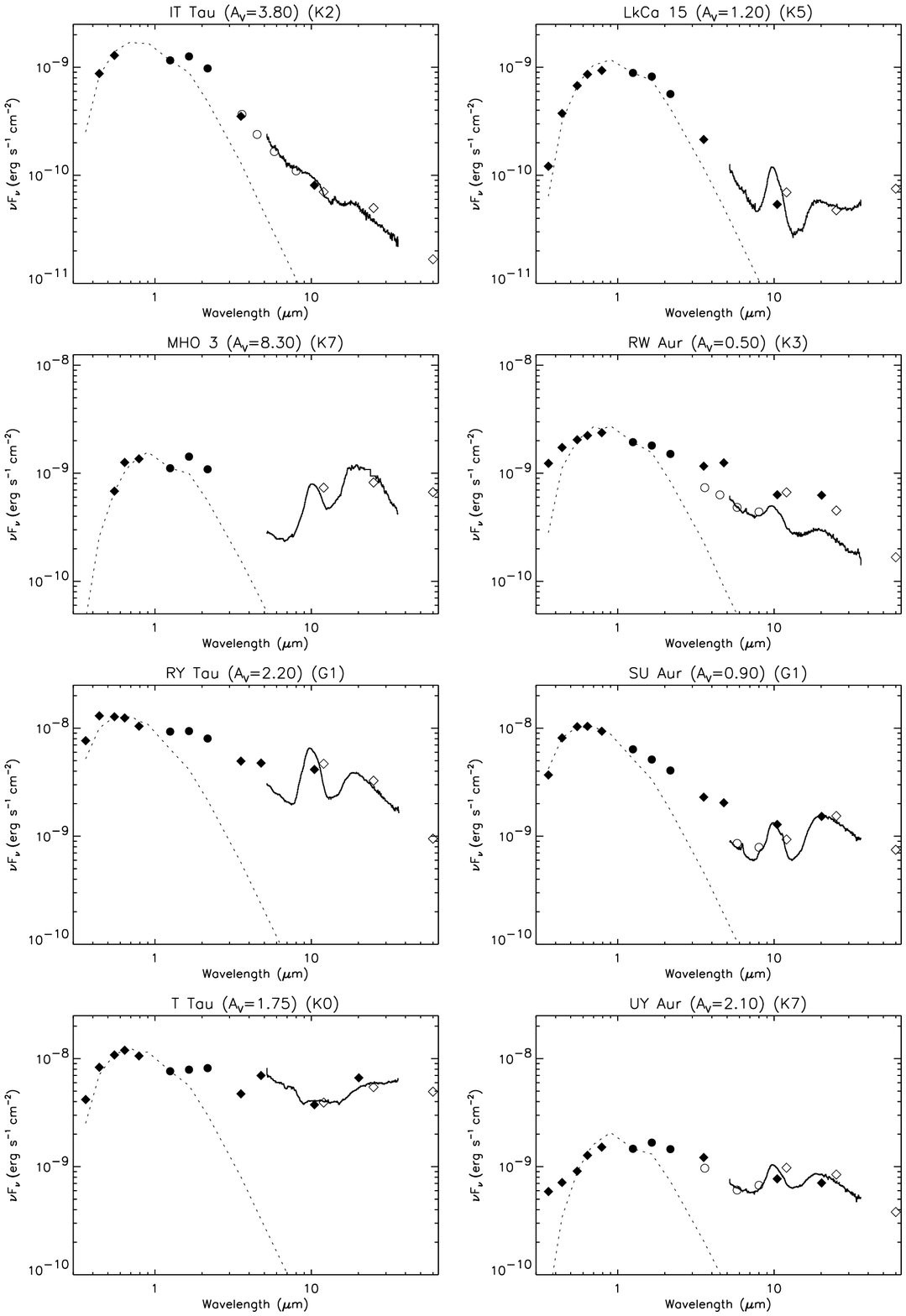}
\figurenum{\ref{fig_classII}}\caption{continued.}
\end{figure}

\clearpage 

\begin{figure}
\epsscale{0.85}
\plotone{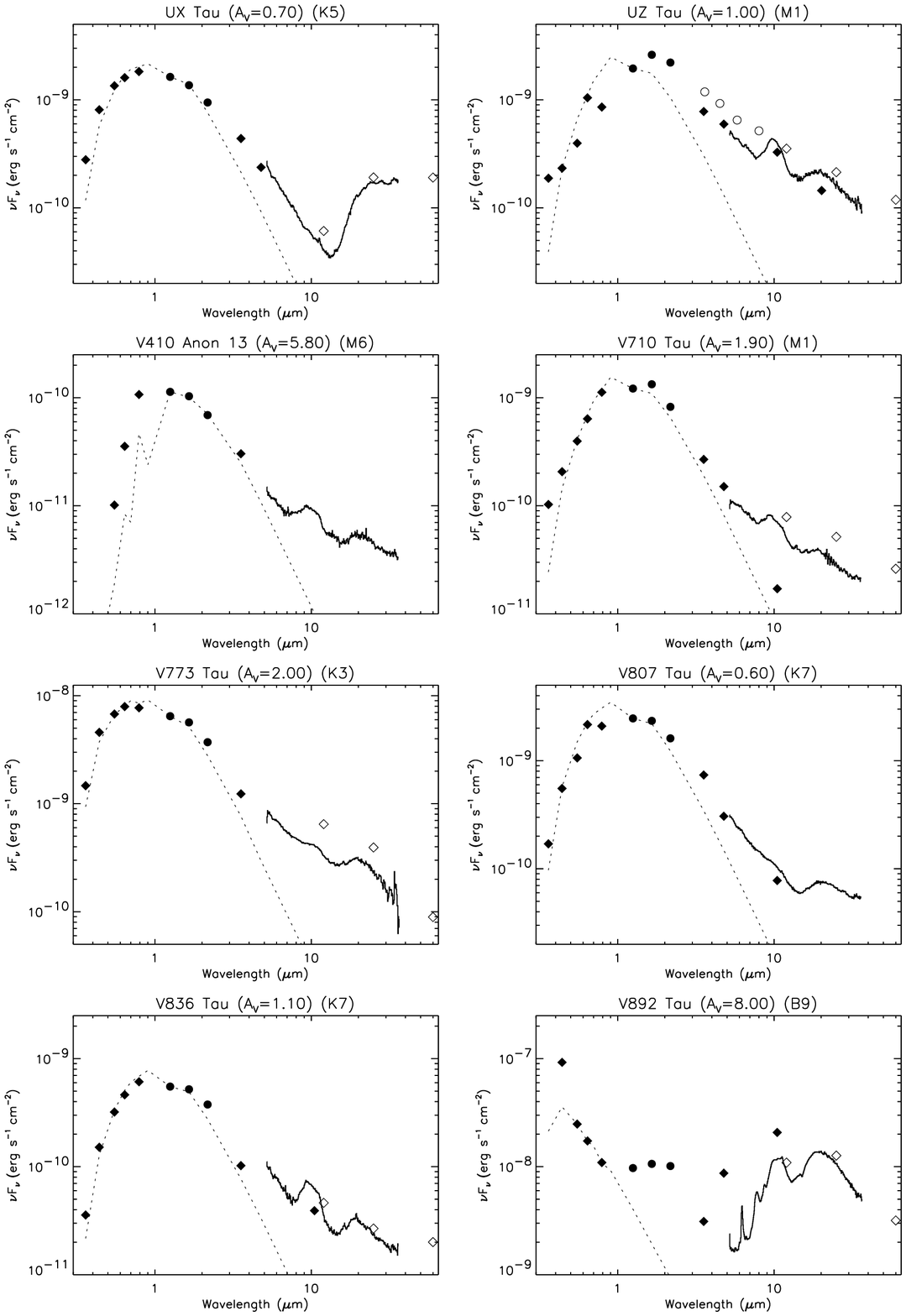}
\figurenum{\ref{fig_classII}}\caption{continued.}
\end{figure}

\clearpage 

\begin{figure}
\epsscale{0.85}
\plotone{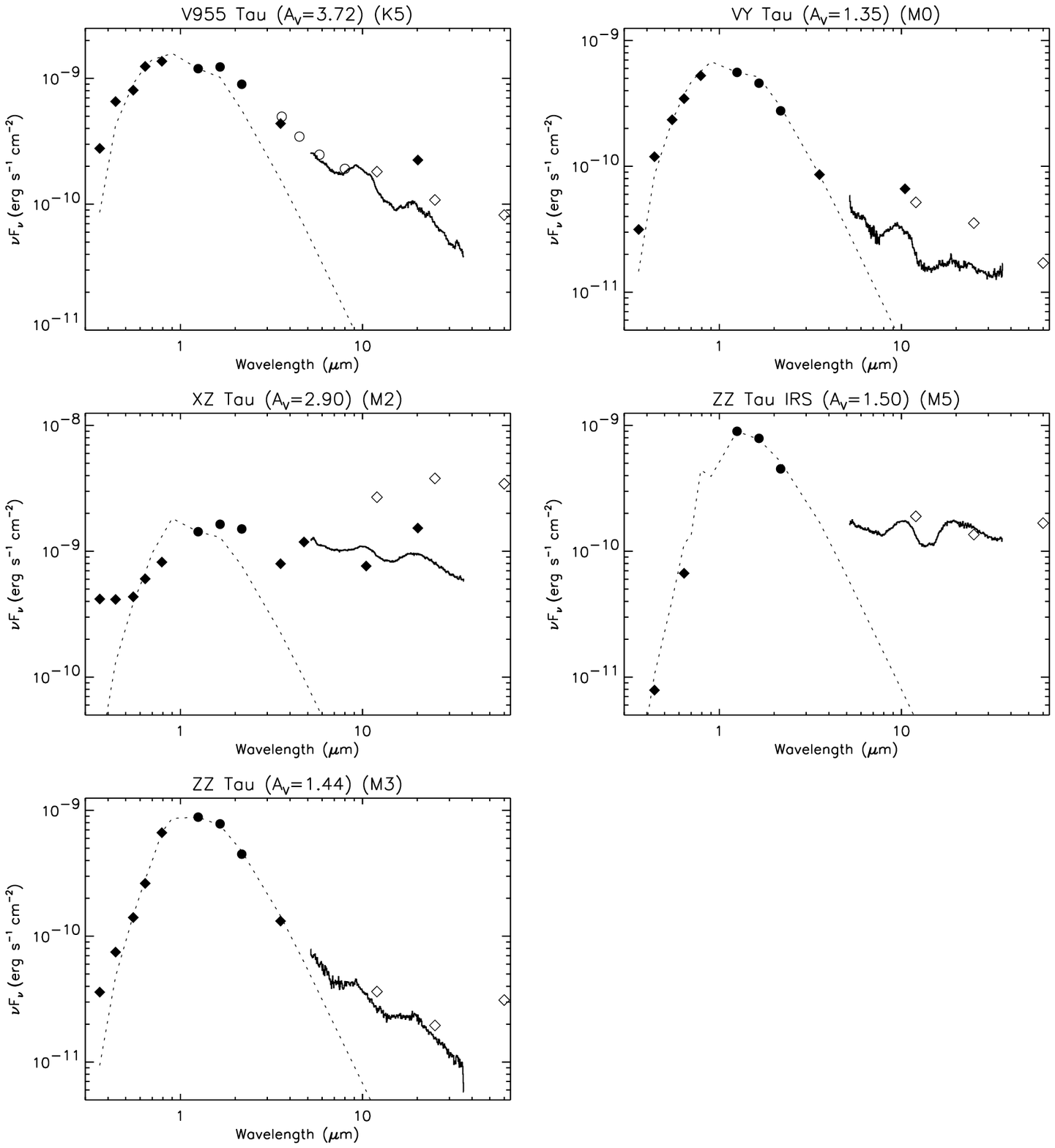}
\figurenum{\ref{fig_classII}}\caption{continued.}
\end{figure}

\clearpage 

\begin{figure}
\epsscale{0.85}
\plotone{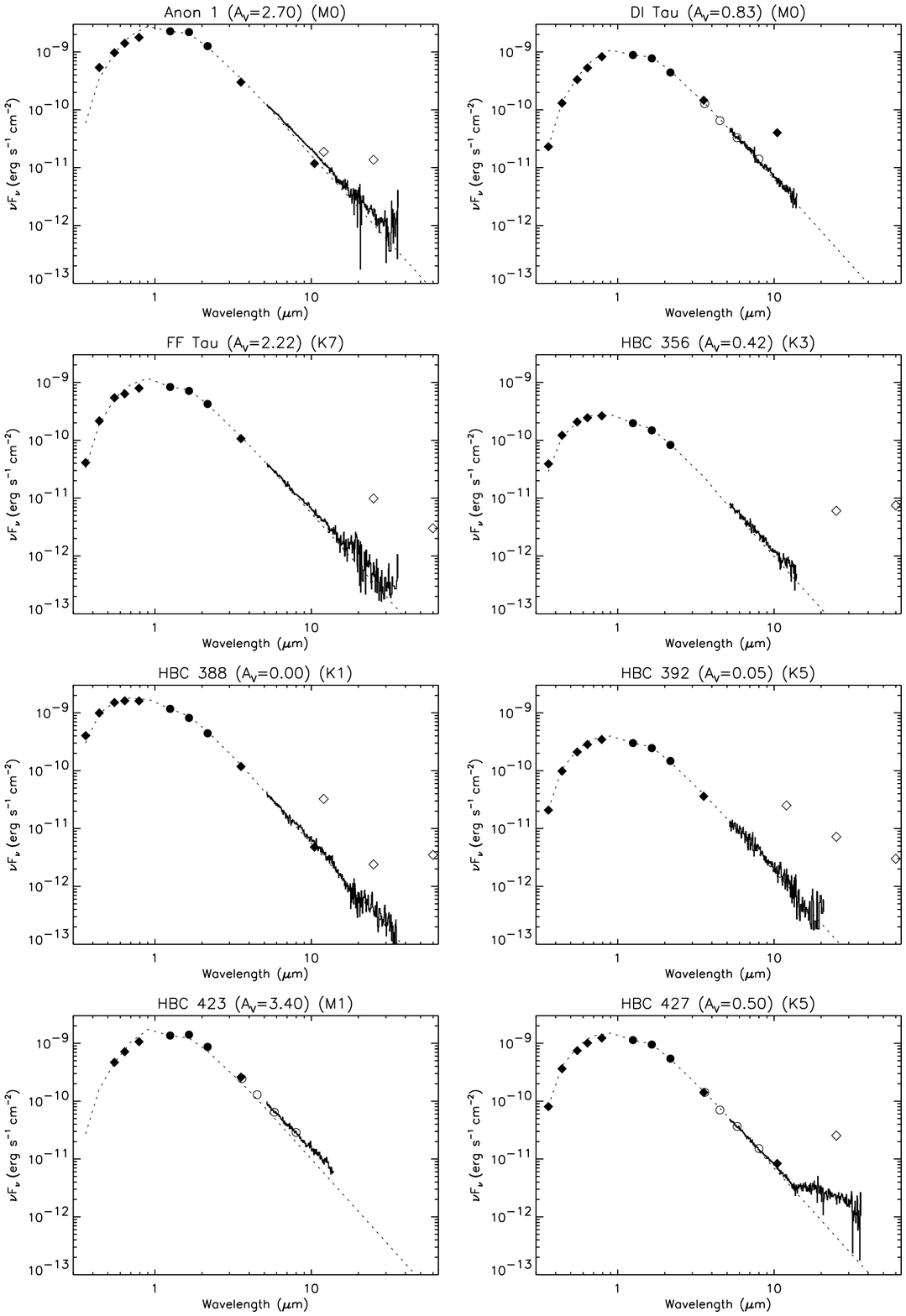}
\caption{SED plots of the Class III objects in our sample, ordered alphabetically by their 
name. All fluxes, including the IRS spectrum, were dereddened using Mathis's reddening 
law \citep{mathis90} and the extinctions listed in table \ref{tab3}. {\it See Figure 
\ref{fig_classII} legend for a description of the plotting symbols}. 
\label{fig_classIII}}
\end{figure}

\clearpage 

\begin{figure}
\epsscale{0.85}
\plotone{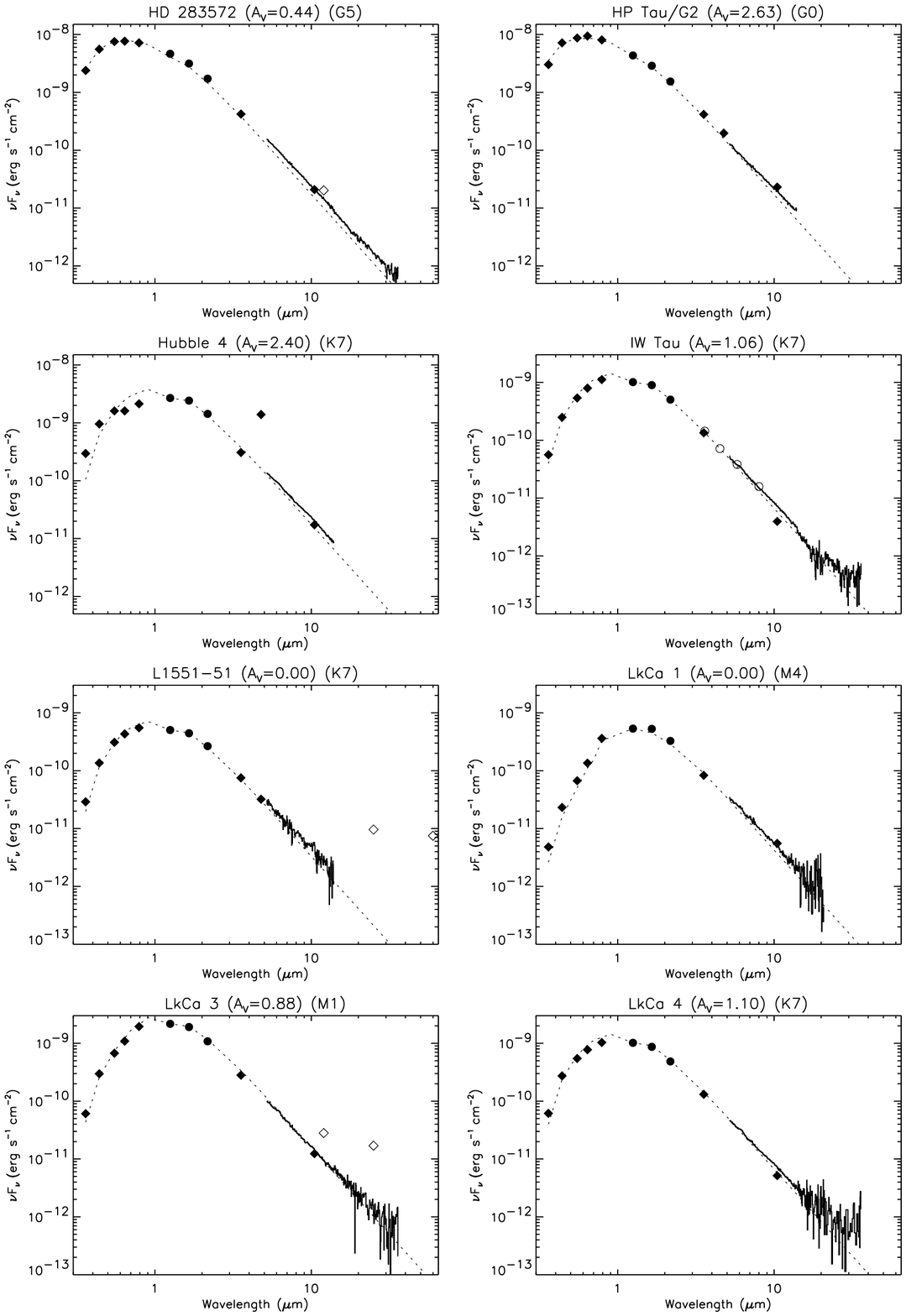}
\figurenum{\ref{fig_classIII}}\caption{continued.}
\end{figure}

\clearpage 

\begin{figure}
\epsscale{0.85}
\plotone{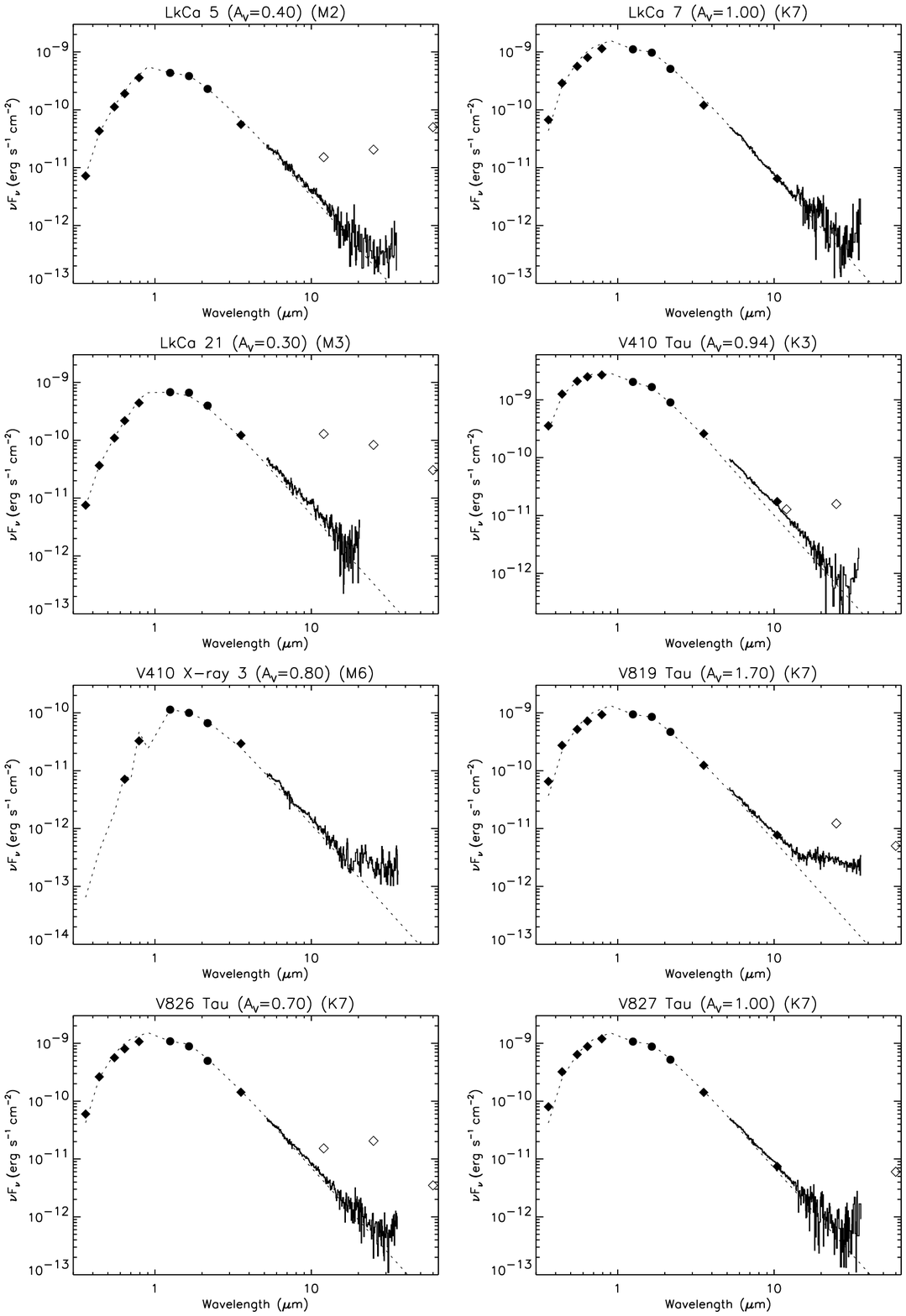}
\figurenum{\ref{fig_classIII}}\caption{continued.}
\end{figure}

\clearpage 

\begin{figure}
\epsscale{0.85}
\plotone{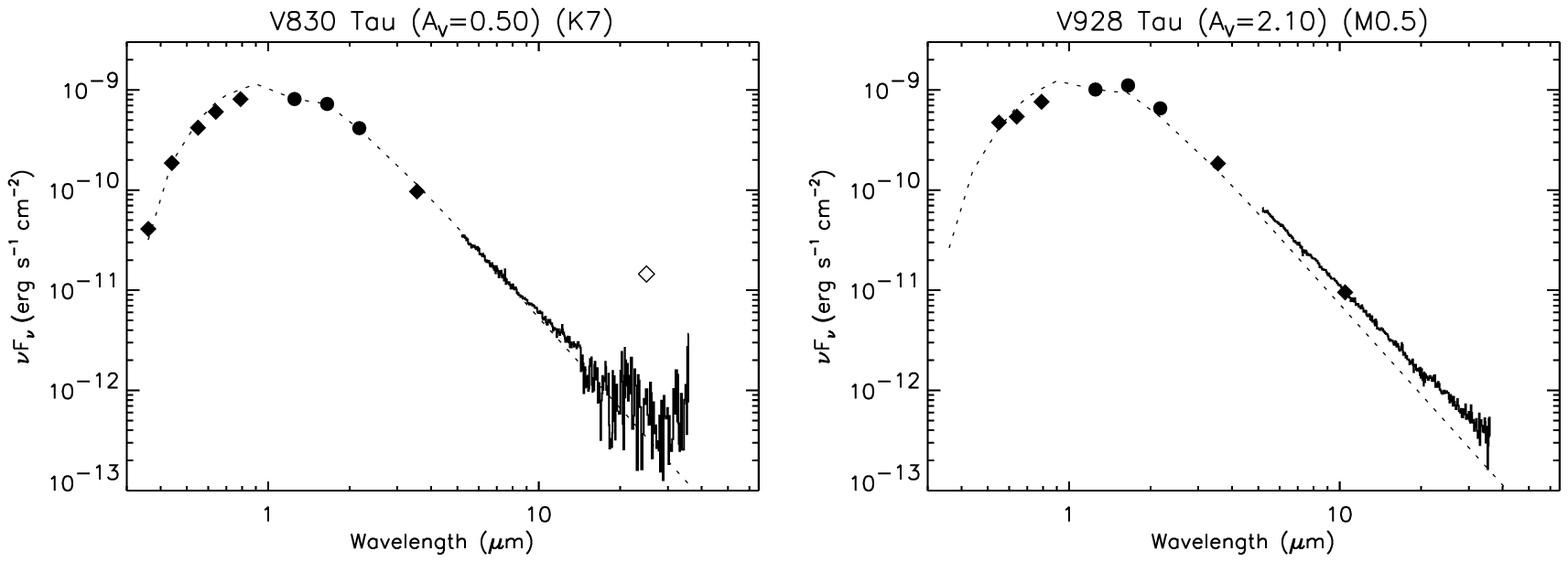}
\figurenum{\ref{fig_classIII}}\caption{continued.}
\end{figure}

\clearpage 

\clearpage 

\clearpage 

\clearpage 

\clearpage 

\clearpage 

\clearpage 

\clearpage

\clearpage

\clearpage

\clearpage 

\clearpage

\clearpage

\begin{deluxetable}{lcccccc}
\tabletypesize{\scriptsize}   
\tablewidth{505pt}
\tablecaption{Properties of Observed Class II Objects in Multiple Systems\label{tab1}}
\tablehead{
\colhead{Name} & \colhead{N} & 
\colhead{Separation [\arcsec]} &\colhead{Spectral} & 
\colhead{CTTS (C)/} & \colhead{Brightness}  &  \colhead{Ref.}  \\
 &  &  & \colhead{Type} & 
\colhead{WTTS (W)} & \colhead{Comparison} & \\
\colhead{(1)} & \colhead{(2)} & \colhead{(3)} & \colhead{(4)} & 
\colhead{(5)} & \colhead{(6)} & \colhead{(7)} 
}
\startdata
04370+2559 (A,B) & 2 & 4.3\arcsec\ & $\sim$ K3--M1 & C & A\,$\gg$\,B & 20,21 \\
04385+2550 (A,B) & 2 & 18.9\arcsec\ & M0  & C & A\,$\gg$\,B & 22,2 \\
CoKu Tau/3 (A,B) & 2 & 2.05\arcsec\ & M1 & C & A\,$>$\,B & 1 \\
CZ Tau (A,B)$^a$ & 2 & 0.32\arcsec\ & M3 & W & A $\sim$ B & 1 \\
DD Tau (A,B)$^a$ & 2 & 0.56\arcsec\ & M3+M3 & C & A $\sim$ B & 1 \\
DF Tau (A,B) & 2 & 0.09\arcsec\  & M0.5+M3 & C & A $\sim$ B & 1 \\
DH Tau$^b$ & 2 & 15\arcsec\ & M2+M2 & C & DH\,$>$\,DI & 7,8 \\
DK Tau (A,B) & 2 & 2.30\arcsec\ & K9+M1 & C & A\,$>$\,B & 1,7 \\
DQ Tau (A,B) & 2 & SB & K5 & C & A $\sim$ B & 12 \\
F04192+2647 (A,B) & 2 & 23.3\arcsec\ & \nodata & \nodata & A\,$>$\,B & 23  \\
F04297+2246 (A,B) & 2 & 6.6\arcsec\ & \nodata & \nodata & A\,$>$\,B & 23 \\
FM Tau$^c$ & 2 & 37.3\arcsec\ & M2 & C & FM\,$<$\,V773 & 7,8 \\
FO Tau (A,B) & 2 & 0.15\arcsec\ & M2+M2  & C &  A\,$\sim$\,B & 1 \\
FQ Tau (A,B) & 2 & 0.76\arcsec\ & M3+M3.5 & C &A\,$\sim$\,B & 1,17 \\
FS Tau (Aa,Ab,B) & 3 & 0.23\arcsec\ (Aa,Ab), 20\arcsec\ (A-B) & 
                             M1+M4 (Aa,Ab) & C & Aa $>$ Ab & 1,24 \\
FV Tau (A,B)$^d$ & 2 & 0.72\arcsec\ & K5+K6 & C & A\,$\sim$\,B, FV\,$>$\,FV/c & 1 \\
FX Tau (A,B) & 2 & 0.89\arcsec\ & M1+M4 & C+W & A\,$>$\,B & 1,6 \\
FZ Tau$^e$ &2 & 16.9\arcsec\ & M0+K5 & C & FZ $>$ FY & 7,8 \\
GG Tau (A,B)$^f$ & 4 & 10.3\arcsec\ & \nodata& C & A\,$\gg$\,B & 1 \\
GG Tau (Aa,Ab) & 2 & 0.25\arcsec\ &  K7+M0.5 & C & Aa\,$\gtrsim$\,Ab & 1 \\
GG Tau (Ba,Bb) & 2 & 1.48\arcsec\ & M5.5+M7.5 & C & Ba\,$>$\,Bb & 1 \\
GH Tau (A,B)$^g$ & 2 & 0.31\arcsec\ & M1.5+M2 & C & A\,$\sim$\,B & 1 \\
GI Tau$^h$ & 2 & 12.9\arcsec\ & K6 & C & GI\,$\sim$\,GK & 5,6 \\
GK Tau (A,B)$^h$ & 2 & 2.5\arcsec\ & K7 & C & A\,$\gg$\,B & 5,6 \\
GN Tau (A,B) & 2 & 0.33\arcsec\ & M2.5 & C & A\,$\sim$\,B & 1,26 \\
Haro 6-37 (Aa,Ab,B) & 3 & 2.62\arcsec (A,B), 0.33\arcsec (Aa,Ab) & 
                                K7+M1 & C & Aa\,$>$\,Ab, A\,$>$\,B & 1,11 \\
HK Tau (A,B) & 2 & 2.34\arcsec\ & M1+M2 & C & A\,$\gg$\,B & 1,7 \\
HN Tau (A,B) & 2 & 3.11\arcsec\ & K5+M4 & C & A\,$\gg$\,B & 1 \\
HP Tau (A,B)$^i$ & 2 & 0.017\arcsec\ & K2 & C & A\,$>$\,B & 7,8,15 \\
IS Tau (A,B) & 2 & 0.22\arcsec\ & K7+M4.5 & C+W & A\,$>$\,B & 1 \\
IT Tau (A,B) & 2 & 2.39\arcsec\ & K3+M4 & C & A\,$\gtrsim$\,B & 1,6 \\
RW Aur (A,B,C)$^j$ & 3 & 1.42\arcsec (A-BC), 0.12\arcsec (B-C) & 
                              K1+K5 (A,B) & C & A\,$>$\,B\,$\gg$\,C & 1,10 \\
T Tau (N,Sa,Sb) & 3 & 0.70\arcsec (N-S), 0.1\arcsec (Sa-Sb) & 
                              K0 & C & N\,$\sim$\,Sa\,$\sim$\,Sb & 1,3 \\ 
UX Tau (A,B,C) & 4 & 5.86\arcsec (A-B), 2.63\arcsec (A-C) & 
                                  K5+M2+M5 & C+W+W & A\,$>$\,B, A\,$\gg$\,C & 1 \\
UX Tau (Ba,Bb) & 2 & 0.138\arcsec & M2 & W & Ba\,$>$\,Bb & 11 \\
UY Aur (A,B) & 2 & 0.88\arcsec\ & M0+M2.5 & C & A\,$\gtrsim$\,B & 1,17 \\
UZ Tau (A,Ba,Bb) & 4 & SB (A), 3.54\arcsec (A-Ba), 0.37\arcsec (Ba-Bb) & 
                              M1+M2+M2 & C & A\,$>$\,B, Ba\,$\sim$\,Bb & 1,13 \\
V710 Tau (A,B) & 2 & 3.17\arcsec\ & M0.5+M2 & C+W & A $\sim$ B & 1 \\
V773 Tau (AB,C,D) & 4 & SB (AB), 0.12\arcsec (AB-C), 0.24\arcsec (AB-D)  & 
                              K2+M0 (AB,C) & W+C & D\,$>$\,C\,$>$\,AB &1,4 \\
V807 Tau (A,Ba,Bb)$^g$ & 3 & 0.30\arcsec (A-B), 0.04\arcsec (Ba-Bb)  & 
              K7+M3 & C+W & A\,$>$\,B, Ba\,$\sim$\,Bb  & 25,1 \\
V892 Tau (Aa,Ab,B) & 3 & 0.06\arcsec, 4.10\arcsec\ & B9+M2 & W &
                              Aa\,$\sim$\,Ab, A\,$\gg$\,B & 16,19 \\
V955 Tau (A,B) & 2 & 0.33\arcsec\ & K5+M1 & C & A\,$>$\,B & 1 \\
VY Tau (A,B) & 2 & 0.66\arcsec\ & M0 & W & A\,$>$\,B & 1 \\
XZ Tau (A,B) & 2 & 0.30\arcsec\ & M3+M1.5 & C & B\,$>$\,A & 1 \\
ZZ Tau IRS$^k$ & 2 & 35\arcsec\ & M4.5 & C & ZZ IRS\,$>$\,ZZ &  2 \\
ZZ Tau (A,B) & 2 & 0.04\arcsec\ & M3 & C & A\,$\gtrsim$\,B & 9 \\
\tableline
\enddata

\tablecomments{
Column (1) gives the name of the object, column (2) the number of components,
column (3) their separation in arcseconds (SB means ``spectroscopic binary''), 
column (4) the spectral type (if only one type is listed, it is usually the one of the 
brighter star in the system), column (5) lists whether the components are classical 
or weak-lined T Tauri stars (if only one identifier is given, it applies to all components), 
column (6) shows how the components compare in brightness, and column (7) 
gives the references for the data listed in the previous columns. \\ 
$^a$ CZ Tau (A,B) is 30{\farcs}1 from DD Tau (A,B). \\
$^b$ The other component in the binary is DI Tau, 
         which is itself a 0{\farcs}12 binary. \\
$^c$ The other component in the binary (if it is a real bound system) is V773 Tau, 
         a quadruple system. \\
$^d$ FV Tau/c (a 0{\farcs}7 binary) is 12{\farcs}3 from FV Tau. \\
$^e$ The other component in the binary is FY Tau. \\
$^f$ We only observed GG Tau Aa+Ab with the IRS. \\
$^g$ GH Tau and V807 Tau are separated by 21{\farcs}5. \\
$^h$ GI Tau and GK Tau are separated by 12{\farcs}9. \\
$^i$ HP Tau might form a triple systems with HP Tau/G2 (21{\farcs}4 away) 
         and HP Tau/G3 (17{\farcs}3 away).  \\
$^j$ RW Aur C could be a false detection \citep{white01}. \\
$^k$ ZZ Tau IRS is 35\arcsec\ from ZZ Tau. \\
}

\tablerefs{
(1) \citet{white01};
(2) \citet{white04};
(3) \citet{koresko00};
(4) \citet{duchene03};
(5) \citet{kenyon95};
(6) \citet{duchene99b};
(7) \citet{monin98};
(8) \citet{hartigan94};
(9) \citet{simon96};
(10) \citet{ghez93};
(11) \citet{duchene99a};
(12) \citet{mathieu97};
(13) \citet{mathieu96};
(14) \citet{white05};
(15) \citet{richichi94};
(16) \citet{leinert97};
(17) \citet{hartigan03};
(18) \citet{padgett99};
(19) \citet{smith05};
(20) \citet{itoh99};
(21) \citet{itoh02};
(22) \citet{duchene04};
(23) 2MASS All-Sky Catalog of Point Sources;
(24) \citet{mundt84};
(25) \citet{schaefer03};
(26) \citet{white03}
}
\end{deluxetable}

\clearpage

\LongTables
\begin{deluxetable}{lccccc}
\tabletypesize{\scriptsize}   
\tablecaption{Properties of Observed Class II Objects \label{tab2}}
\tablehead{
\colhead{Name} & \colhead{Adopted} & \colhead{A$_V$} & 
\colhead{CTTS (C)/} &  \colhead{Ref.} \\
 & \colhead{Spectral Type} &  & \colhead{WTTS (W)} & \\
\colhead{(1)} & \colhead{(2)} & \colhead{(3)} & \colhead{(4)} & \colhead{(5)} 
}
\startdata
 04108+2910 & M0 & 1.40$^a$ & C  & 1,3  \\
 04187+1927 & M0 & \nodata & C  &  1,4  \\
 04200+2759 & \nodata & \nodata & C & 1,5 \\
 04216+2603 & M1 & \nodata  & C & 1,3  \\
 04303+2240 & \nodata & 11.7 & C &  6 \\
 04370+2559 (A,B) & \nodata & 9.82 & C  & 7 \\
 04385+2550 (A,B) & M0 & 7.80$^a$ & C & 6,7  \\
 AA Tau & K7 & 1.75$^a$ & C & 1,3   \\
 AB Aur & A0 & 0.25 & C & 8,3  \\
 BP Tau  & K7 & 1.00$^a$ & C & 1,3  \\
 CI Tau  & K7 & 2.00$^a$ & C & 1,3  \\
 CoKu Tau/3 (A,B) & M1 & 5.00$^a$ & C & 1,2  \\
 CoKu Tau/4 & M1 & 3.0 & W & 9,3  \\
 CW Tau & K3 & 2.75$^a$ & C & 1,3  \\
 CX Tau & M0 & 1.30$^a$ & C &  1,3  \\
 CY Tau & K7 & 1.70$^a$ & C &  1,3  \\
 CZ Tau (A,B) & M1 & 2.45$^a$ & W & 1,3  \\
 DD Tau (A,B) & M3 & 1.00$^a$ & C & 1,2  \\
 DE Tau & M0 & 1.20$^a$ & C & 1,3 \\
 DF Tau (A,B) & M0 & 1.60$^a$ & C & 1,2  \\
 DG Tau & K6 & 1.60 & C & 1,10,3  \\
 DH Tau & M0 & 1.70$^a$ & C & 1,3 \\
 DK Tau (A,B) & M0 & 1.30$^a$ & C & 1,2  \\
 DL Tau & K7 & 1.50$^a$  & C & 1,2  \\
 DM Tau & M1 & 0.72$^a$ & C & 1,2  \\
 DN Tau & M0 & 0.60$^a$ & C & 1,3  \\
 DO Tau & M0 & 2.05$^a$ & C & 1,3 \\
 DP Tau & M0 & 0.60$^a$ & C & 1,3  \\
 DQ Tau (A,B) & M0 & 1.60$^a$ & C & 1,3  \\
 DR Tau & K7 & 1.20 & C & 1,3  \\
 DS Tau & K5 & 1.10$^a$ & C & 1,3  \\
 F04101+3103 & A1 & 1.90$^a$ & C & 1,4 \\
 F04147+2822 & M4 & 2.50$^a$ & C & 1,3  \\
 F04192+2647 A & \nodata & \nodata & \nodata  & \nodata \\
 F04262+2654 & \nodata & \nodata & \nodata & \nodata \\
 F04297+2246 A &\nodata & \nodata & \nodata & \nodata \\
 F04570+2520 & \nodata & \nodata & \nodata & \nodata \\
 FM Tau & M0 & 1.40$^a$ & C & 1,3  \\
 FN Tau & M5 & 1.35 & C & 1,3  \\
 FO Tau (A,B) & M2 & 3.03 & C & 2  \\
 FP Tau & M4 & 0.00 & C & 2 \\
 FQ Tau (A,B) & M2 & 1.87 & C & 1,3  \\
 FS Tau (Aa,Ab) & M1 & 1.43$^a$ & C & 1,2  \\
 FT Tau & C & \nodata & \nodata  & 1 \\
 FV Tau (A,B) & K5 & 5.33 & C & 2  \\
 FX Tau (A,B) & M1 & 2.00$^a$ & C & 1,3  \\ 
 FZ Tau & M0 & 3.70$^a$ & C & 1,3  \\
 GG Tau (Aa,Ab) & M0 & 1.00$^a$ & C & 1,2 \\
 GH Tau (A,B) & M2 & 0.97$^a$ & C & 1,2  \\
 GI Tau & K6 & 2.30$^a$ & C & 1,3  \\
 GK Tau (A,B) & M0 & 1.15$^a$ & C & 1,3  \\
 GM Aur & K3 & 1.21 & C & 2  \\
 GN Tau (A,B) & M2 & 3.50$^a$ & C & 11,12  \\
 GO Tau & M0 & 2.02$^a$ & C & 1,3  \\
 Haro 6-13 & M0 & 11.9 & C & 6  \\
 Haro 6-37 (Aa,Ab,B) & K7 & 3.80$^a$ & C & 1,2 \\
 HK Tau (A,B) & M1 & 2.70$^a$ & C & 6  \\
 HN Tau (A,B) & K5 & 1.50$^a$ & C & 1,2 \\
 HO Tau & M0 & 1.30$^a$ & C & 1,3  \\
 HP Tau (A,B) & K3 & 2.80$^a$ & C & 1,3  \\
 HQ Tau & \nodata & \nodata & \nodata & \nodata  \\
 IP Tau & M0 & 0.51 & C & 2  \\
 IQ Tau & M1 & 1.44 & W & 2  \\
 IS Tau (A,B) & M0 & 3.22$^a$ & C & 1,2  \\
 IT Tau (A,B) & K2 & 3.80$^a$ & C & 1 \\
 LkCa 15 & K5 & 1.20$^a$ & C & 1,3  \\
 MHO 3 & K7 & 8.30$^a$ & C & 1,13  \\
 RW Aur (A,B,C) & K3 & 0.50$^a$ & C & 1,2  \\
 RY Tau & G1 & 2.20 & C & 14,3  \\
 SU Aur & G1 & 0.90 & C & 14 \\
 T Tau (N,Sa,Sb) & K0 & 1.75$^a$ & C & 1,2  \\ 
 UX Tau (A,Ba,Bb,C) & K5 & 0.70$^a$ & C & 2  \\
 UY Aur (A,B) & K7 & 2.10$^a$ & C & 1,3  \\
 UZ Tau (A,Ba,Bb) & M1 & 1.00 & C & 2  \\
 V410 Anon 13 & M6 & 5.80 & C & 1,15,2 \\
 V710 Tau (A,B) & M1 & 1.90$^a$ & C & 1,2  \\
 V773 Tau (AB,C,D) & K3 & 2.00$^a$ & W & 1,2  \\
 V807 Tau (A,Ba,Bb) & K7 & 0.60$^a$ & C & 1,2  \\
 V836 Tau & K7 & 1.10$^a$ & C/W & 1,3  \\
 V892 Tau (Aa,Ab,B) & B9 & 8.00$^a$ & W & 16,3  \\
 V955 Tau (A,B) & K5 & 3.72 & C & 2  \\
 VY Tau (A,B) & M0 & 1.35$^a$ & W & 1,3  \\
 XZ Tau (A,B) & M2 & 2.90$^a$ & C & 1,2  \\
 ZZ Tau IRS & M5 & 1.50$^a$ & C & 1,6  \\
 ZZ Tau (A,B) & M3 & 1.44$^a$ & C & 1,3  \\
\tableline
\enddata

\tablecomments{
Column (1) gives the name of the object, column (2) the adopted spectral type,
column (3) the adopted visual extinction (see also note (a) below), 
column (4) lists whether the object (the primary in case of multiple systems) 
is a classical or weak-lined T Tauri star, and column (5) gives the references 
for the data listed in the previous columns. \\ 
$^a$ Value of $A_V$ derived from the observed V-I colors, spectral types, and intrinsic colors 
for main-sequence stars listed in \citet{kenyon95}, using the \citet{mathis90} reddening law.
In some cases the derived values of $A_V$ were adjusted to yield a good match between
the dereddened and true photospheric colors. }

\tablerefs{
(1) \citet{kenyon95};
(2) \citet{white01};
(3) \citet{kenyon98};
(4) \citet{kenyon90};
(5) \citet{kenyon94a};
(6) \citet{white04};
(7) \citet{itoh02};
(8) \citet{dewarf03};
(9) \citet{dalessio05};
(10) \citet{gullbring00};
(11) \citet{white03};
(12) \citet{luhman04};
(13) \citet{briceno98};
(14) \citet{calvet04};
(15) \citet{furlan05a};
(16) \citet{smith05}
}
\end{deluxetable}

\clearpage

\begin{deluxetable}{lccccc}
\tabletypesize{\scriptsize}   
\tablecaption{Properties of Observed Class III Objects \label{tab3}}
\tablehead{
\colhead{Name} & \colhead{Multiplicity$^a$} & \colhead{Adopted} &
\colhead{A$_V$} & \colhead{CTTS (C)/} & \colhead{Ref.}  \\
 & & \colhead{Spectral Type} & & \colhead{WTTS (W)} &   \\
\colhead{(1)} & \colhead{(2)} & \colhead{(3)} & \colhead{(4)} & 
\colhead{(5)} & \colhead{(6)} 
}
\startdata
 Anon 1 & s & M0 & 2.70$^b$ & W & 1,6 \\
 DI Tau$^c$ & 0.12\arcsec, 15\arcsec & M0 & 0.83$^b$ & W & 7,2,5 \\
 FF Tau (A,B) & 0.026\arcsec\ & K7 & 2.22 & W & 2,4,5 \\
 HBC 356$^d$ & 1.33\arcsec\ & K3 & 0.42$^b$ & W & 3,2 \\
 HBC 388 & s & K1 & 0.0 & W & 1 \\
 HBC 392 & s & K5 & 0.05$^b$ & W & 2,5 \\
 HBC 423$^e$ & 0.24\arcsec, 10.4\arcsec & M1 & 3.40$^b$ & W & 1,2 \\
 HBC 427 & SB ($\sim$ 0.03\arcsec) & K5 & 0.50$^b$ & W & 2,5,9 \\
 HD 283572 & s & G5 & 0.44$^b$ & W & 2,5 \\
 HP Tau/G2$^f$ & 9.9\arcsec\ & G0 & 2.63$^b$ & W & 2,4,5 \\
 Hubble 4 & s & K7 & 2.40$^b$ & W & 1 \\
 IW Tau (A,B) & 0.29\arcsec\ & K7 & 1.06$^b$ & W & 1 \\
 L1551-51 & s & K7 & 0.0 & W & 1,6 \\
 LkCa 1 & s & M4 & 0.0 & W & 1 \\
 LkCa 3 (A,B) & 0.48\arcsec\ & M1 & 0.88$^b$ & W & 2,1 \\
 LkCa 4 & s & K7 & 1.10$^b$ & W & 1 \\
 LkCa 5 & s & M2 & 0.40$^b$ & W & 1 \\
 LkCa 7 (A,B) & 1.02\arcsec\ & K7 & 1.00$^b$ & W & 1 \\
 LkCa 21 & s & M3 & 0.30$^b$ & W & 2,6 \\
 V410 Tau (A,B,C)$^g$ & 0.07\arcsec, 0.29\arcsec\ & K3 & 0.94$^b$ & W & 1 \\
 V410 X-ray 3 & s & M6 & 0.80$^b$ & W & 1 \\ 
 V819 Tau & 10.5\arcsec\ & K7 & 1.70$^b$ & W & 1,8 \\
 V826 Tau & SB & K7 & 0.70$^b$ & W & 8,5 \\
 V827 Tau & s & K7 & 1.00$^b$ & W & 1 \\
 V830 Tau & s & K7 & 0.50$^b$ & W & 1 \\
 V928 Tau (A,B) & 0.19\arcsec\  & M0.5 & 2.10$^b$ & W & 2,6 \\
\tableline
\enddata

\tablecomments{
Column (1) gives the name of the object, column (2) the multiplicity of the object
(see footnote a), column (3) the adopted spectral type, column (4) the adopted 
visual extinction (see also note (b) below), column (5) lists whether the object 
(the primary in case of multiple systems) is a classical or weak-lined T Tauri star, 
and column (6) gives the references for the data listed in the previous columns. \\ 
$^a$ The letter ``s'' means single star; for multiple systems, the separation between 
the components in arcseconds is listed; SB means ``spectroscopic binary''. \\
$^b$ Value of $A_V$ derived from the observed V-I colors, spectral types, and intrinsic colors 
for main-sequence stars listed in \citet{kenyon95}, using the \citet{mathis90} reddening law.
In some cases the derived values of $A_V$ were adjusted to yield a good match between
the dereddened and true photospheric colors. \\
$^c$ DI Tau is separated by 15\arcsec\ from DH Tau; DI Tau is a 0{\farcs}12
binary.\\
$^d$ The other component is HBC 357. \\
$^e$ HBC 423 is also knows as LkHa 332/G1. The separation between components A and 
B of HBC 423 is 0{\farcs}24; the separation between HBC 423 and V955 Tau is
10{\farcs}4. \\
$^f$ HP Tau/G2 is likely part of a triple system consisting of HP Tau, HP Tau/G2,
and HP Tau/G3; the separation listed is the distance between HP Tau/G2 and
HP Tau/G3 (also a WTTS). \\
$^g$ The value of 0{\farcs}07 is the separation between A and B, 0{\farcs}29 
between A and C. 
}

\tablerefs{
(1) \citet{white01}
(2) \citet{kenyon95}
(3) \citet{duchene99b}
(4) \citet{simon95a}
(5) \citet{kenyon98}
(6) \citet{martin94}
(7) \citet{ghez93}
(8) \citet{leinert93}
(9) \citet{steffen01}
}
\end{deluxetable}

\clearpage

\begin{deluxetable}{lcc|lcc}
\tabletypesize{\scriptsize}   
\tablewidth{380pt}
\tablecaption{Spectral Index $n_{6-25}$ and 10-$\mu$m feature strength
for the Class II Objects in our Sample  
\label{tab_sp_ind_10mic}}
\tablehead{
\colhead{Name} & \colhead{$n_{6-25}$} & 
\colhead{(F$_{10}-$F$_{\mathrm{cont}}$)/F$_{\mathrm{cont}}$} & 
\colhead{Name} & \colhead{$n_{6-25}$} & 
\colhead{(F$_{10}-$F$_{\mathrm{cont}}$)/F$_{\mathrm{cont}}$}
}
\startdata
04108+2910 & -0.90 & 0.01 & FS Tau & -0.15 & 0.26 \\
04187+1927 & -0.56 & 0.28 & FT Tau & -0.48 & 0.43  \\ 
04200+2759 & -0.32 & 0.44 & FV Tau & -0.62 & 0.17 \\
04216+2603 & -0.57 & 0.27 & FX Tau & -0.58 & 0.81 \\
04303+2240 & -0.70 & 0.21 & FZ Tau & -0.86 & 0.18 \\
04370+2559 & -0.24 & 0.61 & GG Tau & -0.47 & 0.64 \\
04385+2550 & 0.35 & 0.31 & GH Tau & -0.75 & 0.26 \\
AA Tau & -0.65 & 0.35 & GI Tau & -0.47 & 0.51 \\
AB Aur & -0.05 & 1.08 & GK Tau & -0.20 & 1.03 \\
BP Tau & -0.48 & 0.55 & GM Aur & 0.46 & 1.19 \\
CI Tau & -0.54 & 0.54 & GN Tau & -0.75 & 0.59 \\
CoKu Tau/3 & -0.95 & 0.41 & GO Tau & -0.37 & 0.40 \\
CoKu Tau/4 & 0.95 & 0.85 & Haro 6-13 & 0.15 & 0.35 \\
CW Tau & -0.83 & 0.16 & Haro 6-37 & -0.87 & 0.18 \\
CX Tau & -0.14 & 0.48 & HK Tau & 0.09 & 0.39 \\
CY Tau & -1.22 & 0.12 & HN Tau & -0.45 & 0.51 \\
CZ Tau & -0.01 & 0.53 & HO Tau & -0.66 & 0.58 \\
DD Tau & -0.64 & 0.21 & HP Tau & -0.07 & 0.48 \\
DE Tau & -0.44 & 0.43 & HQ Tau & -0.50 & 0.70 \\
DF Tau & -1.13 & 0.17 & IP Tau & -0.45 & 0.99 \\
DG Tau & 0.00 & 0.00 & IQ Tau & -1.12 & 0.35 \\
DH Tau & -0.67 & 0.70 & IS Tau & -0.85 & 0.44 \\
DK Tau & -0.72 & 0.83 & IT Tau & -1.05 & 0.14 \\
DL Tau & -0.67 & 0.09 & LkCa 15 & -0.36 & 1.47 \\
DM Tau & 0.72 & 0.96 & MHO 3 & 0.94 & 0.69 \\
DN Tau & -0.45 & 0.22 & RW Aur & -0.50 & 0.32 \\
DO Tau & -0.27 & 0.18 & RY Tau & 0.05 & 1.36 \\
DP Tau & -0.23 & 0.35 & SU Aur & 0.37 & 0.98 \\
DQ Tau & -0.38 & 0.14 & T Tau & -0.04 & -0.10 \\
DR Tau & -0.61 & 0.27 & UX Tau A & -0.08 & 0.15 \\
DS Tau & -0.90 & 0.55 & UY Aur & 0.08 & 0.40 \\
F04101+3103 & 0.75 & 1.03 & UZ Tau & -0.66 & 0.45 \\
F04147+2822 & -0.75 & 0.56 & V410 Anon13 & -0.61 & 0.20 \\
F04192+2647 & -0.67 & 0.28 & V710 Tau & -0.87 & 0.21 \\
F04262+2654 & -0.38 & 0.19 & V773 Tau & -0.76 & 0.09 \\
F04297+2246 A & -0.24 & 0.98 & V807 Tau & -0.91 & 0.16 \\
F04570+2520 & -1.04 & 0.25 & V836 Tau & -0.86 & 0.79 \\
FM Tau & -0.19 & 0.72 & V892 Tau & 1.28 & 0.65 \\
FN Tau & -0.04 & 0.39 & V955 Tau & -0.78 & 0.24 \\
FO Tau & -0.52 & 0.26 & VY Tau & -0.63 & 0.49 \\
FP Tau & -0.39 & 0.22 & XZ Tau & -0.23 & 0.09 \\
FQ Tau & -0.78 & 0.19 & ZZ Tau IRS & -0.02 & 0.31 \\
 & & & ZZ Tau & -0.94 & 0.26 \\
\tableline
\enddata

\tablecomments{
Spectral indices are defined as \begin{math} 
n \equiv d \log(\lambda  F_{\lambda})/d \log (\lambda). 
\end{math}
See text for the definition of the strength of the 10-$\mu$m feature, 
(F$_{10}-$F$_{cont}$)/F$_{cont}$.
}
\end{deluxetable}

\clearpage

\begin{deluxetable}{lcc|lcc}
\tabletypesize{\scriptsize}   
\tablewidth{300pt}
\tablecaption{Spectral Indices of the Class II and III Objects in our Sample  
\label{tab_sp_ind}}
\tablehead{
\colhead{Name} & \colhead{$n_{6-13}$} & \colhead{$n_{13-25}$} & 
\colhead{Name} & \colhead{$n_{6-13}$} & \colhead{$n_{13-25}$} 
}
\startdata
{\it Class II Objects} & & \\
    04108+2910  &  -1.17  &  -0.55  &  GN Tau  &  -0.89  &  -0.63  \\
   04187+1927  &  -0.49  &  -0.66  & GO Tau  &  -0.80  &   0.17  \\
   04200+2759  &  -0.47  &  -0.11  & Haro 6-13  &  -0.56  &   0.90  \\
   04216+2603  &  -0.75  &  -0.33  & Haro 6-37  &  -1.17  &  -0.53  \\
   04303+2240  &  -0.80  &  -0.74  & HK Tau  &  -0.72  &   1.12 \\
   04370+2559  &  -0.37  &  -0.22  & HN Tau  &  -0.70  &  -0.14  \\
   04385+2550  &   0.13  &   0.53  & HO Tau  &  -0.89  &  -0.37  \\
       AA Tau  &  -0.87  &  -0.37  & HP Tau  &  -0.43  &   0.37  \\
       AB Aur  &  -0.79  &   0.93  & HQ Tau  &  -0.76  &  -0.17  \\
       BP Tau  &  -0.67  &  -0.24  & IP Tau  &  -0.94  &   0.20 \\
       CI Tau  &  -0.95  &  -0.02  & IQ Tau  &  -1.30  &  -0.90  \\
    CoKu Tau/3  &  -1.04  &  -0.91  &  IS Tau  &  -0.87  &  -0.89  \\
    CoKu Tau/4  &  -0.60  &   2.98  & IT Tau  &  -1.38  &  -0.67  \\
       CW Tau  &  -1.26  &  -0.30  & LkCa 15  &  -1.30  &   0.87  \\
       CX Tau  &  -0.37  &   0.14  & MHO 3  &   0.72  &   1.10  \\
       CY Tau  &  -1.45  &  -0.95  & RW Aur  &  -0.67  &  -0.27 \\
       CZ Tau  &   0.42  &  -0.61  &  RY Tau  &  -0.16  &   0.30  \\
       DD Tau  &  -0.73  &  -0.53  & SU Aur  &  -0.32  &   1.26  \\
       DE Tau  &  -0.87  &   0.12  & T Tau  &  -0.56  &   0.63  \\
       DF Tau  &  -1.36  &  -0.85  & UX Tau A  &  -1.99  &   2.45  \\
       DG Tau  &  -0.20  &   0.24  & UY Aur  &  -0.01  &   0.17  \\
       DH Tau  &  -1.69  &   0.66  & UZ Tau  &  -0.92  &  -0.34  \\
       DK Tau  &  -1.00  &  -0.37  & V410 Anon 13  &  -0.91  &  -0.31  \\
       DL Tau  &  -0.78  &  -0.54  & V710 Tau  &  -1.07  &  -0.63 \\
       DM Tau  &  -0.11  &   1.82  & V773 Tau  &  -1.10  &  -0.34   \\
       DN Tau  &  -0.60  &  -0.27  & V807 Tau  &  -1.61  &   0.01   \\
       DO Tau  &  -0.56  &   0.08  & V836 Tau  &  -1.37  &  -0.20  \\
       DP Tau  &  -0.43  &   0.02  & V892 Tau  &   1.75  &   0.53  \\
       DQ Tau  &  -0.47  &  -0.28  & V955 Tau  &  -0.89  &  -0.68  \\
       DR Tau  &  -0.88  &  -0.27  & VY Tau  &  -1.11  &  -0.02  \\
       DS Tau  &  -1.03  &  -0.73  & XZ Tau  &  -0.35  &  -0.10   \\
  F04101+3103  &   0.97  &   0.43  & ZZ Tau IRS  &  -0.42  &   0.49  \\
  F04147+2822  &  -0.83  &  -0.70  &  ZZ Tau  &  -1.15  &  -0.68  \\
  F04192+2647  &  -1.10  &  -0.10  &   \\
  F04262+2654  &  -0.89  &   0.30  & {\it Class III Objects} & &   \\
 F04297+2246 A  &  -0.41  &  -0.01  & Anon 1  &  -2.78  &  -2.71   \\
  F04570+2520  &  -0.79  &  -1.37  &  FF Tau  &  -2.76  &  -2.70   \\
       FM Tau  &  -0.45  &   0.14  &  HBC 388  &  -2.78  &  -3.02 \\
       FN Tau  &  -0.29  &   0.27  & HBC 427  &  -2.85  &  -0.70  \\
       FO Tau  &  -0.82  &  -0.18  & HD 283572  &  -2.85  &  -2.95   \\
       FP Tau  &  -0.84  &   0.21  & IW Tau  &  -2.82  &  -2.69   \\
       FQ Tau  &  -1.12  &  -0.36  & LkCa 3 &  -2.81  &  -2.65    \\
       FS Tau  &  -0.56  &   0.37  & LkCa 4  &  -2.74  &  -2.72    \\
       FT Tau  &  -0.62  &  -0.31  & LkCa 5  &  -2.77  &  -3.06    \\
       FV Tau  &  -0.83  &  -0.42  & LkCa 7  &  -2.94  &  -2.46   \\
       FX Tau  &  -1.05  &   0.02   & V410 Tau  &  -2.78  &  -5.19  \\
       FZ Tau  &  -1.14  &  -0.56   & V410 Xray 3  &  -2.89  &  -1.61  \\
       GG Tau  &  -0.79  &  -0.05  & V819 Tau  &  -2.57  &  -0.67  \\
       GH Tau  &  -1.26  &  -0.10  & V826 Tau  &  -2.80  &  -2.68  \\
       GI Tau  &  -0.55  &  -0.40  & V827 Tau  &  -2.77  &  -3.13   \\
       GK Tau  &  -0.45  &   0.12  & V830 Tau  &  -2.65  &  -2.99    \\
       GM Aur  &  -0.91  &   2.28  & V928 Tau  &  -2.76  &  -2.72   \\
\tableline
\enddata

\tablecomments{
Spectral indices are defined as \begin{math} 
n \equiv d \log(\lambda  F_{\lambda})/d \log (\lambda).
\end{math}
}
\end{deluxetable}

\clearpage

\begin{deluxetable}{cccc}
\tabletypesize{\scriptsize}   
\tablewidth{260pt}
\tablecaption{IRS Taurus Median (K5-M2 spectral types) \label{tab_median}}
\tablehead{
\colhead{Wavelength} & \colhead{Median} & 
\colhead{Lower Quartile} & \colhead{Upper Quartile} \\
\colhead{[$\mu$m]} & \colhead{[$\log({\nu} F_{\nu}$)]$^a$} &
\colhead{[$\log({\nu} F_{\nu}$)]$^a$} & 
\colhead{[$\log({\nu} F_{\nu}$)]$^a$} 
}
\startdata
 1.25$^b$  &  -9.039  &  -9.092  &  -9.013  \\
 1.65$^b$  &  -9.035  &  -9.035  &  -9.035  \\
 2.17$^b$  &  -9.163  &  -9.195  &  -9.114  \\
 5.70  &  -9.764  &  -9.879  &  -9.614  \\
 7.10  &  -9.873  & -10.060  &  -9.735  \\
 8.00  &  -9.905  & -10.081  &  -9.712  \\
 9.20  &  -9.789  &  -9.967  &  -9.552  \\
 9.80  &  -9.793  &  -9.942  &  -9.519  \\
11.30  &  -9.902  & -10.060  &  -9.615  \\
12.30  & -10.034  & -10.234  &  -9.763  \\
13.25  & -10.060  & -10.304  &  -9.787  \\
16.25  & -10.066  & -10.237  &  -9.777  \\
18.00  & -10.030  & -10.182  &  -9.725  \\
21.00  & -10.037  & -10.204  &  -9.722  \\
25.00  & -10.093  & -10.278  &  -9.778  \\
30.00  & -10.185  & -10.404  &  -9.839  \\
34.00  & -10.229  & -10.461  &  -9.865  \\
\tableline
\enddata

\tablecomments{
$^a$ Units of median, lower and upper quartiles are ergs s$^{-1}$ cm$^{-2}$. \\
$^b$ The median at these 3 wavelengths was constructed from reddening-corrected
2MASS data.
}
\end{deluxetable}

\end{document}